\documentclass[11pt,twoside,a4paper,cmspaper,final,collab]{cms-tdr}
\def\svnVersion{b5895a3-D}\def\svnDate{2026/08/12}\def\cmsCernNoTag{CERN-EP-2026-212}\def\cmsCernDate{\today}\def\cmsMessage{Submitted to the Journal of High Energy Physics}
\begin{document}\cmsNoteHeader{BTV-25-002}

\newcommand{\mhh}{\ensuremath{m_{\PH\PH}}\xspace}
\newcommand{\hhbb}{\ensuremath{{\PH\PH \to 4\PQb}}\xspace}
\newcommand{\hhbbtt}{\ensuremath{{\PH\PH \to \PQb\PAQb \PGtm\PGtp}}\xspace}
\newcommand{\hbb}{\ensuremath{{\PH \to \PQb\PAQb}}\xspace}
\newcommand{\hcc}{\ensuremath{{\PH \to \PQc\PAQc}}\xspace}
\newcommand{\htt}{\ensuremath{{\PH \to \PGtm\PGtp}}\xspace}
\newcommand{\zbb}{\ensuremath{{\PZ \to \PQb\PAQb}}\xspace}
\newcommand{\zcc}{\ensuremath{{\PZ \to \PQc\PAQc}}\xspace}
\newcommand{\ztt}{\ensuremath{{\PZ \to \PGtm\PGtp}}\xspace}
\newcommand{\probb}{\ensuremath{\mathcal{P}_{\PQb}}\xspace}
\newcommand{\probc}{\ensuremath{\mathcal{P}_{\PQc}}\xspace}
\newcommand{\probuds}{\ensuremath{\mathcal{P}_{\PQq}}\xspace}
\newcommand{\probg}{\ensuremath{\mathcal{P}_{\Pg}}\xspace}
\newcommand{\probbtr}{\ensuremath{\mathcal{P}_{\PQb}^{\mathrm{Tr}}}\xspace}
\newcommand{\probctr}{\ensuremath{\mathcal{P}_{\PQc}^{\mathrm{Tr}}}\xspace}
\newcommand{\deepjet}{\textsc{DeepJet}\xspace}
\newcommand{\deepcsv}{\textsc{DeepCSV}\xspace}
\newcommand{\pnet}{\textsc{PNet}\xspace}
\newcommand{\DoubleB}{\textsc{Double-B}\xspace}
\newcommand{\probXbb}{\ensuremath{\mathcal{P}_{\mathrm{X\to\PQb\PAQb}}}\xspace}
\newcommand{\probXbbtr}{\ensuremath{\mathcal{P}_{\mathrm{X\to\PQb\PAQb}}^{\mathrm{Tr}}}\xspace}
\newcommand{\ptH}{\ensuremath{\pt^{\PH}}\xspace}
\newcommand{\Vjets}{\ensuremath{\PV\text{+jets}}\xspace}
\newcommand{\mSD}{\ensuremath{m_{\mathrm{SD}}}\xspace}
\newcommand{\mX}{\ensuremath{m_{\mathrm{X}}}\xspace}
\newcommand{\Linst}{\ensuremath{\mathcal{L}_{\text{inst}}}\xspace}
\newcommand{\ttH}{\ensuremath{\ttbar\PH}\xspace}
\newcommand{\ttHbb}{\ensuremath{\ttbar (\PH \to \PQb\PAQb)}\xspace}
\newcommand{\ttHcc}{\ensuremath{\ttbar (\PH \to \PQc\PAQc)}\xspace}
\newcommand{\xbb}{\ensuremath{{\PX \to \PQb\PAQb}}\xspace}
\providecommand{\PY}{\HepParticle{Y}{}{}\xspace}
\newcommand{\ybb}{\ensuremath{{\PY \to \PQb\PAQb}}\xspace}
\newcommand{\bb}{\ensuremath{\PQb\PAQb}\xspace}
\newcommand{\xcc}{\ensuremath{{\PX \to \PQc\PAQc}}\xspace}
\newcommand{\xqq}{\ensuremath{{\PX \to \PQq\PAQq}}\xspace}
\newcommand{\xgg}{\ensuremath{{\PX \to \Pg\Pg}}\xspace}
\newcommand{\xtt}{\ensuremath{{\PX \to \PGtm\PGtp}}\xspace}
\newcommand{\mjj}{\ensuremath{m_{\mathrm{jj}}}\xspace}
\newcommand{\detajj}{\ensuremath{\Delta\eta_{\mathrm{jj}}}\xspace}
\newcommand{\hy}{\ensuremath{\PH\PY}\xspace}
\newcommand{\hh}{\ensuremath{\PH\PH}\xspace}
\providecommand{\fbinvns}{\ensuremath{\text{fb}^{-1}}\xspace}
\providecommand{\cmsTable}[1]{\resizebox{\textwidth}{!}{#1}}

\ifthenelse{\boolean{cms@external}}{\providecommand{\cmsLeft}{upper\xspace}}{\providecommand{\cmsLeft}{left\xspace}}
\ifthenelse{\boolean{cms@external}}{\providecommand{\cmsRight}{lower\xspace}}{\providecommand{\cmsRight}{right\xspace}}

\cmsNoteHeader{BTV-25-002}
\title{Performance of heavy-flavour jet identification in the CMS high-level trigger in proton-proton collisions at \texorpdfstring{$\sqrt{s}=13.6\TeV$}{sqrt(s) = 13.6 TeV}}
\date{\today}

\abstract{The CMS trigger system plays a crucial role during data taking, reducing the large collision rate delivered by the LHC to a few kHz for data storage and subsequent off\-line analysis. The system aims to maintain a high selection efficiency for a wide range of processes, including those involving jets originating from heavy-flavour quarks (\PQb and \PQc), which provide a distinctive signature in many physics analyses. To achieve this while maintaining a sustainable trigger output rate, dedicated  jet flavour identification methods are developed and optimized for use in the high-level trigger (HLT). This paper presents the design, commissioning, and performance of deep-learning-based jet identification algorithms deployed in the HLT during 2022--2024, for proton-proton collisions at ${\sqrt{s}=13.6\TeV}$. The new algorithms enabled significant improvements in signal efficiency for a variety of key physics processes, including the non-resonant production of Higgs boson pairs decaying to four \PQb quarks, as well as Higgs boson production via both vector boson fusion and in association with a \ttbar pair, in the \hbb and \hcc decay channels.}

\hypersetup{%
pdfauthor={CMS Collaboration},%
pdftitle={Performance of heavy-flavour jet identification in the CMS high-level trigger in proton-proton collisions at sqrt(s) = 13.6 TeV},%
pdfsubject={CMS},%
pdfkeywords={CMS, BTV, HLT, JME, TSG, Higgs}
}

\maketitle 

\section{Introduction}\label{sec:introduction}

The identification of jets originating from heavy quarks (\PQb and \PQc) is a cornerstone of the CMS physics programme~\cite{CMS:2008xjf}, playing a key role in both searches for new phenomena and precision measurements of standard model (SM) processes. The efficient selection of events containing jets originating from \PQb or \PQc quarks (heavy-flavour jets) is particularly important when dealing with fully hadronic final states, where no distinctive experimental signatures such as high transverse momentum (\pt) leptons or large missing transverse momentum are available to isolate signal events.

Several important analyses probe such final states, including searches for Higgs boson pair production with decay in the 4\PQb channel~\cite{CMS:2022cpr,CMS:2022gjd}, measurements of Higgs boson production via vector boson fusion (VBF) and in association with a \ttbar pair using \hbb decays~\cite{CMS:2023tfj,CMS:2024ddc,CMS:2024fdo}, analyses probing the Yukawa coupling between the Higgs boson and \PQc quarks through \hcc decays~\cite{CMS:2025dsh}, searches for heavy resonances (\eg ${\PX\to\hh}$ or ${\PX \to \hy}$ with \hbb and \ybb decays)~\cite{CMS:2017aza,CMS:2024phk}, and measurements of fully hadronic \ttbar production cross sections~\cite{CMS:2020tvq}. These analyses are experimentally challenging due to the large cross section at hadron colliders of events with multiple jets produced by the strong interaction, referred to as quantum chromodynamics (QCD) multijet events. A key handle to suppress this often dominant QCD multijet background is the identification of jets from \PQb and \PQc quark fragmentation, referred to as \PQb and \PQc jets, which in the case of \PQb jets are abundant in the signal processes but are relatively rare in QCD multijet events. Similarly, for many other measurements involving, \eg top quarks, the identification of \PQb jets is an important tool for suppressing background contributions while retaining a high signal efficiency.

\begin{figure*}[!htb]
  \centering
  \includegraphics[width=0.55\textwidth]{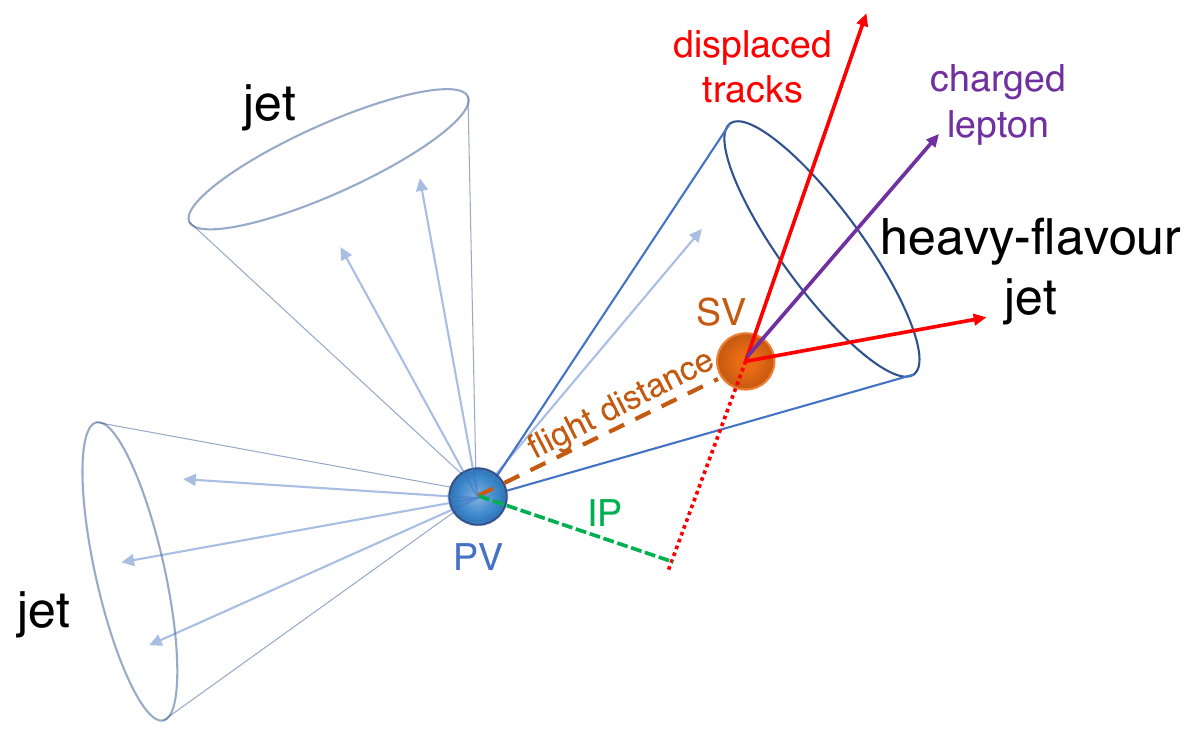}
  \caption{Schematic diagram of a collision event with two light jets and one \PQb jet. The finite and large lifetime of heavy-flavour hadrons (in particular \PQb hadrons) leads to a displaced SV as well as tracks or leptons with large impact parameters. In contrast, light jets originate from the PV and contain predominantly prompt tracks.}
  \label{fig:bhadron_secondary_vertex}
\end{figure*}

The discrimination of \PQb jets from those initiated by light quarks or gluons (light jets) relies on both jet fragmentation structure and the distinctive properties of \PQb hadron decays. In particular, the relatively long lifetime of \PQb hadrons (${\approx}1.5\unit{ps}$)~\cite{ParticleDataGroup:2024cfk} leads to displaced secondary vertices (SVs), located a few millimetres away from the primary interaction point (referred to as the primary vertex, PV). Charged particles produced by these decays exhibit large transverse impact parameters (IPs), defined as the distance of closest approach to the PV in the transverse plane. In addition, semileptonic \PQb hadron decays, with a branching fraction of about 20\%~\cite{ParticleDataGroup:2024cfk}, produce electrons or muons within the jet. The presence of such displaced leptons provides an additional handle for \PQb jet identification (\PQb tagging). A schematic example of these features is shown in Fig.~\ref{fig:bhadron_secondary_vertex}, taken from Ref.~\cite{CMS:2017wtu}.

The identification of jets originating from \PQc quarks (\PQc tagging) relies on similar principles, but is experimentally more challenging. Compared to \PQb hadrons, \PQc hadrons generally have  shorter lifetimes, with a typical value of about 0.5\unit{ps}~\cite{ParticleDataGroup:2024cfk}, and lower masses, producing SVs that are less displaced from the PV and decay tracks with smaller IPs. In addition, the decay products of \PQc hadrons typically carry lower \pt, making them more difficult to distinguish from light jets. Nevertheless, efficient \PQc tagging plays an important role in several analyses, including measurements of Higgs boson decays to charm quarks~\cite{CMS:2022psv,CMS:2025dsh} and searches for physics beyond the SM involving charm-enriched final states. This has recently become feasible at the trigger level with the development of modern machine-learning-based tagging algorithms.

The efficient identification of \PQb jets in the CMS trigger system, which reduces the CERN LHC 40\unit{MHz} proton-proton ($\Pp\Pp$) collision rate to a few~\unit{kHz}, is essential to retain high efficiency for hadronic signal processes while keeping the trigger output within resource constraints. This task relies on precise tracking to reconstruct SVs and measure track IPs relative to the PV, which is particularly challenging at the trigger level because of limited computing resources and time available in the online trigger farm.

During LHC Run~2 (2015--2018, ${\sqrt{s}=13\TeV}$), the first trigger algorithms for fully hadronic signatures based on deep learning were deployed~\cite{CMS:2024aqx}. These relied on the \textsc{DeepCSV} classifier~\cite{CMS:2017wtu}, which combined information from track impact parameters, SVs, and jet kinematics to identify \PQb jets.

During Run~3 (2022--2026), the LHC delivered $\Pp\Pp$ collisions at ${\sqrt{s}=13.6\TeV}$, and the increased instantaneous luminosity and higher in-time pileup (additional $\Pp\Pp$ interactions occurring in the same bunch or adjacent crossing as the interaction of interest) made these methods insufficient to maintain high signal efficiency at an acceptable trigger rate for key hadronic processes in the CMS physics programme. To address this, a new generation of trigger algorithms was developed to optimize the balance between the signal efficiency and output rate, while respecting the timing and bandwidth constraints of the CMS trigger system. The new triggers perform jet flavour identification using an algorithm based on dynamic graph convolutional neural network (DGCNN), \textsc{ParticleNet} (\pnet)~\cite{Qu:2019gqs}, tailored for use at the trigger level. Compared to previous approaches, \pnet significantly reduces the light-jet misidentification rate at the trigger level, while maintaining high \PQb tagging efficiency. In addition, it provides excellent performance in the identification of \PQc jets, enabling, for the first time in the CMS trigger menu, the implementation of dedicated \PQc jet triggers targeting processes such as VBF Higgs boson production with \hcc decays.

In addition, new trigger algorithms aiming to identify high-\pt Higgs boson decays to bottom quarks (\hbb) were also developed, tailored to a topology in which the fragmentation products of the two \PQb quarks are reconstructed as a single large-radius jet. A \pnet-based tagger was specifically developed to identify highly energetic heavy particles decaying to \bb (\xbb), enabling general-purpose triggers targeting signatures such as merged high-\pt \hbb topologies.

This paper presents the performance and commissioning with ${\sqrt{s}=13.6\TeV}$ data of \PQb and \PQc jet triggers developed for Run~3, including those targeting merged \hbb decays. Improvements in trigger acceptance are discussed for key analyses such as \hhbb, \hbb and \hcc production via VBF, and $\ttbar$-associated Higgs boson production in the ${\ttbar\PH(\hbb)}$ and ${\ttbar\PH(\hcc)}$ channels. The majority of the triggers described in this paper were recently adopted in the search for the \hhbb process with 61\fbinv of Run~3 $\Pp\Pp$ collision data~\cite{CMS:2026znu}, contributing significantly to an improvement in the search sensitivity compared with previously published results. Tabulated results from this paper are provided in a corresponding HEPData record~\cite{hepdata}.

\section{The CMS detector}
\label{sec:detector}

The CMS apparatus~\cite{CMS:2008xjf,CMS:2023gfb} is a multipurpose, nearly hermetic detector, designed to trigger on~\cite{CMS:2020cmk,CMS:2016ngn,CMS:2024aqx} and identify electrons, muons, photons, and (charged and neutral) hadrons~\cite{CMS:2020uim,CMS:2018rym,CMS:2014pgm}. Its central feature is a superconducting solenoid of 6\unit{m} internal diameter, providing a magnetic field of 3.8\unit{T}. Within the solenoid volume are a silicon pixel and strip tracker, a lead tungstate crystal electromagnetic calorimeter (ECAL), and a brass and scintillator hadron calorimeter (HCAL), each composed of a barrel and two endcap sections. Forward calorimeters extend the pseudorapidity  ($\eta$) coverage provided by the barrel and endcap detectors. Muons are reconstructed using gas-ionization detectors embedded in the steel flux-return yoke outside the solenoid. More detailed descriptions of the CMS detector, together with a definition of the coordinate system used and the relevant kinematic variables, can be found in Refs.~\cite{CMS:2008xjf,CMS:2023gfb}.

\section{Data sets, simulated samples, and offline reconstruction}\label{sec:dataset}

The results presented in this paper are based on $\Pp\Pp$ collision data at ${\sqrt{s}=13.6\TeV}$, collected with the CMS detector during 2022--2024. The performance of \PQb, \PQc, and \xbb tagging triggers, along with that of the corresponding jet flavour identification efficiencies, was continuously monitored online to promptly identify problems and improve the overall data quality.

During Run~3, the peak instantaneous luminosity and the maximum in-time pileup~\cite{CMS:2018elu} increased by about a factor of two with respect to Run~2, reaching up to ${\approx}80$ simultaneous interactions per bunch crossing. A summary of the Run~3 data-taking conditions is reported in Table~\ref{tab:run3_summary}, indicating the average pileup, and the total delivered and validated integrated luminosity, with all sub-detectors of the CMS experiment operating satisfactorily, measured by CMS using a combination of different subdetectors~\cite{CMS:2021xjt}. All results presented in this paper use validated data set.

\begin{table*}[!htb]
    \topcaption{LHC operating parameters during 2022--2024. Pileup is computed from physics fills with the nominal filling scheme, assuming a total inelastic cross section of 80~\unit{mb}. The reported integrated luminosities correspond to both the LHC delivered and CMS-validated values for physics analyses.}
    \centering
    \begin{tabular}{l c c c} 
    &  2022 & 2023 & 2024 \\
    \hline
    Average pileup & 46 & 52 & 57 \\
    Integrated luminosity delivered by LHC (\fbinvns) & 41.5 &  32.7  & 122 \\
    Integrated luminosity validated by CMS (\fbinvns) & 34.7 &  27.9  & 109 \\ 
  \end{tabular}
  \label{tab:run3_summary}
\end{table*}

Simulated samples are used for the training of the \pnet taggers and for their performance studies. The QCD multijet production is simulated at leading order (LO) using version 2.9.13 of \MGvATNLO~\cite{Alwall:2014hca}, including matrix elements with up to four outgoing partons. The production of top quark pair (\ttbar) events is simulated at next-to-LO (NLO) in QCD with \POWHEG~v2.0~\cite{Nason:2004rx,Frixione:2007vw,Alioli:2010xd,Jezo:2016ujg}, and normalized to the cross section calculated at next-to-NLO (NNLO) precision using {{\textsc{Top++} v2.0}}~\cite{Czakon:2011xx}. The differential \ttbar cross section as a function of the top quark \pt is further corrected to the NNLO QCD and NLO electroweak (EW) precision~\cite{Czakon:2017wor}. Single top quark and \Vjets ($\PV=\PZ,\PW$) processes are generated at NLO in QCD with \POWHEG~v2.0 and \MGvATNLO~v2.9.13, respectively. For \Vjets events, matrix elements with up to two additional outgoing partons are considered. Higgs boson production via gluon fusion, VBF, in association with a vector boson ($\PV\PH$), and $\ttbar\PH$ are simulated at NLO in QCD using \POWHEG~v2.0~\cite{Bagnaschi:2011tu,Nason:2009ai,Luisoni:2013cuh,Hartanto:2015uka}. Finally, the non-resonant Higgs boson pair production in the \hhbb and \hhbbtt channels is simulated at NLO in QCD using \POWHEG~v2.0~\cite{Bagnaschi:2011tu,Heinrich:2017kxx,Jones:2017giv,Buchalla:2018yce,Heinrich:2019bkc,Heinrich:2020ckp}, and is reweighted as a function of \mhh according to Ref.~\cite{Davies:2019dfy}. Higgs boson decays to \PQb or \PQc quarks, and to \PGt leptons, are simulated with \PYTHIA~v8.309~\cite{Sjostrand:2014zea}. For processes simulated at NLO in QCD with \MGvATNLO, events from the matrix element (ME) calculations with different parton multiplicities are merged with those from the parton shower using the FxFx prescription~\cite{Frederix:2012ps}.

For the training of the \pnet \xbb tagger, signal samples are obtained by simulating the decay of a Kaluza--Klein graviton (G), predicted in the Randall--Sundrum model~\cite{Randall:1999ee,Randall:1999vf}, into a pair of scalar heavy resonances (X). These simulated events are generated at LO using \PYTHIA v8.309. The graviton mass ranges between 500\GeV and 4.5\TeV, while the mass of the scalar bosons (\mX) is randomly sampled in the 30-200\GeV range in steps of 2\GeV. This strategy enables the development of a mass-decorrelated tagger over a wide kinematic range~\cite{CMS-DP-2020-002}. Finally, the scalar boson is decayed with equal branching fractions in the \xbb, \xcc, \xqq, and \xtt channels.

Parton showering, hadronization, and the underlying event are simulated with \PYTHIA~v8.309 using the CP5 tune~\cite{CMS:2019csb}. The simulated samples are generated using either the NNPDF~3.0 or 3.1 parton distribution functions (PDFs)~\cite{Ball:2014uwa,Ball:2017lmj}, depending on the data-taking period for which they are produced. The CMS detector response is modelled with \GEANTfour~\cite{GEANT4:2002zbu}, and simulated minimum bias interactions are overlaid on the hard scattering to reproduce the additional $\Pp\Pp$ collisions occurring within the same or nearby bunch crossings. The simulated events are then weighted to match the pileup distribution measured in data.

Offline reconstruction is based on the particle-flow (PF) algorithm~\cite{CMS:2017yfk}, which aims to reconstruct and identify each individual particle (photons, electrons, muons, charged and neutral hadrons, referred to as PF candidates) in an event, with an optimized combination of information from the various elements of the CMS detector. The PF candidates are clustered into jets using the anti-\kt algorithm~\cite{Cacciari:2008gp} implemented in \FASTJET~\cite{Cacciari:2011ma} with a distance parameter of 0.4 (small-radius jets, AK4) or 0.8 (large-radius jets, AK8). Pileup interactions contribute to the jet momentum by adding both charged and neutral particles, whose effect is mitigated using the PUPPI algorithm~\cite{Bertolini:2014bba,CMS:2020ebo}. Jet momentum, defined as the vectorial sum of all particle momenta in the jet, is found from simulation to be, on average, within 5 to 10\% of the true momentum over the whole \pt spectrum and detector acceptance. Jet energy corrections are derived from simulation so that the average measured energy of jets becomes identical to that of particle-level jets. In situ measurements of the momentum balance in multijet, $\gamma+\mathrm{jets}$, and $\PZ+\mathrm{jets}$ events are used to determine any residual differences between the jet energy scale in data and in simulation~\cite{CMS:2016lmd}. Additional selection criteria are applied to each jet to remove those altered by instrumental effects or reconstruction failures~\cite{CMS:2020ebo}.

Muon candidates, within the geometrical acceptance of the muon detectors (${\abs{\eta}< 2.4}$), are reconstructed by combining the information from the silicon tracker and the muon detector~\cite{CMS:2018rym}. These candidates are required to satisfy a set of quality criteria based on the number of hits measured in the tracker and muon systems, the properties of the fitted muon track, and the impact parameters of the track relative to the PV. Electron candidates within ${\abs{\eta}< 2.5}$ are reconstructed using an algorithm that associates fitted tracks in the silicon tracker with electromagnetic energy clusters in the ECAL detector~\cite{CMS:2020uim}. To reduce the misidentification rate, these candidates are required to satisfy identification criteria based on the shower shape of the energy deposit, the matching of the electron track to the ECAL energy cluster, the relative amount of energy deposited in the HCAL detector, and the consistency of the electron track with the PV. Identified muons and electrons are also required to be isolated from hadronic activity in the event. The isolation sum is defined by summing the \pt of all the PF candidates in a cone of radius $\Delta R=\sqrt{\smash[b]{(\Delta \eta)^{2}+(\Delta \phi)^{2}}} = 0.4~(0.3)$ around the muon (electron) track, where $\phi$ is the azimuthal angle in radians, and is corrected for the contribution of neutral particles from pileup interactions~\cite{CMS:2018rym,CMS:2020uim}.

\section{The CMS trigger system}\label{sec:cms_trigger_system}

The CMS trigger system selects physics events of interest from the large data flux produced by $\Pp\Pp$ collisions at the LHC. During Run~3, the LHC operated at instantaneous luminosities of up to ${2.2\times 10^{34}\percms}$ with a bunch spacing of 25\unit{ns}, corresponding to an average collision rate of about 30\unit{MHz} for fills with $\sim$2500 bunches. To match the limited resources for storage and offline processing, the trigger system reduces this rate in two stages: the level-1 trigger (L1T)~\cite{CMS:2020cmk}, implemented in custom hardware with a latency of about 4\unit{$\mu$s}, which uses coarse information from the calorimeter and muon systems, and the high level trigger HLT~\cite{CMS:2024aqx, CMS:2016ngn}, which runs in software on a farm of commercial processors. The HLT processes events accepted by the L1T at rates up to 110\unit{kHz} and reduces them to a few \unit{kHz} for the primary physics data sets. The HLT output rate  increased during Run~3 due to progressively higher instantaneous luminosities reached by the LHC, as well as the continued expansion of the trigger menu. 

\subsection{The HLT system}\label{sec:hlt_system}

The events selected by the L1T are sent to the HLT farm, where the raw detector data are first unpacked and then reconstructed using the same modular software framework~\cite{Jones:2014uza,Jones:2015soc,Dagenhart:2020igp} as in the offline reconstruction, with algorithms optimized for fast processing. The HLT then selects a subset of these events using a configuration composed of hundreds of trigger algorithms designed to identify physics signatures of interest. Selections within a trigger are applied in stages: early steps use fast calorimeter and muon information to reject background events, and more computationally intensive algorithms---such as PF reconstruction and jet tagging---are deferred to later stages to save computing resources. This staged approach maintains high trigger efficiency across a broad range of physics signatures, including those with \PQb jets, whereas the modular design facilitates the development of specialized triggers. Events accepted by the HLT are transferred to the Tier 0 computing centre for offline processing and permanent storage. In a typical 2024 fill, the average luminosity was about ${2\times 10^{34}\percms}$, with an average HLT output rate for physics data sets of roughly 7\unit{kHz}.

\subsection{Event reconstruction in the HLT}\label{sec:hlt_reconstruction}

This section describes the subset of event reconstruction algorithms employed at the HLT that are most relevant to jet flavour tagging: charged-particle tracks, PV and SV reconstruction, and jet clustering from PF candidates.

At the HLT, charged particle tracks are reconstructed from pixel and strip detector hits using a Kalman-filter-based approach~\cite{FRUHWIRTH1987444}, starting from initial estimates of the track parameters (seeds) built from pixel hits. During Run~2, tracking followed an iterative procedure similar to offline reconstruction~\cite{CMS:2014pgm}. Starting from stringent seed requirements, including at least three or four pixel hits, pixel tracks were first reconstructed~\cite{CMS:2024aqx}. Vertices (``pixel vertices'') were then reconstructed by clustering pixel tracks---with at least four hits and ${\pt>0.5\GeV}$---based on the $z$-coordinate of their closest approach to the beam axis. To satisfy tight latency constraints, only seeds compatible with these pixel vertices, whose $\sum\pt^2$ exceeds 30\% of  that of the leading-$\sum\pt^2$ pixel vertex, were used for the full track reconstruction. All reconstructed tracks were then used for the ultimate PV reconstruction, where vertex positions were determined using a deterministic annealing algorithm~\cite{726788} followed by an adaptive vertex fit~\cite{Fruhwirth:2007hz}. Tracks and PVs were subsequently used to reconstruct SVs with the inclusive vertex finder (IVF) algorithm~\cite{CMS:2017wtu}.

In Run~3, tracking at the HLT is performed in a single global iteration seeded by a subset of pixel tracks reconstructed with the \textsc{PataTrack} algorithm~\cite{Bocci:2020pmi}, requiring at least three pixel hits and ${\pt>0.3\GeV}$. Since 2024, following issues with the pixel tracker, a recovery iteration seeded with pixel doublets was added, active only in regions of the detector with a nonnegligible fraction of inactive modules. Despite employing fewer iterations, the Run~3 strategy achieves higher tracking efficiency and a lower fake rate than that obtained during Run~2, while significantly reducing the processing time~\cite{CMS:2023gfb}. Similarly to Run~2, only pixel tracks compatible with the subset of vertices whose $\sum\pt^2$ exceeds 30\% of that of the leading-$\sum\pt^2$ pixel vertex are used as seeds to the first HLT tracking iteration. Then, the full collection of reconstructed tracks is used to reconstruct PVs, following the same methods used in Run~2.

Triggers exploiting jet activity within the tracker acceptance employ a two-stage jet clustering strategy. In the first stage, jets are clustered either from calorimeter energy deposits (calo jets) or from PF candidates reconstructed using only pixel tracks (pixel jets) rather than the full track reconstruction. While calo jets were predominantly used in Run~2, pixel jets are used more extensively in Run~3 to pre-filter events and seed the full PF reconstruction because they provide better energy resolution. This enables a reduction in the event rate requiring full PF reconstruction without any loss in trigger efficiency. In the second stage, both AK4 and AK8 PF jets are used in the final trigger selection, including \PQb jet and \xbb tagging. Since 2024, for jets within the tracker acceptance, the event selection relies solely on PF jets.

thin the jet. Deployed at the HLT in 2017, it served as the reference online \PQb tagger through the end of Run~2. The identification of high-\pt resonances decaying into heavy-quark pairs, such as \hbb, was instead performed using the \DoubleB tagger~\cite{CMS:2017wtu}: a boosted decision tree (BDT) trained on AK8 jets to identify the decay products of two \PQb hadrons from a \xbb decay, exploiting jet substructure variables, leading track properties, SVs, and PF candidate information.

During the shutdown between Run~2 and Run~3, novel tagging strategies were developed and validated for both \PQb tagging and high-\pt \xbb identification, bringing substantial improvements over the Run~2 methods with significant impact on several key physics channels, as discussed in Section~\ref{sec:physics_impact}.

\section{Jet flavour identification in the HLT}\label{sec:jet_tagging}

As introduced in Section~\ref{sec:introduction}, a number of key measurements and searches in the CMS physics programme rely on the ability to identify the parton or hadronically decaying resonance that initiated a jet observed in the detector. During Run~2, \PQb tagging of AK4 jets at the HLT was performed using \deepcsv~\cite{CMS:2017wtu,CMS:2024aqx}, a feed-forward deep neural network (DNN) with four fully connected layers of approximately 100 neurons each. Although deployed online at the HLT, the \deepcsv model was trained using offline-reconstructed input features available in Run~2. The \deepcsv algorithm is a multi-class classifier assigning, for each jet, output scores representing the probability that it originates from a bottom quark, a charm quark, or a light jet (u, d, s quark or gluon).  The input features include jet kinematics and its constituents, track quality and impact parameters, and the properties of any SV present within the jet. Deployed at the HLT in 2017, it served as the reference online \PQb tagger through the end of Run~2. The identification of high-\pt resonances decaying into heavy-quark pairs, such as \hbb, was instead performed using the \DoubleB tagger~\cite{CMS:2017wtu}: a boosted decision tree (BDT) trained on AK8 jets to identify the decay products of two \PQb hadrons from a \xbb decay, exploiting jet substructure variables, leading track properties, SVs, and PF candidate information.

During the shutdown between Run~2 and Run~3, novel tagging strategies were developed and validated for both \PQb tagging and high-\pt \xbb identification, bringing substantial improvements over the Run~2 methods with significant impact on several key physics channels, as discussed in Section~\ref{sec:physics_impact}.

\subsection{Small-radius jet tagging in Run~3}\label{sec:ak4_jet_tagging}

Two new AK4 jet tagging algorithms, based on the \deepjet~\cite{Bols:2020bkb} and \pnet~\cite{Qu:2019gqs} models, were developed and deployed for \PQb and \PQc tagging at the HLT from the start of Run~3 in 2022.

The \deepjet convolutional neural network processes PF candidates and SVs within a small-radius through three blocks of convolutional layers---one each for charged PF candidates, neutral PF candidates, and SVs---with kernel of dimension 1 and a decreasing numbers of filters to produce compact feature embeddings. These are then passed to dedicated long short-term memory (LSTM) layers, whose outputs are merged with global jet properties and fed into a final feed-forward DNN returning six output scores, each corresponding to the probability that the jet originates from hadronic or semileptonic \PQb decays, two \PQb quarks, a \PQc quark, a light-quark (uds), or a gluon. A dedicated \deepjet training was performed on simulated HLT-reconstructed jets to maximise trigger performance. Several \PQb jet triggers based on \deepjet were deployed in 2022 and remained in use until their progressive replacement in 2024.

\begin{figure*}[!htb]
  \centering
  \includegraphics[width=0.45\textwidth]{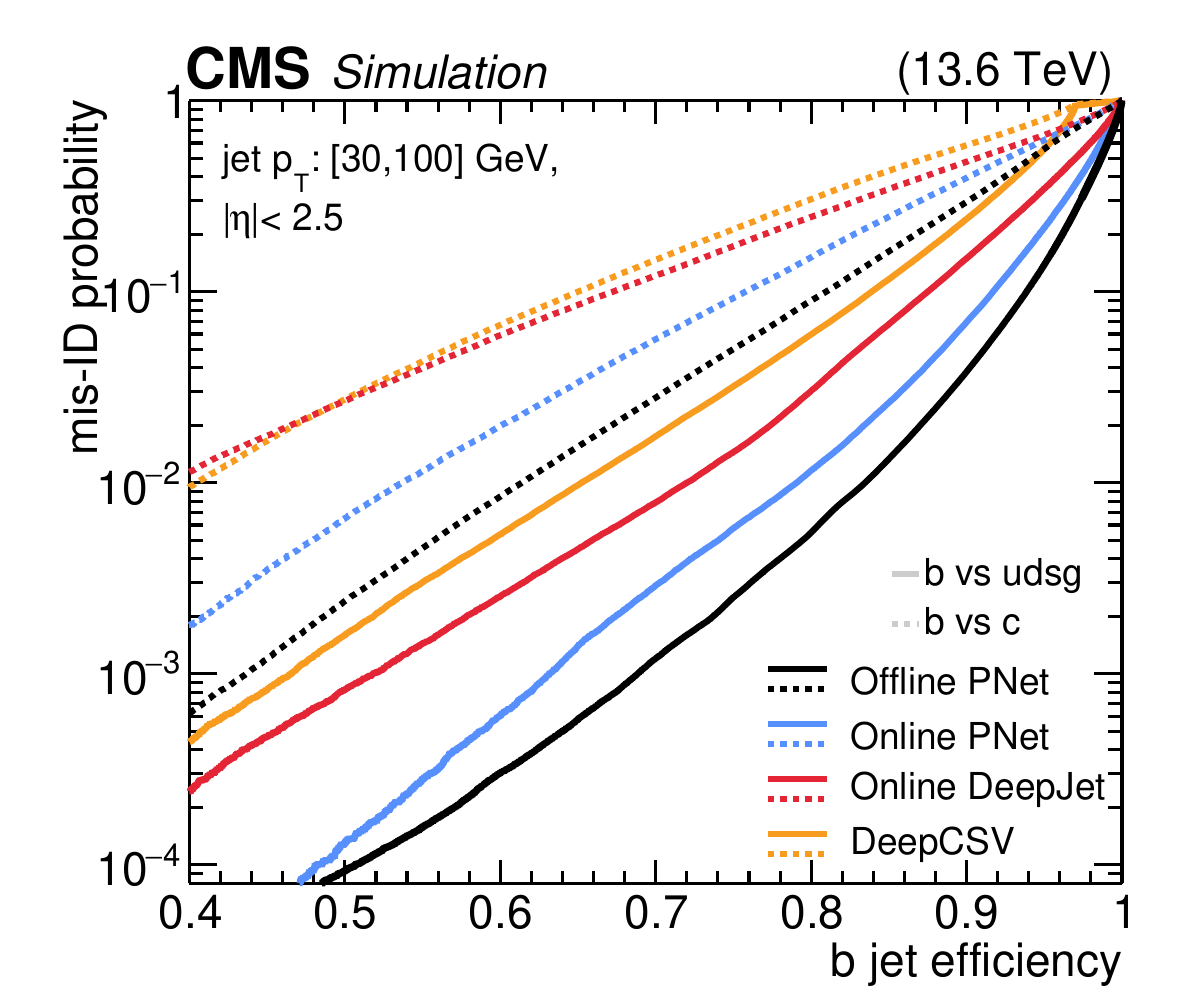}
  \includegraphics[width=0.45\textwidth]{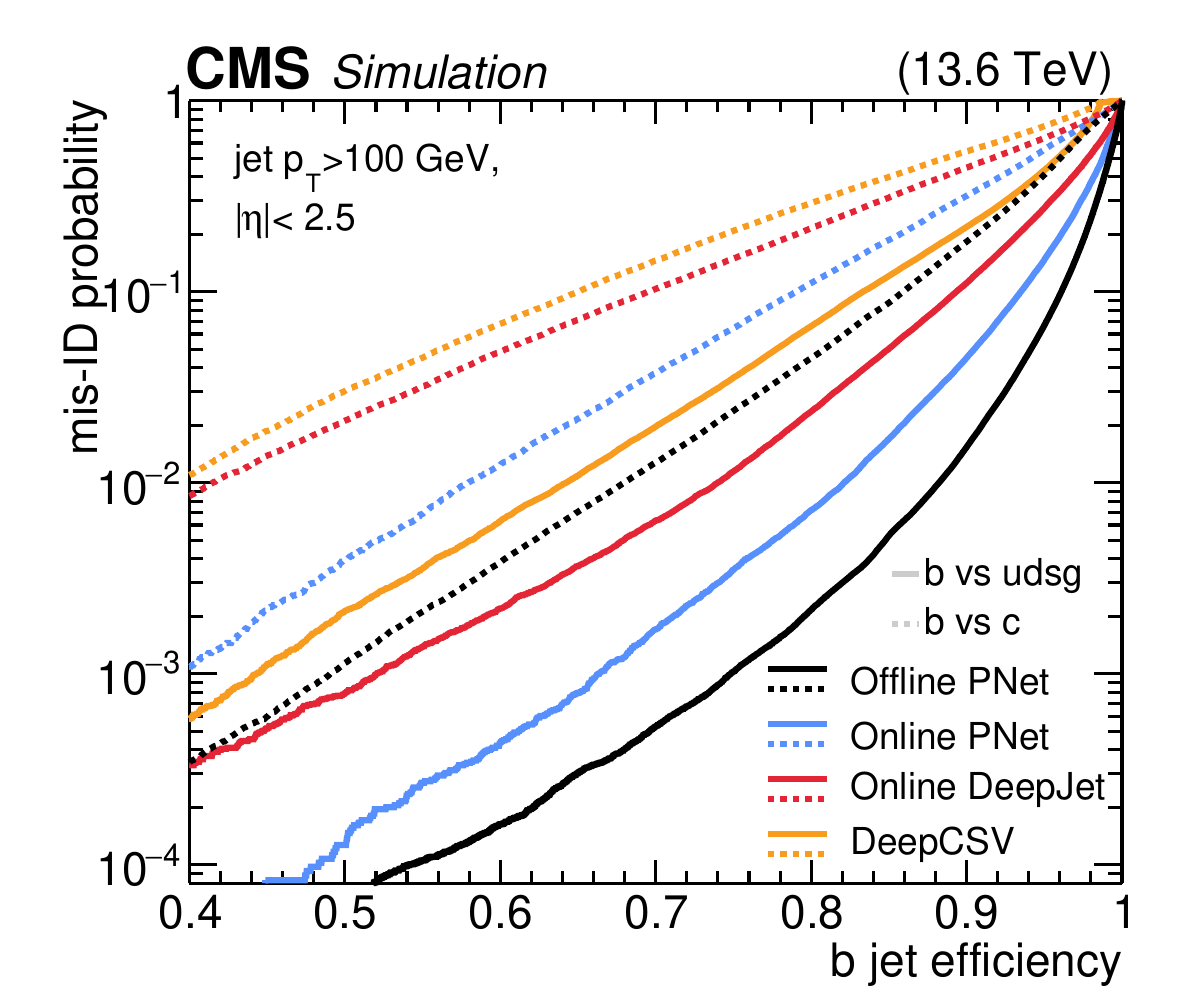}
  \caption{Misidentification probability for light jets (solid curves) and \PQc (dashed curves) as a function of \PQb jet identification efficiency for various jet tagging algorithms applied to AK4 jets with ${\abs{\eta} < 2.5}$ and ${30 < \pt < 100\GeV}$ (\cmsLeft) or ${\pt > 100\GeV}$ (\cmsRight) from simulated \ttbar events. Results are shown for the \pnet (blue) and \deepjet (red) taggers trained for Run~3, and for the \deepcsv (orange) discriminator used during Run~2. The performance of jet flavour taggers developed for the HLT can be compared with that obtained from the offline \pnet algorithm (black)~\cite{CMS-DP-2024-066}, evaluated on jets from the same selected events.}
  \label{fig:roc_performance_btag}
\end{figure*}

A major advancement in the CMS trigger capability came with the development of an HLT-customized version of \pnet. The \pnet algorithm is a DGCNN representing each jet as a cloud of particles in a multidimensional feature space. Rather than processing constituents sequentially, \pnet constructs a graph from PF candidates and SVs (nodes), dynamically updating it through edge convolution operations that capture local and global interactions between nodes. At each stage, every node is connected to its nearest neighbours in feature space, with connections redefined within each edge convolution block, enabling \pnet to learn complex non-linear dependencies among jet constituents in a permutation-invariant manner. This architecture yields rich high-level jet representations, significantly improving flavour tagging performance in the real-time HLT environment and substantially reducing the gap with offline reconstruction performance. As \pnet demonstrated superior performance for both \PQb and \PQc tagging (Fig.~\ref{fig:roc_performance_btag}), it was progressively adopted as the reference AK4 jet tagger at the HLT. Starting in 2024, all \PQb tagging triggers were fully migrated to \pnet, which became the standard tagger for the remainder of Run~3. A key byproduct of this improvement was the deployment of the first dedicated HLT triggers targeting \hcc decays using an efficient \PQc jet tagger.

The \pnet model designed for the HLT takes as input twenty-three features per PF candidate and eleven per SV, and derives an optimal discriminant by minimising the categorical cross-entropy loss over \PQb, \PQc, uds, gluon, or hadronically decaying \PGt lepton jet classes. While the improved capabilities shown by \pnet in the online tagging of hadronically decaying \PGt leptons have been used in redesigning all \PGt-based HLT triggers starting from 2024~\cite{CMS-DP-2026-008}, \PGt performance studies are not presented in this paper. The architecture comprises three stacked edge convolution blocks, each operating on the eight nearest neighbours per node, and processes up to 50 nodes per jet (zero-padded when fewer constituents are present). Only PF candidates with ${\pt>0.4\GeV}$ are used. For charged PF candidates, the computation of track-based IP and flight distance observables require ${\pt>0.9\GeV}$ and loose track quality selections.

The most discriminating features for heavy-flavour jet tagging exploit the displaced-decay topology of \PQb or \PQc hadrons. For each charged PF candidate, the IP vector---defined as the vector from the PV to the point of closest track approach---is signed positive if the angle with the jet axis is less than $\pi/2$ (track produced ``upstream'' of the jet) and negative otherwise. The IP value can be defined either in spatial dimensions ($d_{xyz}$), in the plane transverse to the beam line ($d_{xy}$), or along the beam line ($d_{z}$). For each IP definition, the corresponding significance is defined as the magnitude of the IP divided by its uncertainty. Additional discrimination comes from the larger mass and harder fragmentation of \PQb and \PQc quarks compared to lighter quarks and gluons. This causes their decay products to have, on average, larger \pt relative to the jet axis than the other jet constituents~\cite{Cacciari:2002xb}. Jets from \PQb quarks also exhibit differences in charged-to-neutral and hadronic-to-electromagnetic energy balance. Finally, the presence of muons produced in semileptonic \PQb hadron decays is also exploited by \pnet.

\begin{figure*}[!htb]
  \centering
  \includegraphics[width=0.45\textwidth]{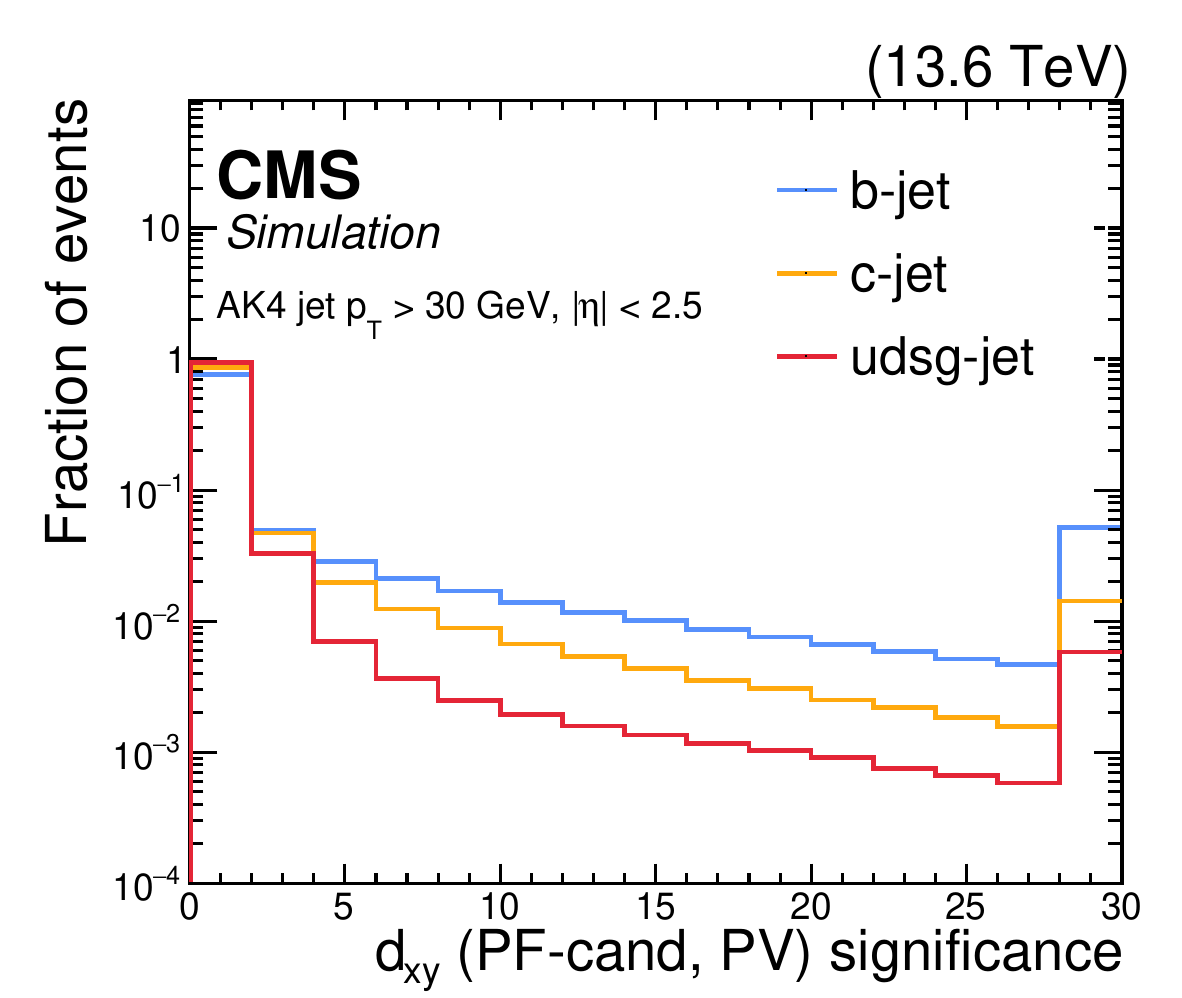}
  \includegraphics[width=0.45\textwidth]{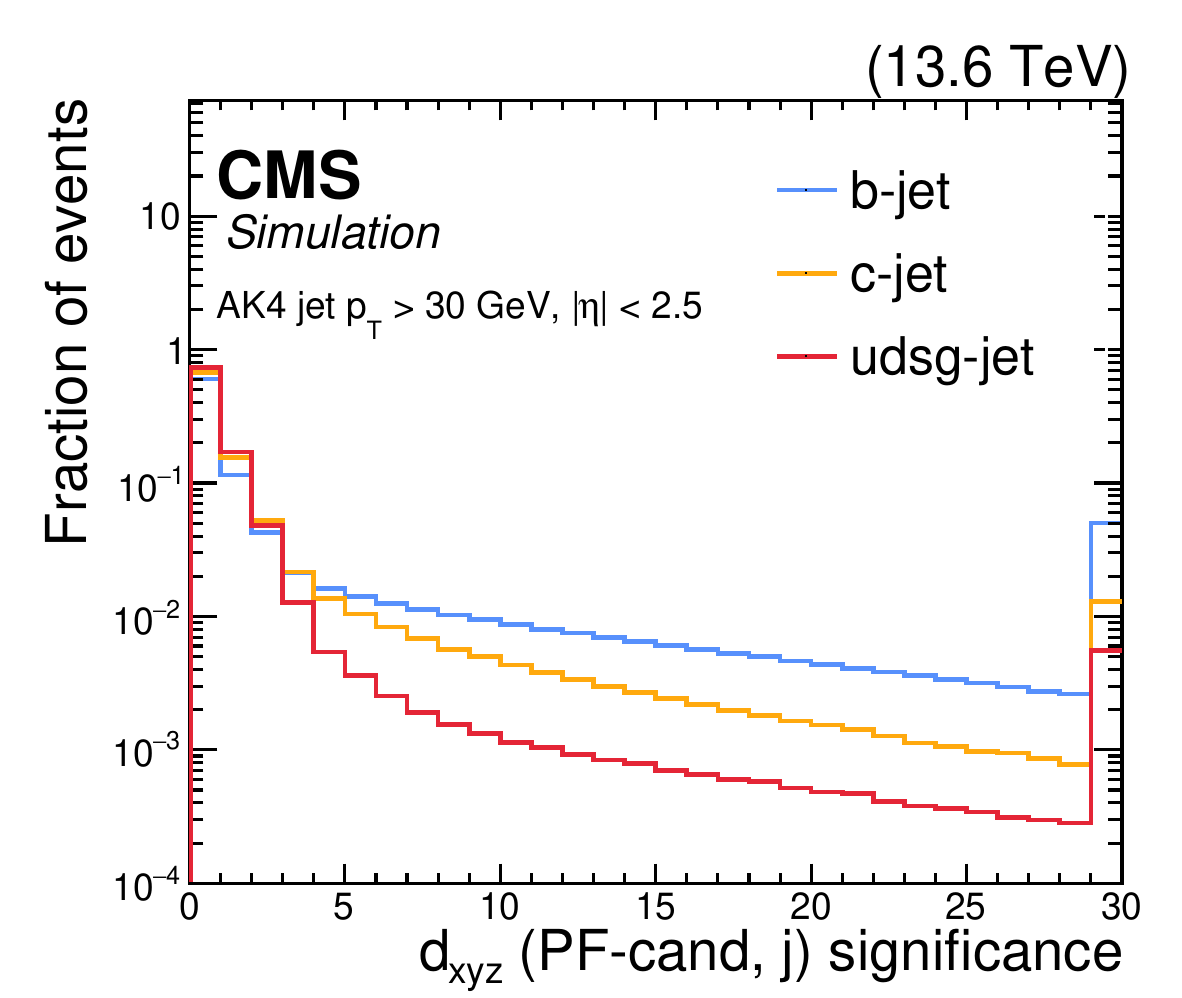}\\
  \includegraphics[width=0.45\textwidth]{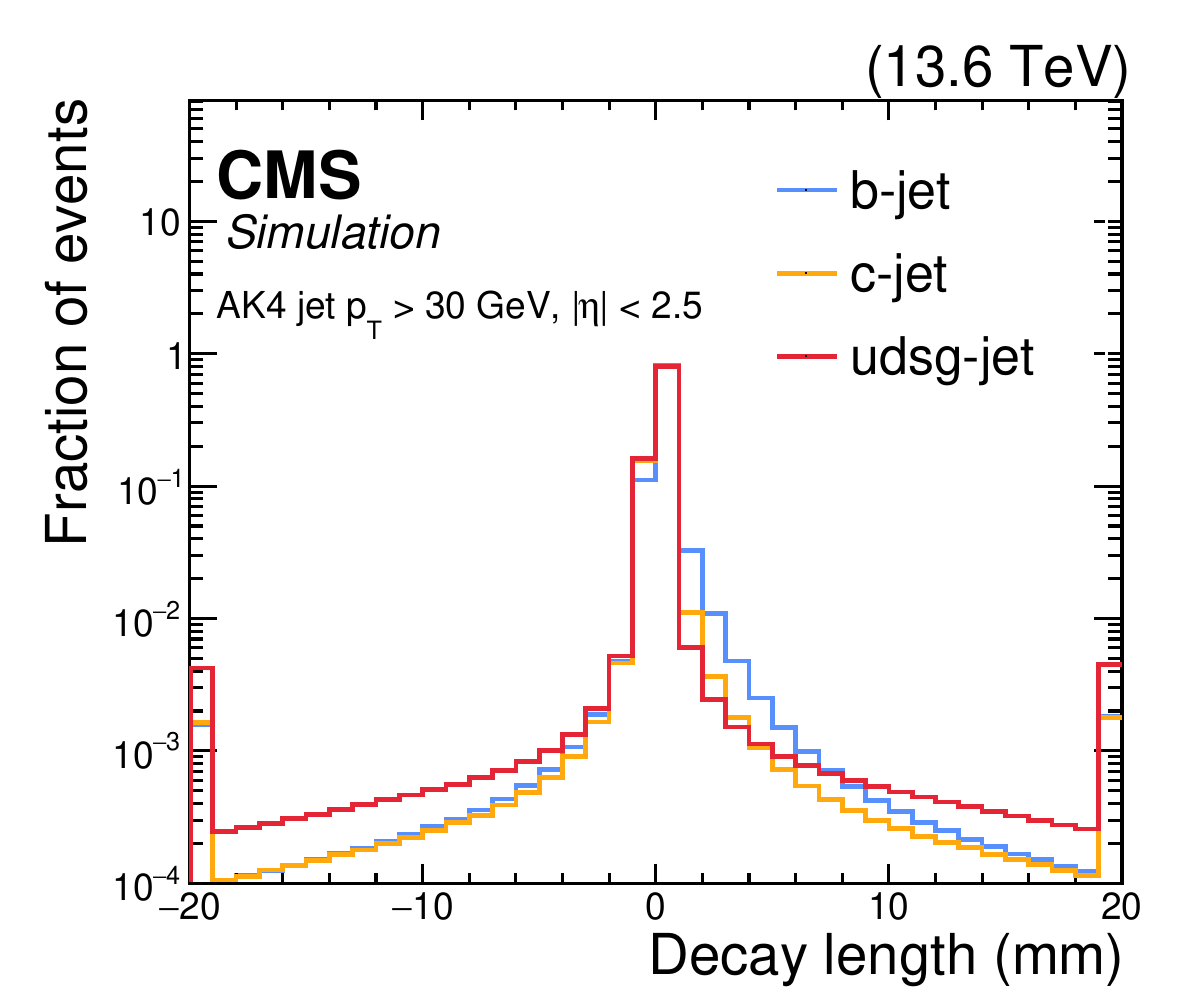}
  \includegraphics[width=0.45\textwidth]{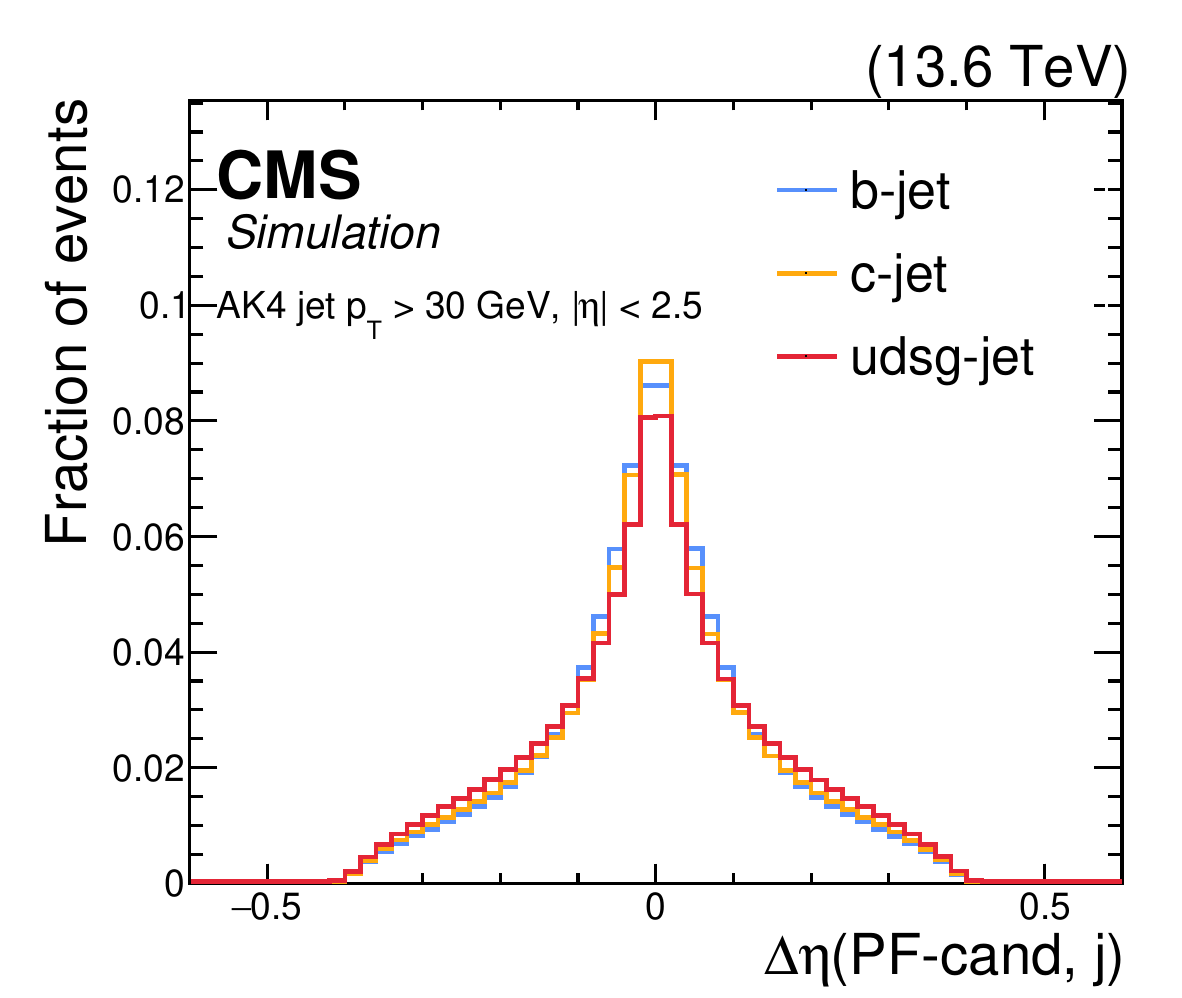}
  \caption{Distributions of \pnet input variables related to PF candidates in \PQb (blue), \PQc (orange), and light (red) AK4 jets with  ${\pt>30\GeV}$ and ${\abs{\eta} < 2.5}$ from simulated \ttbar events. All distributions are normalized to unit area. Upper: $d_{xy}$ (\cmsLeft) and $d_{xyz}$ (\cmsRight) IP significances. Lower: decay length with respect to the jet axis (\cmsLeft) and the $\Delta\eta$ between  the PF candidate and the jet axis (\cmsRight). The first (last) bin includes underflow (overflow) entries.}
  \label{fig:pnet_variable_pfcand}
\end{figure*}

\begin{figure*}[!htb]
  \centering
  \includegraphics[width=0.45\textwidth]{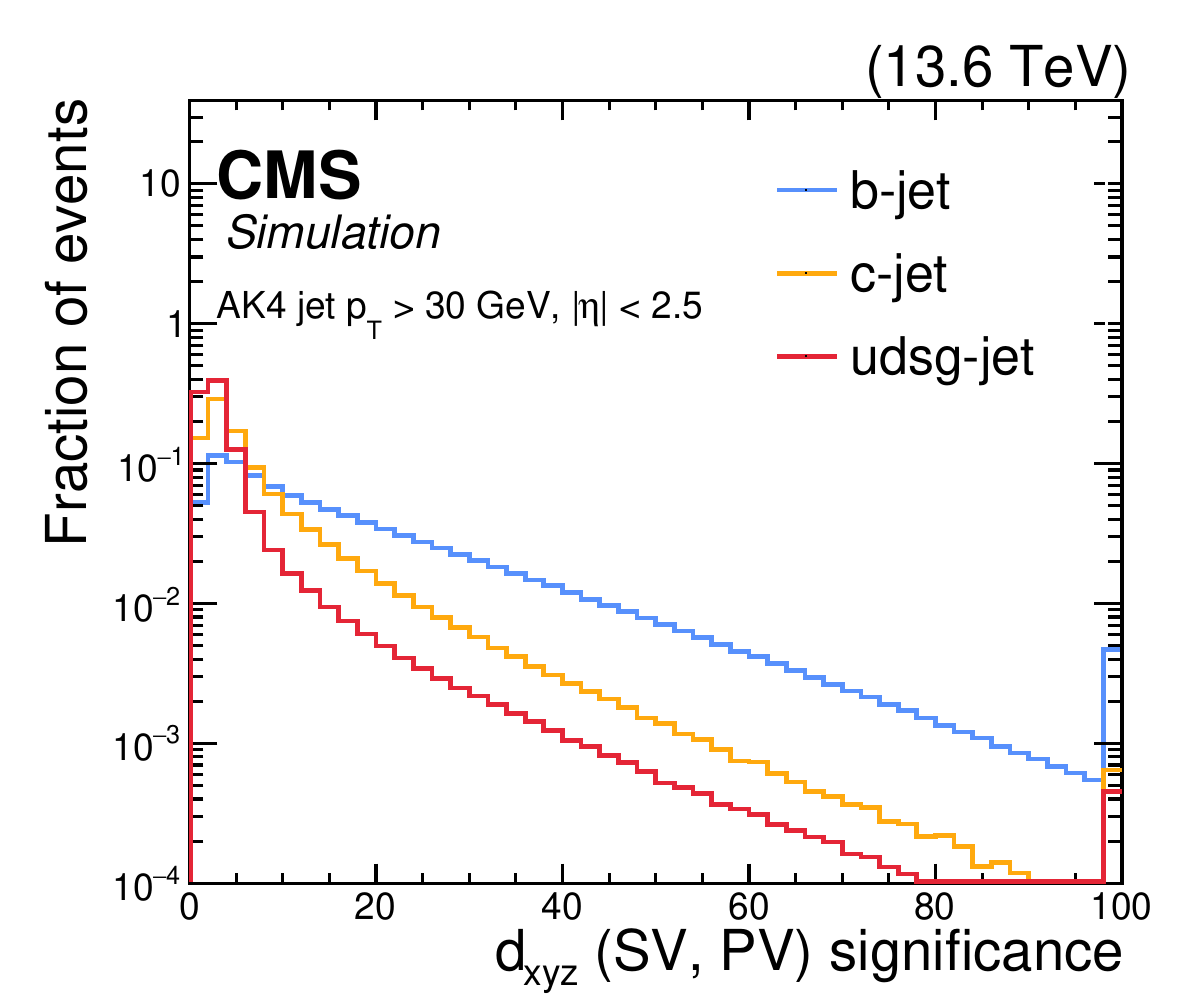}
  \includegraphics[width=0.45\textwidth]{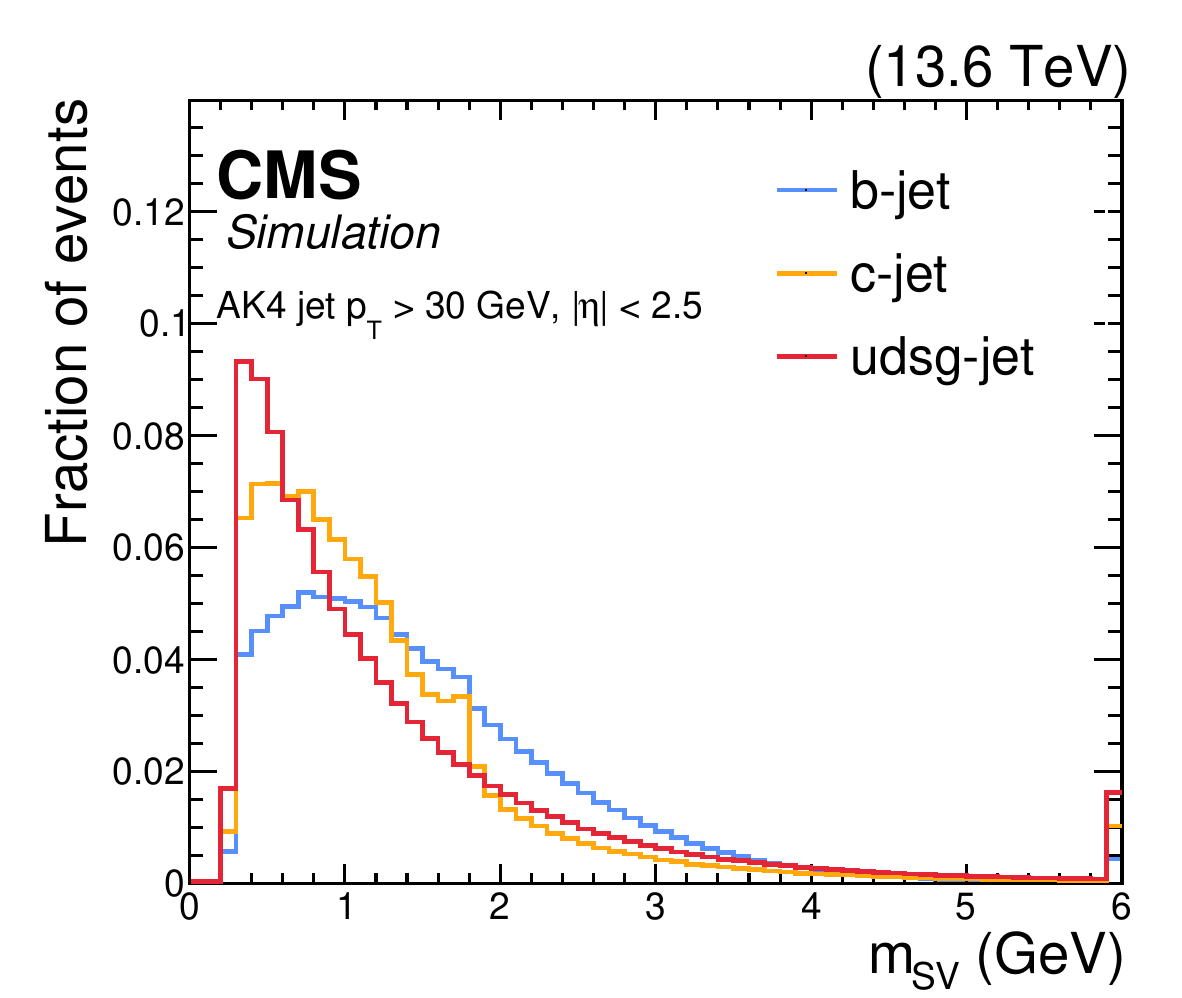}\\
  \includegraphics[width=0.45\textwidth]{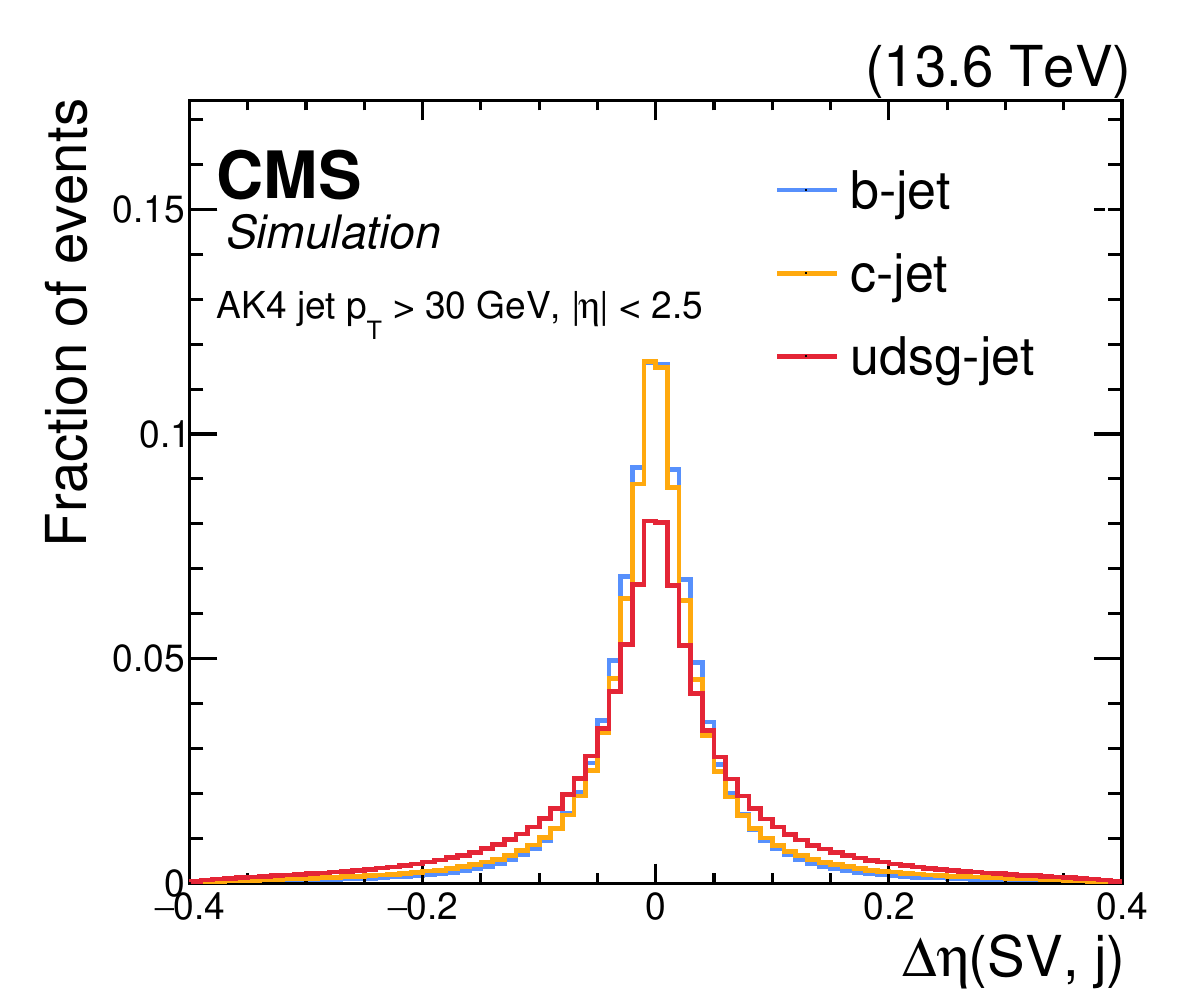}
  \includegraphics[width=0.45\textwidth]{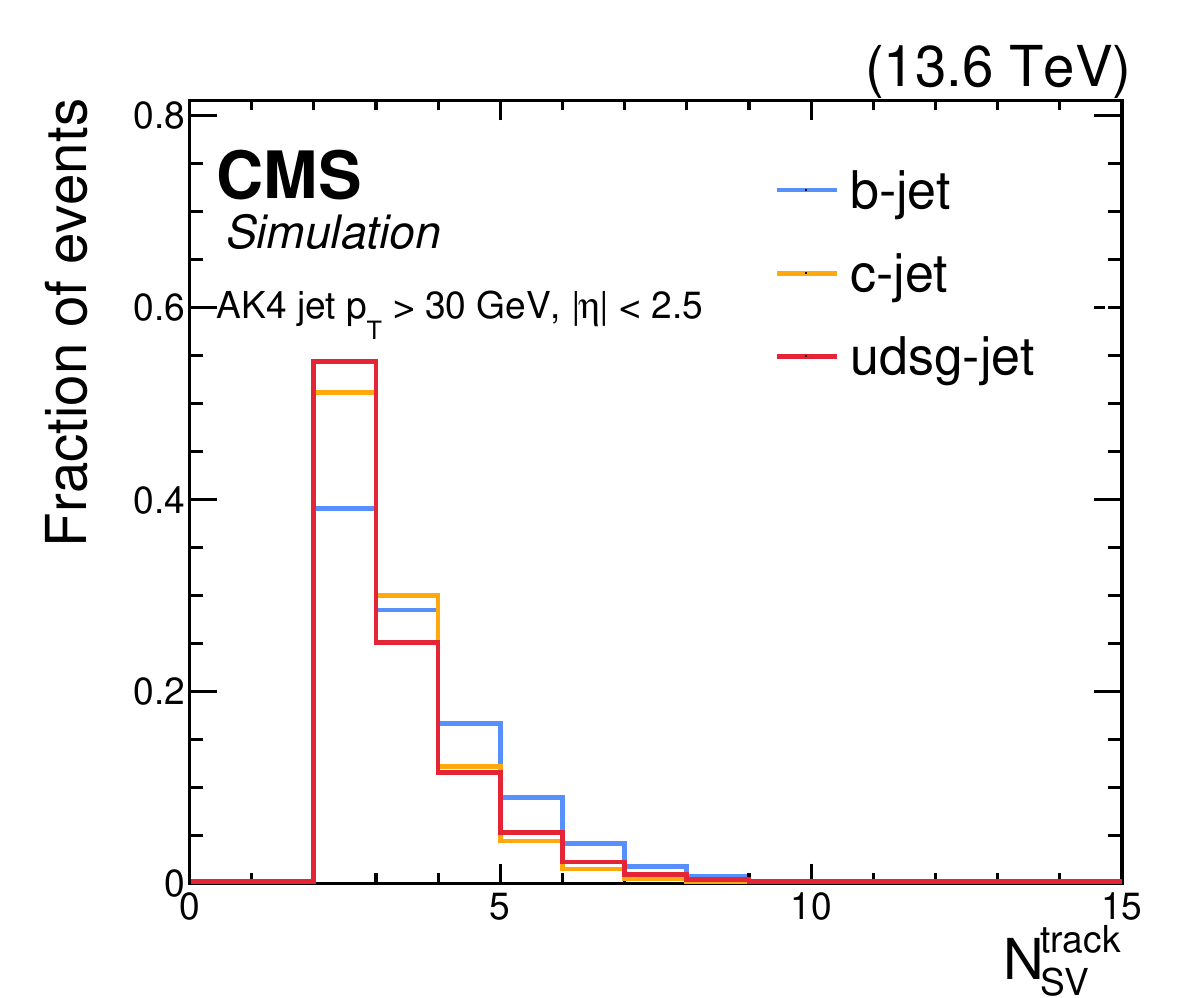}
  \caption{Distributions of \pnet input variables related to SVs in \PQb (blue), \PQc (orange), and light (red) AK4 jets with ${\pt>30\GeV}$ and ${\abs{\eta} < 2.5}$ from simulated \ttbar events. All distributions are normalized to unit area. Upper: $d_{xyz}$ significance (\cmsLeft) and SV invariant mass (\cmsRight). Lower: $\Delta\eta$ between the SV and the jet axis (\cmsLeft) and the SV track multiplicity (\cmsRight). The first (last) bin includes underflow (overflow) entries. The feature present in the SV invariant mass distribution of \PQb and \PQc jets (top right) in the 1.8--2\GeV range is due to the reconstruction, as a SV, of a D-meson from its decay products.}
  \label{fig:pnet_variable_sv}
\end{figure*}

The SVs are reconstructed using the IVF algorithm, which selects track seeds from tracks with ${\pt > 0.8\GeV}$ and a longitudinal IP smaller than 0.3\unit{cm} with respect to the PVs. Such displaced vertices typically originate from the decays of long-lived particles such as \PQb and \PQc hadrons. SVs offer powerful discriminating features for \PQb and \PQc tagging as 3D ($d^{\mathrm{SV}}_{xyz}$) and 2D ($d^{\mathrm{SV}}_{xy}$) displacements from the PV, their invariant mass ($m_{\mathrm{SV}}$, defined as the invariant mass of all tracks associated with the SV), angular separation from the jet axis ($\Delta\eta (j,\mathrm{SV})$, $\Delta\phi (j,\mathrm{SV})$), and track multiplicity. Figures~\ref{fig:pnet_variable_pfcand} and~\ref{fig:pnet_variable_sv} show distributions in simulated events of the most important PF candidate and SV input features for \PQb tagging, for AK4 jets from \ttbar events with ${\pt>30\GeV}$ and ${\abs{\eta}<2.5}$.  The decay length distribution shown in Fig.~\ref{fig:pnet_variable_pfcand} (lower left) is asymmetric for both \PQb and \PQc jets. This asymmetry arises because the decay vertices of \PQb and \PQc hadrons are displaced along the jet flight direction with respect to the PV, resulting predominantly in positive values of the signed decay length.

The \pnet model is trained on AK4 jets with ${\pt>30\GeV}$ and ${\abs{\eta}<2.5}$, geometrically matched within ${\Delta R <0.4}$ to a generator-level jet. This requirement excludes pileup jets and enables correct flavour assignment via the ghost-association method~\cite{Cacciari:2007fd}. To avoid biases from specific processes or phase space regions, jets are collected from diverse simulated samples, including \ttbar, QCD multijet, \Vjets, and Higgs boson production (single-$\PH$ and non-resonant $\PH\PH$ via gluon fusion and VBF) with the decays \hbb, \hcc, and \htt. This data set is then resampled so that all jet flavours are uniformly distributed in the {\pt--$\eta$} plane, preventing overall jet kinematic information from contributing to the jet flavour identification performance.

The trained \pnet model predicts, for each AK4 jet, probabilities for the jet to originate from \PQb, \PQc, uds, gluon, or hadronically decaying \PGt leptons. The training minimizes the categorical cross-entropy loss over up to 30 epochs,  with the input jet data set split into 60\% for training, 20\% for validation, and 20\% for testing. The learning rate remains constant for the first 10 epochs and then decreases linearly until convergence. The final model is selected based on the epoch yielding the highest validation accuracy.

\begin{figure*}[!htb]
  \centering
  \includegraphics[width=0.45\textwidth]{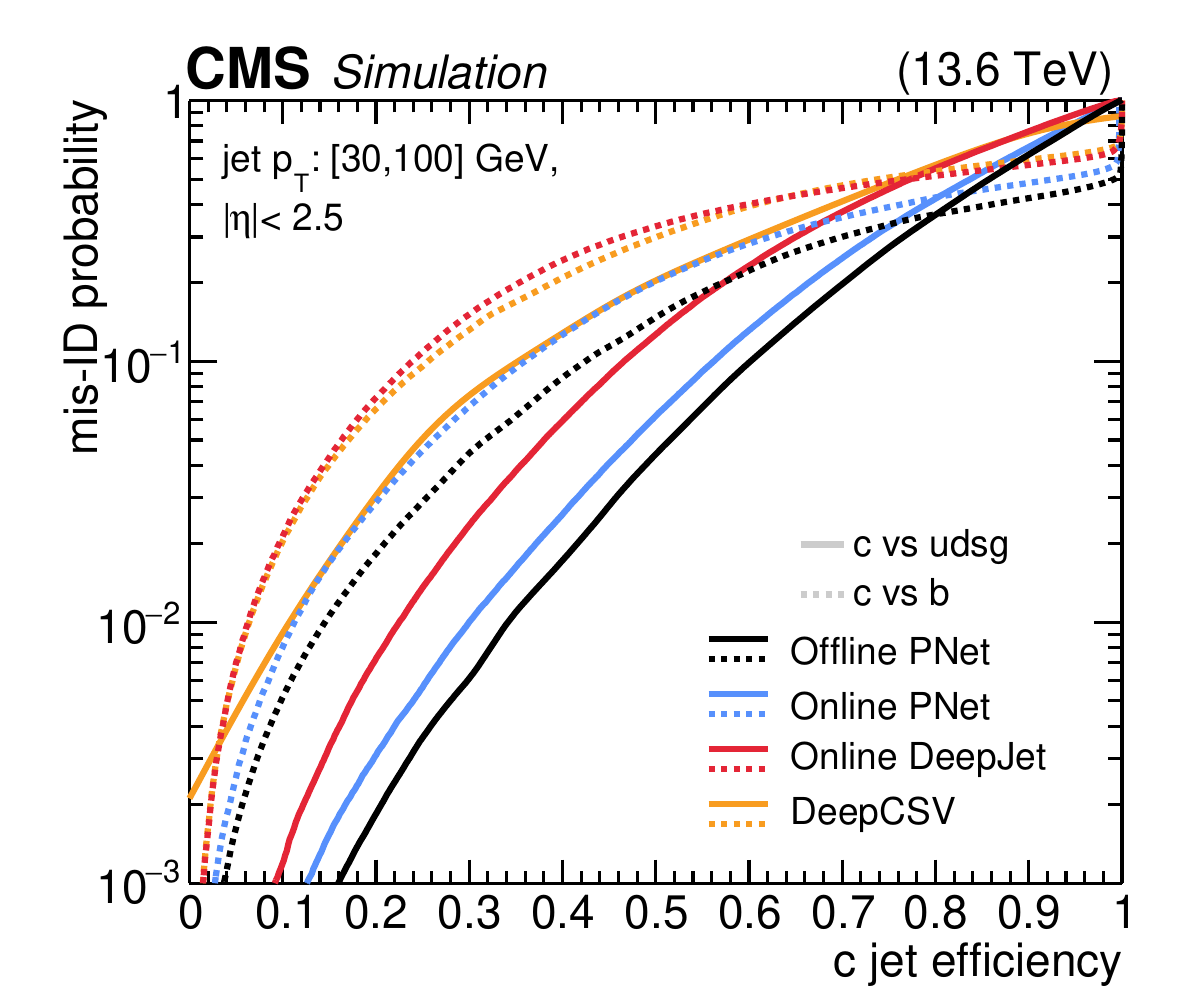}
  \includegraphics[width=0.45\textwidth]{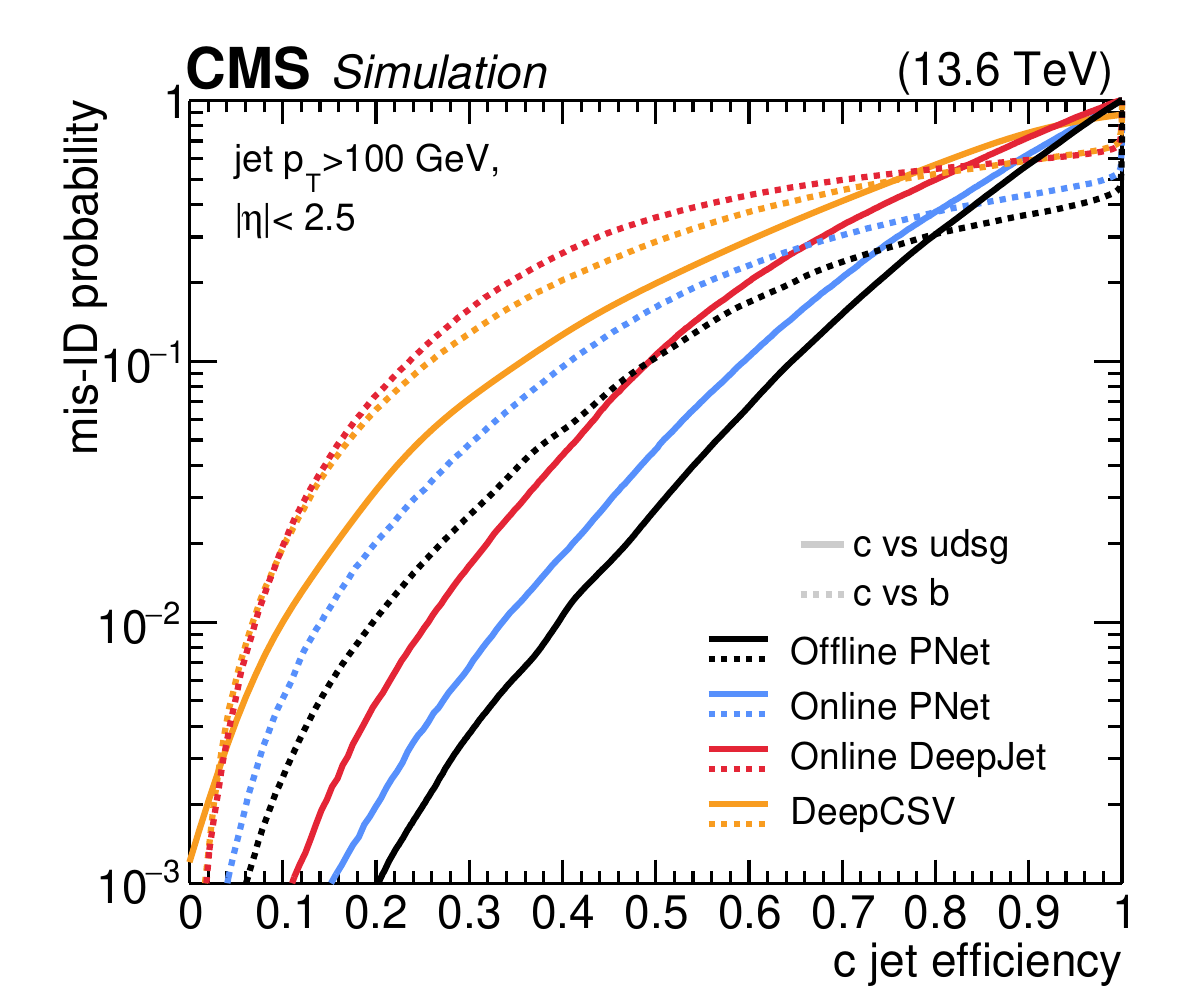}
  \caption{Misidentification probability for light jets (solid curves) and \PQb (dashed curves) as a function of \PQc jet identification efficiency for various jet tagging algorithms applied to AK4 jets with ${\abs{\eta} < 2.5}$ and ${30 < \pt < 100\GeV}$ (\cmsLeft) or ${\pt > 100\GeV}$ (\cmsRight) from simulated \ttbar events. Results are shown for the \pnet (blue) and \deepjet (red) taggers trained for Run~3, and for the \deepcsv (orange) discriminator used during Run~2. The performance of jet flavour taggers developed for the HLT can be compared with that obtained from the offline \pnet algorithm (black)~\cite{CMS-DP-2024-066}, evaluated on jets from the same selected events.}
  \label{fig:roc_performance_ctag}
\end{figure*}

The efficiency (misidentification probability) to correctly (incorrectly) tag a jet with flavour $f$ (with flavour different from $f$) is defined as the fraction of jets of flavour $f$ passing (not passing) the discriminator output score ${\mathcal{P}_{f}>X}$. Figure~\ref{fig:roc_performance_btag} shows the \PQb tagging efficiency versus misidentification probability for \PQc and light jets in simulated fully hadronic \ttbar events. AK4 jets with $\abs{\eta}<2.5$ are considered in two regions:
${30<\pt<100\GeV}$ (\cmsLeft) and $\pt>100\GeV$ (\cmsRight). Results for \pnet (blue), \deepjet (red), and \deepcsv (orange) show a clear improvement from \pnet across the full \pt range in both \PQb vs. \PQc and \PQb vs light jet discrimination. The average gain is about $15\%$~($10\%$) in \PQb tagging efficiency at misidentification rates of $0.1\%$~($1\%$). Figure~\ref{fig:roc_performance_ctag} presents analogous results for \PQc tagging efficiency versus the misidentification probability for either \PQb or light jets, showing similar improvements of \pnet over \deepjet and \deepcsv~\cite{CMS:2021scf} also for this task.

The offline \pnet algorithm, designed for jet flavour tagging in the offline reconstruction~\cite{CMS-DP-2024-066}, achieves better performance on the same selected \ttbar simulated events. This improvement arises from a more precise event reconstruction, leading to improved track impact parameter resolution, additional input features (\eg track-to-vertex association, track hits, and calorimetric cluster properties), and the relaxed timing constraints that allow a more complex and therefore expressive \pnet architecture.

\subsection{Large-radius jet tagging in Run~3}\label{sec:ak8_jet_tagging}

\begin{figure*}[!htb]
  \centering
  \includegraphics[width=0.45\textwidth]{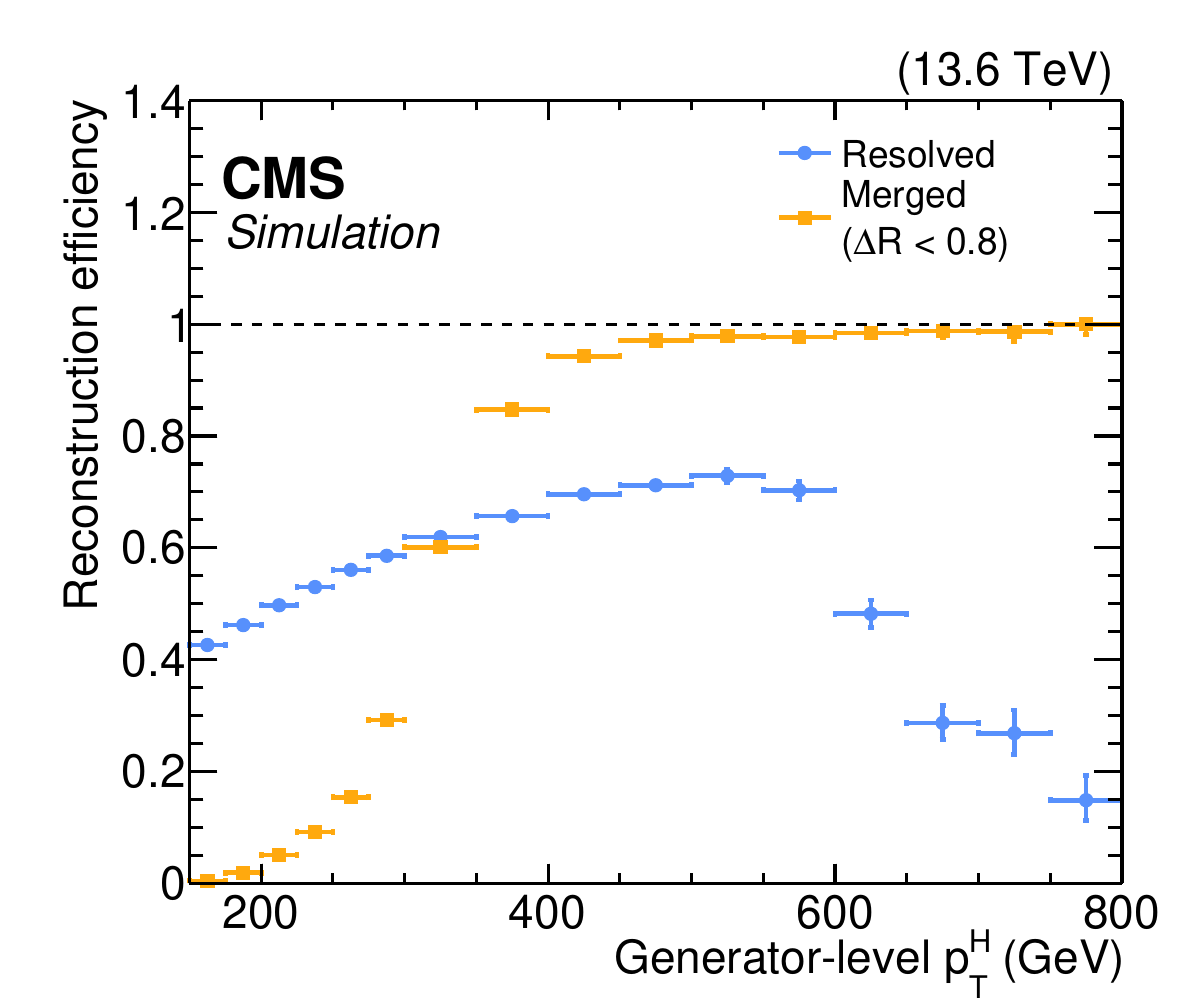}
  \includegraphics[width=0.45\textwidth]{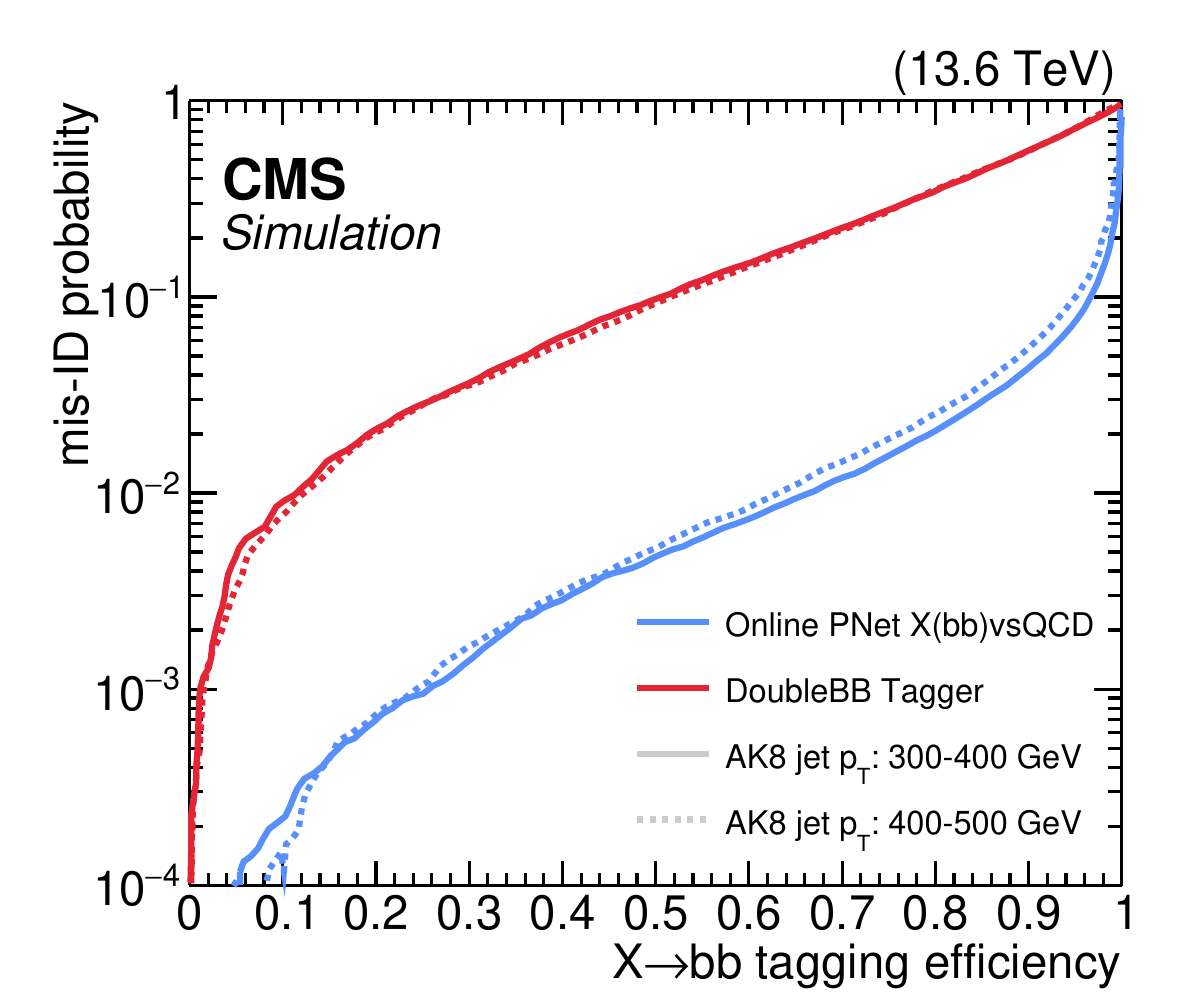}
  \caption{Left: Reconstruction efficiency of \hbb candidates, as a function of the its generator-level \pt, as two AK4 jets (resolved approach) in azure or as a single AK8 jet (merged approach) in orange. The vertical bars indicate the statistical uncertainty in the reconstruction efficiency. The \hbb candidates are obtained from simulated \hhbb events. Right: Misidentification probability for QCD jets versus the \xbb tagging efficiency for the \pnet (blue) and \DoubleB (red) algorithms, applied to simulated jets reconstructed at the HLT with ${\abs{\eta}<2.5}$, ${\mSD>40\GeV}$, and ${300<\pt<400\GeV}$ (solid) or ${400<\pt<500\GeV}$ (dashed).}
  \label{fig:trigger_merged_xbb}
\end{figure*}

At the high centre-of-mass energy of the LHC, massive resonances (X) produced in $\Pp\Pp$ collisions can be highly boosted in the laboratory frame. When such resonances decay hadronically into two-body systems, \eg a high-\pt Higgs or \PZ boson decaying to $\PQb\PAQb$, their decay products often overlap in the detector. In these cases, it is more efficient to reconstruct all decay products within a single AK8 jet rather than two AK4 jets. For the standard CMS AK8 jet size ($R=0.8$), most \hbb~(\zbb) decays are reconstructed as a single AK8 jet when the resonance \pt exceeds about 350~(250)\GeV. Figure~\ref{fig:trigger_merged_xbb} (\cmsLeft) shows the reconstruction efficiency for \hbb decays in simulated \hhbb events as a function of the generator-level \ptH, comparing the merged (single AK8 jet) and resolved (two AK4 jets) approaches. The efficiencies are computed by matching the generator-level Higgs boson (or the generator-level jets from the \PQb quarks) to the reconstructed AK8 jet (or AK4 jets) within $\Delta R<0.1$. The merged approach becomes more efficient than the resolved one for ${\ptH>350\GeV}$.

Selecting events with high-\pt \xbb candidates is challenging due to the overwhelming hadronic backgrounds, mainly from QCD multijet and \ttbar production. These processes can mimic the expected signal (\xbb) through gluon splitting to $\PQb\PAQb$ or hadronic \PW decays produced in association with \PQb jets. The primary discriminating observable between signal and background is the jet mass. In QCD jets, radiated quarks and gluons from the initiating parton are typically soft and collimated, producing a characteristic Sudakov peak in the jet mass distribution~\cite{CMS:2018vzn}. Additional contributions from initial-state radiation, underlying event, and pileup further broaden the mass spectrum, by adding energetic particles at large angles relative to the jet axis. As a result, QCD jets can acquire relatively large masse values. Grooming techniques are designed to mitigate these effects by removing soft and uncorrelated radiation within the jet. In particular, the soft-drop (SD) algorithm~\cite{Larkoski:2014wba} is widely used to suppress such contributions, with the resulting groomed jet mass, referred to as the soft-drop mass (\mSD), providing an improved separation between signal jets (\xbb, \xcc, etc.) and the QCD background. Further discrimination can be achieved using jet substructure observables and correlations among constituents. During Run~2, the \DoubleB tagger was deployed at the HLT for high-\pt \hbb signatures, but its performance is significantly lower than that of modern offline algorithms~\cite{CMS:2025kje}, especially \pnet.

To reduce the gap between HLT and offline performance in \xbb tagging, a \pnet tagger for AK8 jets was developed using HLT-reconstructed jets. The network architecture, inputs features, and selections on PF candidates and SVs follow those described in Section~\ref{sec:ak4_jet_tagging}. The \pnet model for AK8 jets allows for a larger graph (up to 100 nodes), more nearest neighbours (16 per node), and about twice as many trainable parameters compared to the AK4 jet model. This network targets the following jet classes: \xbb, \xcc, \xqq, \xgg, \xtt, and jets from QCD multijet events. Unlike the offline \pnet~\cite{CMS:2025kje}, QCD jets are not subdivided in exclusive classes based on the \PQb and \PQc content, since the main goal at trigger level is to maximize the inclusive background suppression.

The training uses jets with ${\pt>250\GeV}$, ${\abs{\eta}<2.5}$, and ${\mSD>40\GeV}$, matched within ${\Delta R < 0.8}$ to generator-level jets. As outlined in Section~\ref{sec:dataset}, signal jets are obtained from simulated graviton resonances, with variable mass, decaying into scalar boson (X) pairs with ${30<m_{X}<200\GeV}$, ensuring a mass-decorrelated tagger over a wide \pt range. Background jets are taken from QCD multijet simulation. The data set is resampled to enforce a uniform distribution in the {\pt--\mSD} plane, preventing the use of global kinematics for discrimination among jet classes. The training procedure, as well as the definition of the training, validation, and test data sets, follows the strategy described in Section~\ref{sec:ak4_jet_tagging}.

Figure~\ref{fig:trigger_merged_xbb} (\cmsRight) shows the \xbb tagging efficiency versus the misidentification probability for QCD jets in the test sample, requiring ${\abs{\eta}<2.5}$ and ${\mSD>40\GeV}$ in two \pt ranges: ${300<\pt<400\GeV}$ (solid) and ${400<\pt<500\GeV}$ (dashed). Results for the Run~2 \DoubleB tagger (red) and the Run~3 \pnet discriminator (blue) demonstrate a significant improvement in \xbb tagging at the HLT with \pnet. This major improvement enabled a redesign of Run~3 triggers targeting high-\pt \hbb and \zbb signatures, substantially increasing the acceptance for key analyses such as \hhbb, as discussed in Section~\ref{sec:monitoring}.

\section{Jet tagging performance in data}\label{sec:monitoring}

The jet tagging algorithms described in Section~\ref{sec:jet_tagging} developed for HLT applications show significant performance gains over previous methods in simulated events. However, these improvements, enabled by machine learning techniques exploiting low-level features (\eg IP variables, SVs, track properties, and calorimetric information of PF candidates), must be validated using observed events. In addition, the \pnet taggers for both AK4 and AK8 jets are trained on simulated samples corresponding to ideal detector conditions at the start of Run~3. Since detector performance degrades over time due to radiation damage and hardware failures, it is essential to validate these algorithms with early Run~3 data and monitor their stability throughout data taking. In this section, the performance of \pnet for heavy-flavour tagging of AK4 and AK8 jets is presented using collision data collected in 2024, corresponding to an integrated luminosity of 109\fbinv. The performance observed in data is compared with  expectations from simulation as a function of several key observables. End-to-end trigger efficiency measurements for the triggers described in Section~\ref{sec:trigger_menu} are not included in this paper, as they depend on the specific analysis phase space and are therefore left to individual analyses. Mistagging rate measurements are also not presented, as the primary goal of this section is the data-to-simulation comparison for signal jets.

\subsection{Measurement of online \texorpdfstring{\PQb}{b} tagging efficiency}\label{sec:monitor_btag}

Top quark pairs are abundantly produced at the LHC, with a cross section of about 924\unit{pb} at ${\sqrt{s}=13.6\TeV}$, computed with {{\textsc{Top++} v2.0}}~\cite{Czakon:2011xx}. Since the top quark decays to $\PW\PQb$ nearly 100\% of the time~\cite{ParticleDataGroup:2024cfk}, \ttbar events provide a large and pure source of \PQb jets. While the largest yield of \ttbar events can be obtained by selecting either fully hadronic or lepton + jets final states, the highest purity is achieved in $\ttbar \to 2\Pell 2\nu \PQb\PAQb$ events, where both \PW bosons decay leptonically ($\Pell=\Pe,\PGm$). This sample is therefore used to monitor the \PQb tagging performance. To suppress the Drell--Yan background (${\PZ/\PGgst \to \Pellp\Pellm}$), events with at least two AK4 jets and one opposite-sign electron--muon pair are selected.

Events are selected at the L1T stage using the logical OR of several electron--muon and single-muon triggers: two requiring $\pt>5\GeV$ (muon) and $\pt>23\GeV$ (electron), and $\pt>7\GeV$ (muon) and $\pt>20\GeV$ (electron), respectively, as well as additional seeds requiring either $\pt>20\GeV$ (muon) and $\pt>10\GeV$ (electron), or a single muon with $\pt>22\GeV$. At the HLT, events are required to contain at least one isolated muon (electron) with $\pt>8~(23)\GeV$ or $\pt>23~(12)\GeV$, depending on the trigger. Lepton isolation is characterized as the scalar sum of the \pt of additional tracks associated with the PV and calorimeter energy deposits within a cone of radius ${\Delta R=0.3}$ around the lepton direction. Identification and isolation efficiencies exceed 95\% (90\%) for muons (electrons) in simulated \ttbar events. Offline, an opposite-sign $\Pe\PGm$ pair with $m_{\Pe\PGm}>20\GeV$ is required to suppress contributions from low-mass resonances and non-prompt leptons originating from heavy-flavour decays. Both leptons must satisfy ${\pt>25\GeV}$ and ${\abs{\eta}<2.5}$, with tight identification and isolation criteria applied. Events with additional loosely identified leptons with ${\pt>10\GeV}$ and ${\abs{\eta}<2.5}$ are vetoed. Selected events must contain at least two AK4 jets with ${\pt>25\GeV}$ and ${\abs{\eta}<2.5}$, separated from the selected leptons by ${\Delta R(j,\Pell)>0.4}$. This defines a control region in data with $\ttbar(\Pe\PGm)$ purity above 95\%. Residual backgrounds arise mainly from single top (dominated by the $\PQt\PW$ channel) and ${\PWpm\PWmp}$ production. The two offline jets with the highest \PQb tagging discriminator value (\probb) are required to match (${\Delta R<0.4}$) two HLT jets with ${\pt>30\GeV}$ and ${\abs{\eta}<2.5}$. A loose requirement on the offline \pnet \probb score is applied to suppress non-\PQb jets, yielding a highly pure \PQb sample for efficiency measurements.

\begin{figure*}[!htb]
    \centering
     \includegraphics[width=0.45\textwidth]{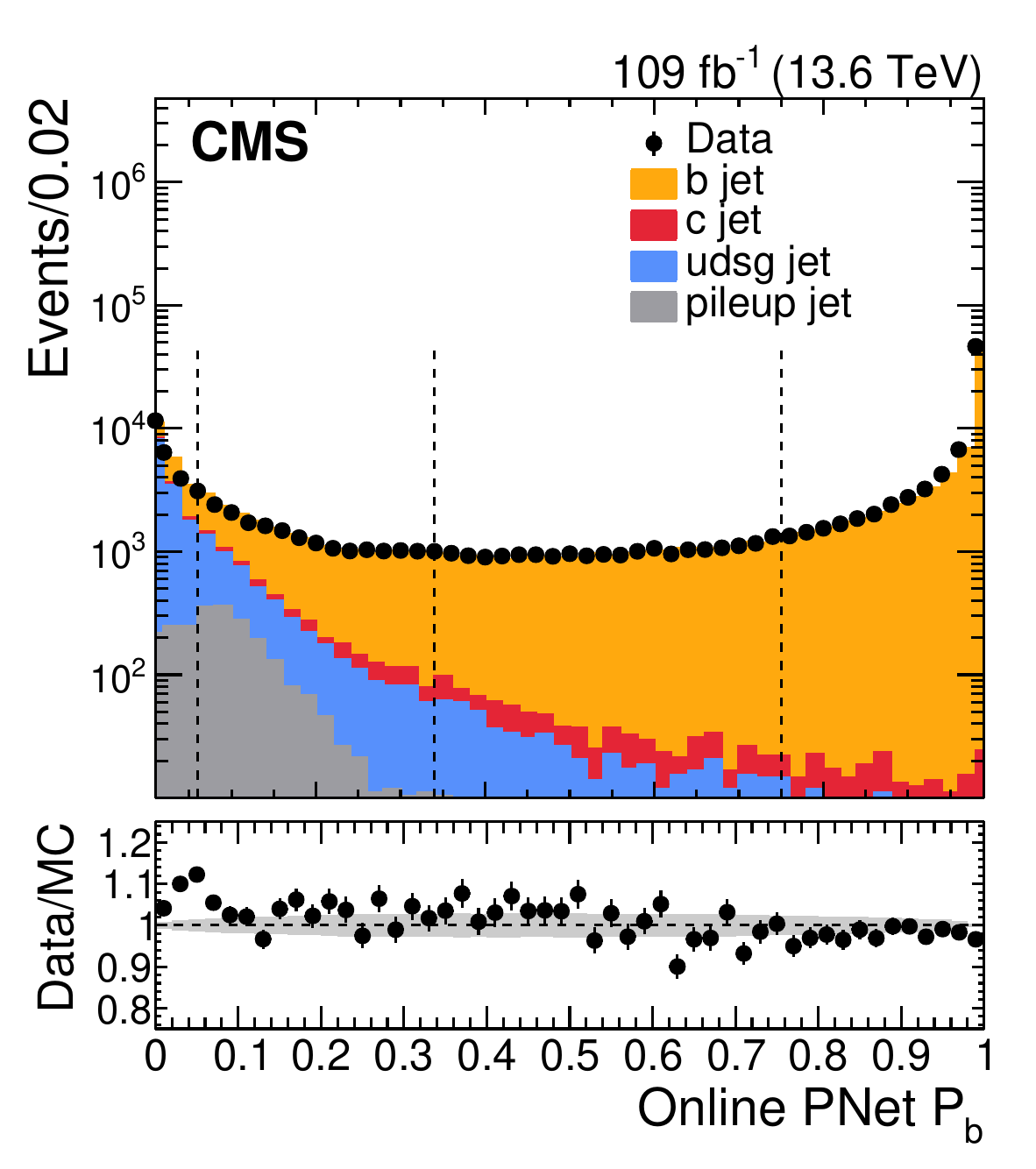}
     \includegraphics[width=0.45\textwidth]{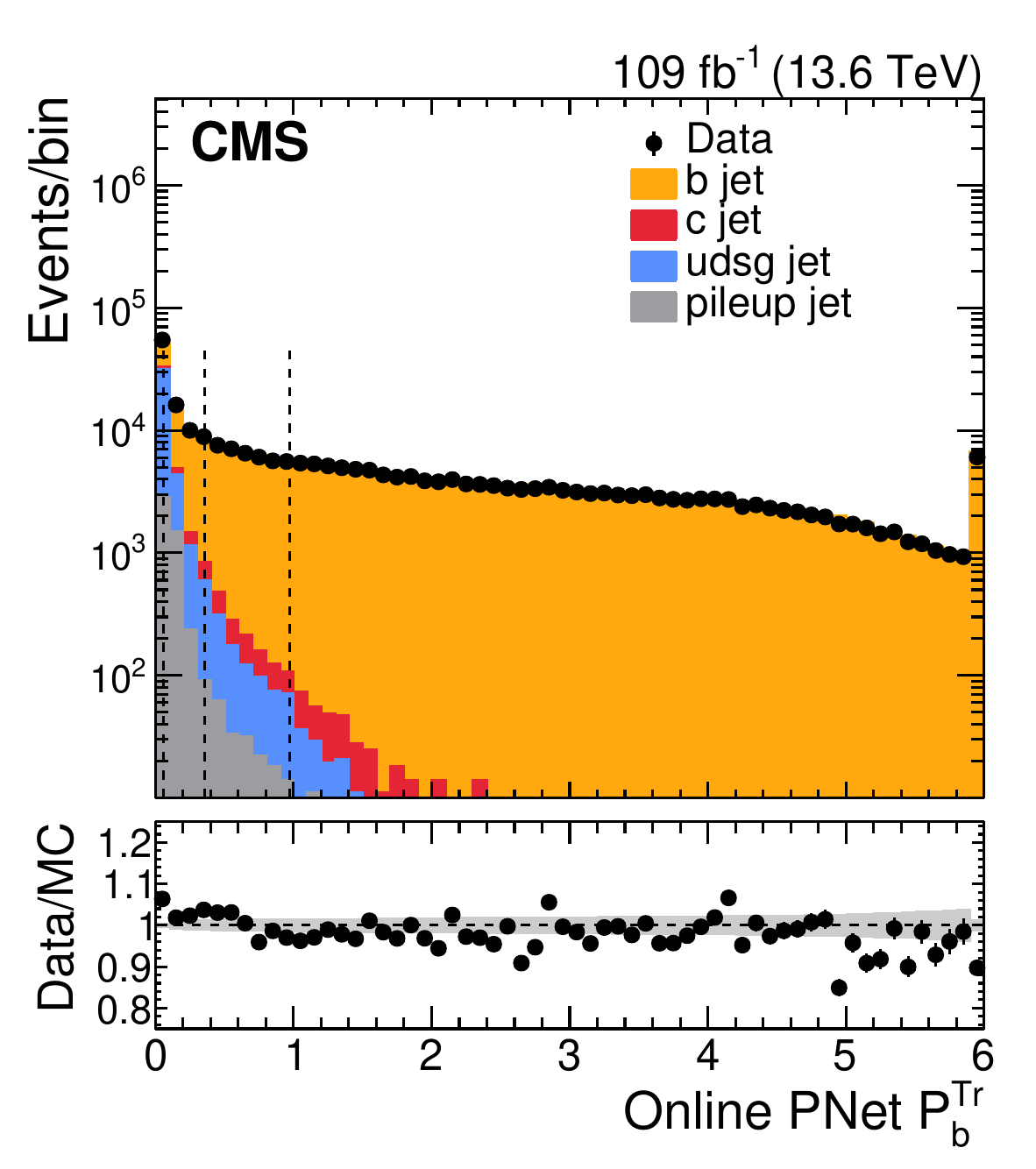}	
    \caption{Left: distribution of the online \pnet \probb score for AK4 jets with online ${\pt>30\GeV}$ and ${\abs{\eta}<2.5}$ in the ${\ttbar(\Pe\PGm)}$ region, shown for data (black points) and the MC prediction for SM processes. The simulated contribution is separated into exclusive jet-flavour categories: \PQb (orange), \PQc (red), and light-flavour quarks plus gluons (blue). Jets originating from pileup interactions are shown in gray. The lower panel displays the ratio of data to the MC prediction as a function of \probb, and the grey error band displays the statistical uncertainty of the simulation. The vertical bars on the data points represent the statistical uncertainty. Right: distribution of the transformed \PQb tagging score (\probbtr), defined as $\probbtr = \tanh^{-1}(\probb)$. The black dashed vertical lines indicate, from left to right, the three \PQb tagging working points (L, M, and T) used in the efficiency studies.}
    \label{fig:data_mc_pnet}
\end{figure*}

Figure~\ref{fig:data_mc_pnet} (\cmsLeft) shows the distribution of the online \probb score of the selected AK4 jets in the ${\ttbar(\Pe\PGm)}$ region, comparing data and simulation. Simulated contributions from SM processes are categorized by jet flavour, whereas jets without a matched truth flavour are attributed to pileup interactions. Good agreement is observed in both normalization and shape between data and simulation. As expected, many \PQb jets populate the region near ${\probb=1}$, leading to a concentration of jets in the highest score bin of Fig.~\ref{fig:data_mc_pnet} (\cmsLeft). To better resolve this region, a transformation ${\tanh^{-1}(\probb)}$ is applied. The transformed score, \probbtr, is shown in Fig.~\ref{fig:data_mc_pnet} (\cmsRight).

\begin{figure*}[!htb]
    \centering
    \includegraphics[width=0.32\textwidth]{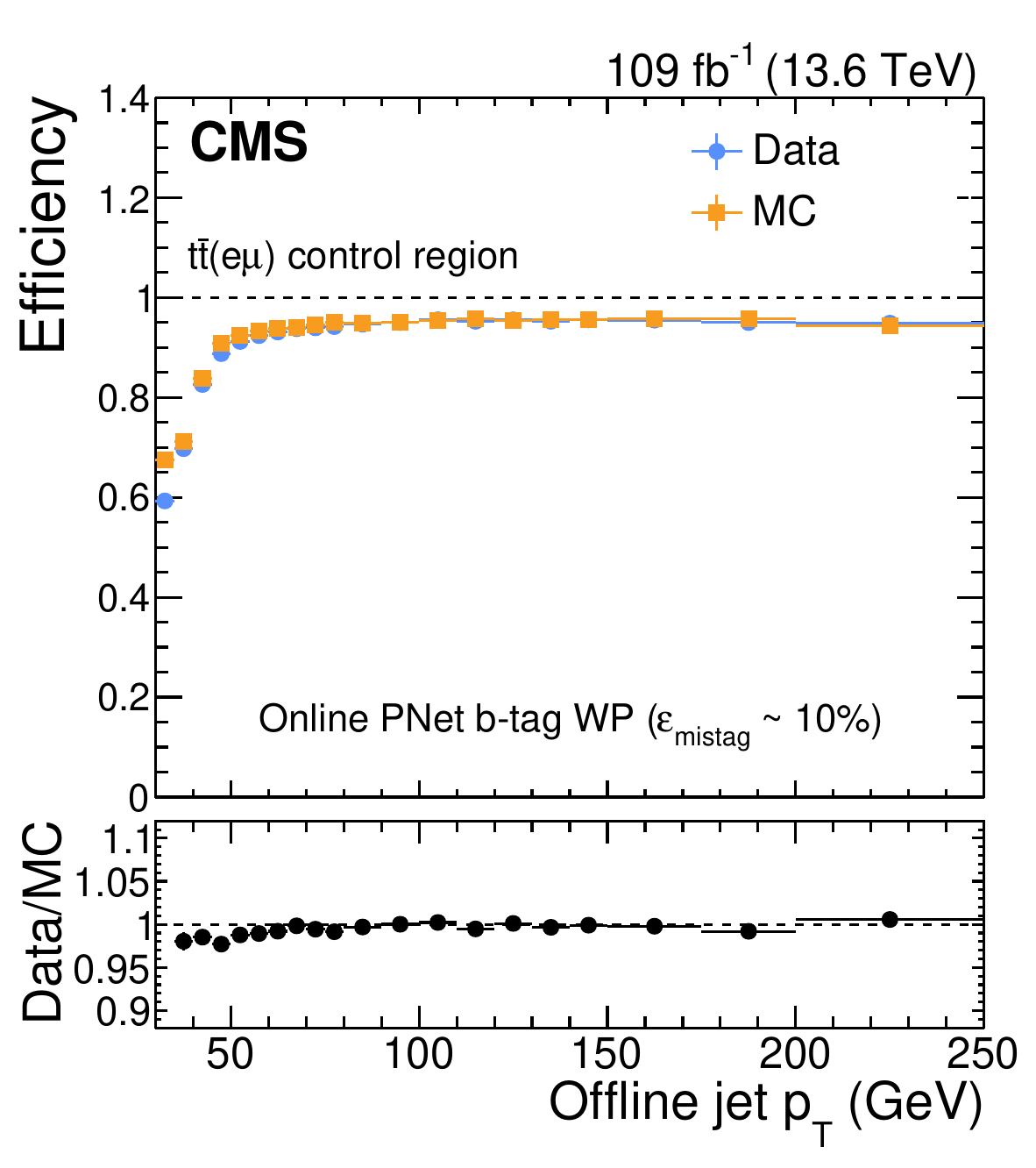}
    \includegraphics[width=0.32\textwidth]{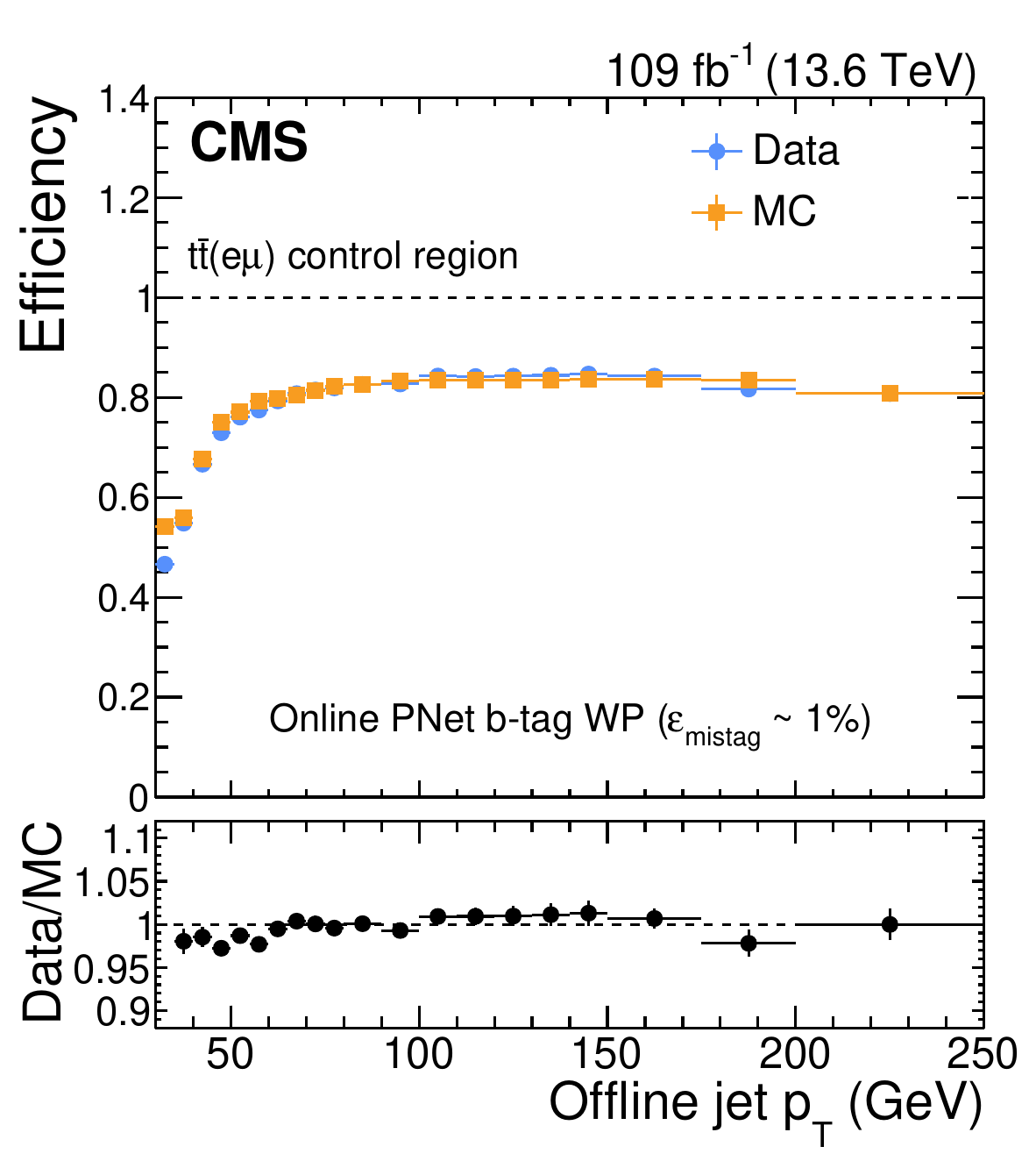}
    \includegraphics[width=0.32\textwidth]{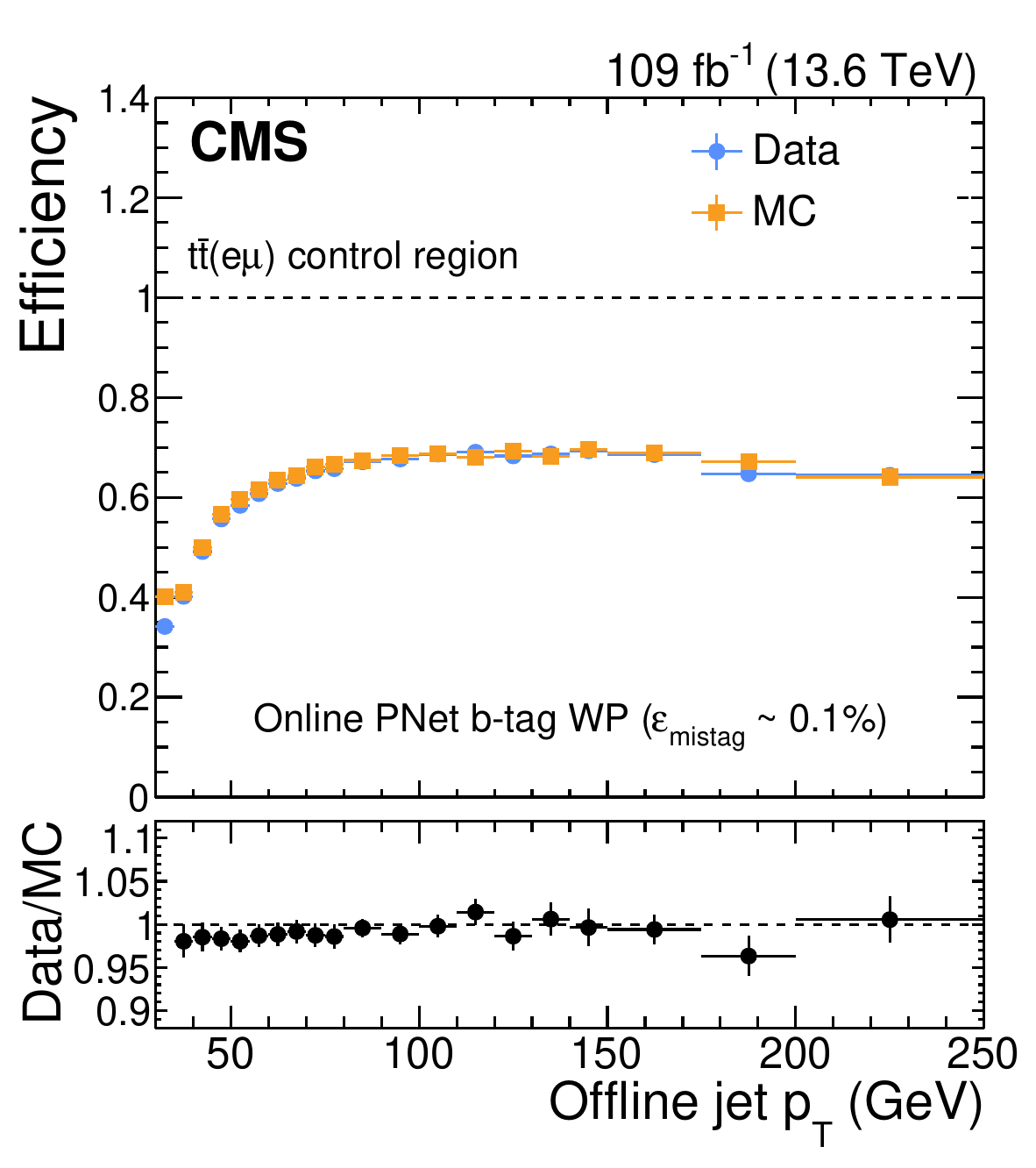}
    \caption{Efficiency of the online \pnet \PQb tagging algorithm as a function of the offline Ak4 jet \pt in the $\ttbar(\Pe\PGm)$ region, measured in data (blue) and simulation (orange). Results are shown for the loose (\cmsLeft), medium (middle), and tight (\cmsRight) \probb working points (WP), corresponding to mistag rates ($\varepsilon_{\mathrm{mistag}}$) of 10\%, 1\%, and 0.1\%, respectively. The vertical bars on the data and simulation points represent the statistical uncertainty.}
    \label{fig:btag_efficiency_hlt_pt}
\end{figure*}

\begin{figure*}[!htb]
    \centering
    \includegraphics[width=0.32\textwidth]{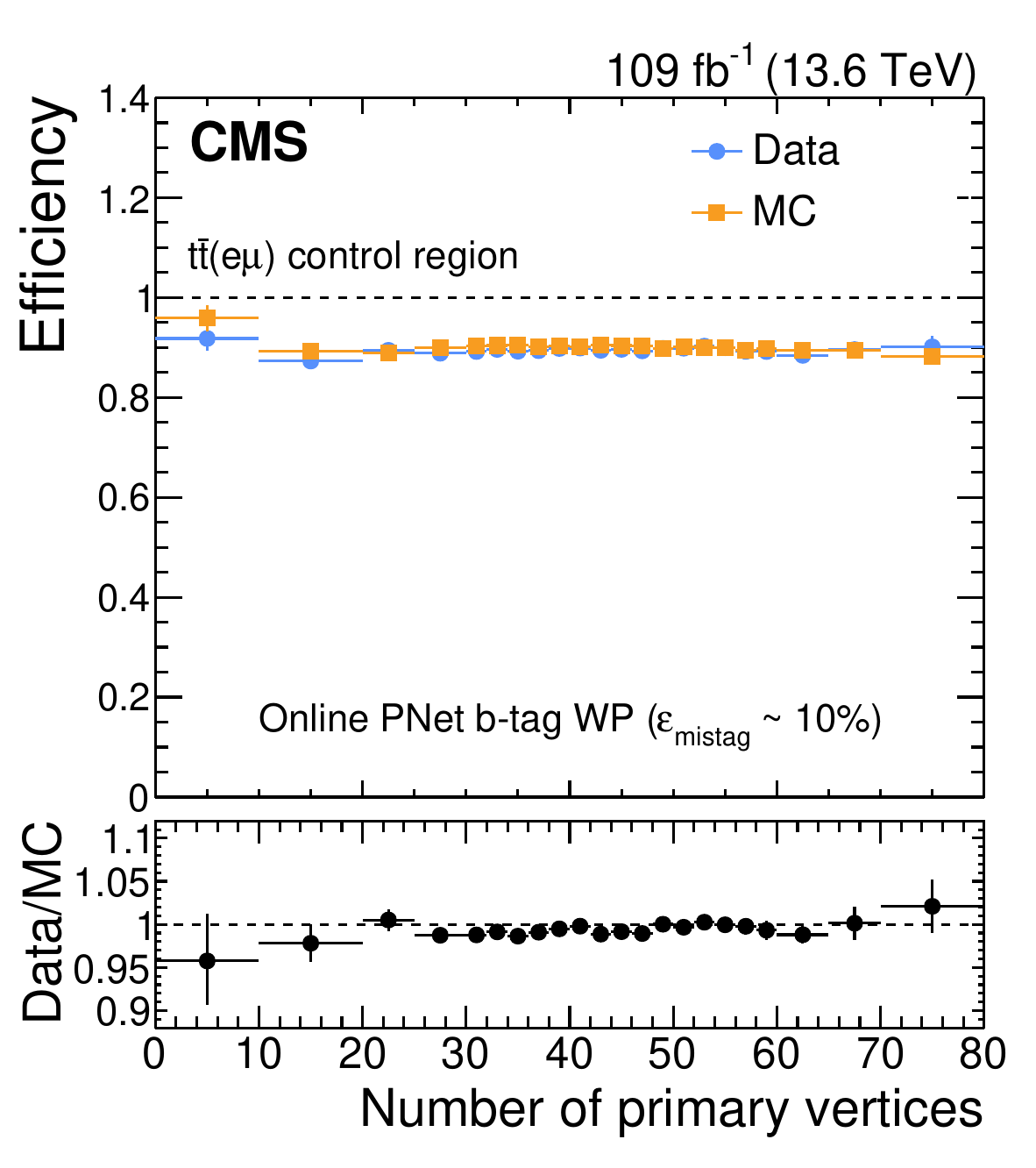}
    \includegraphics[width=0.32\textwidth]{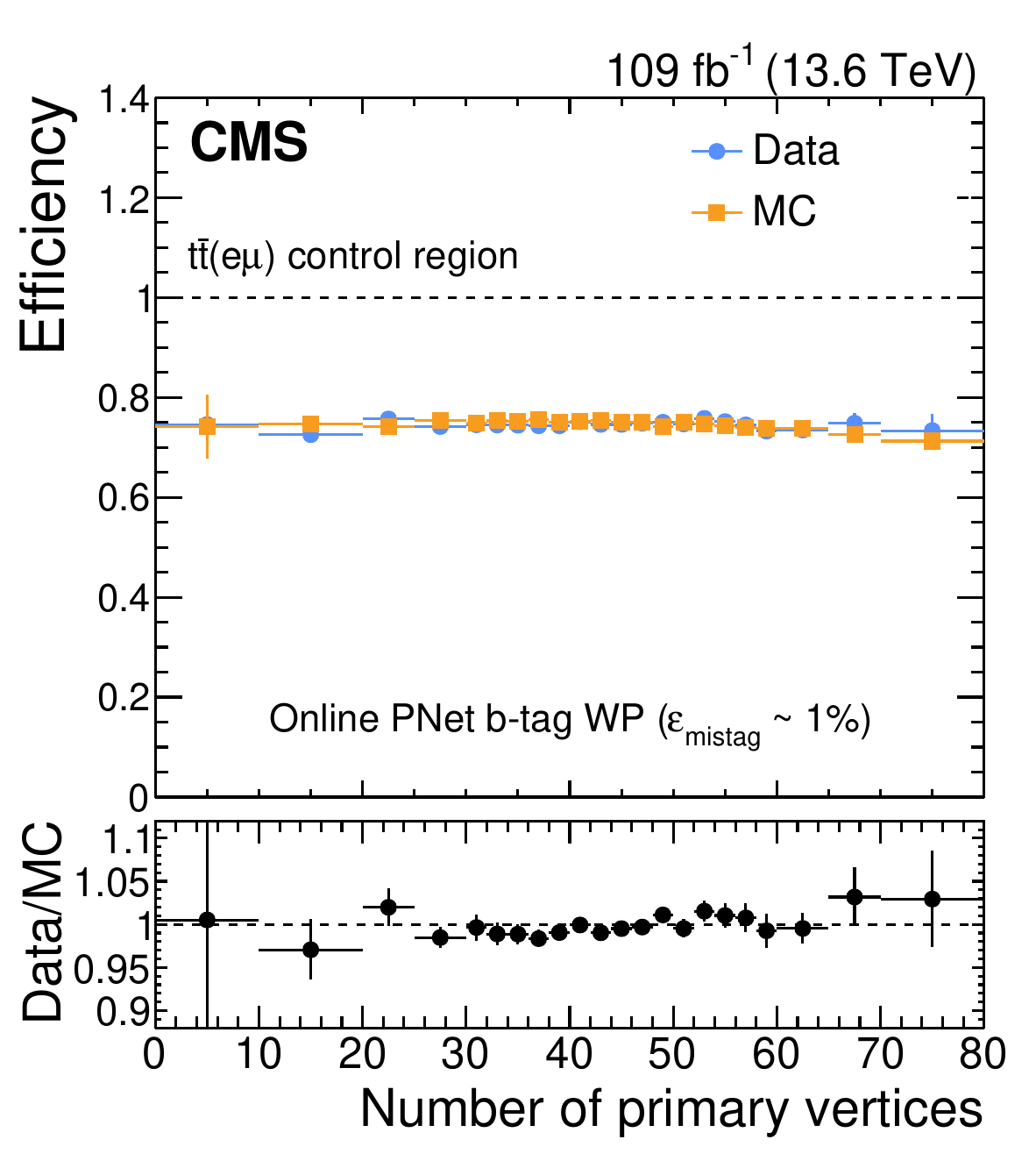}
    \includegraphics[width=0.32\textwidth]{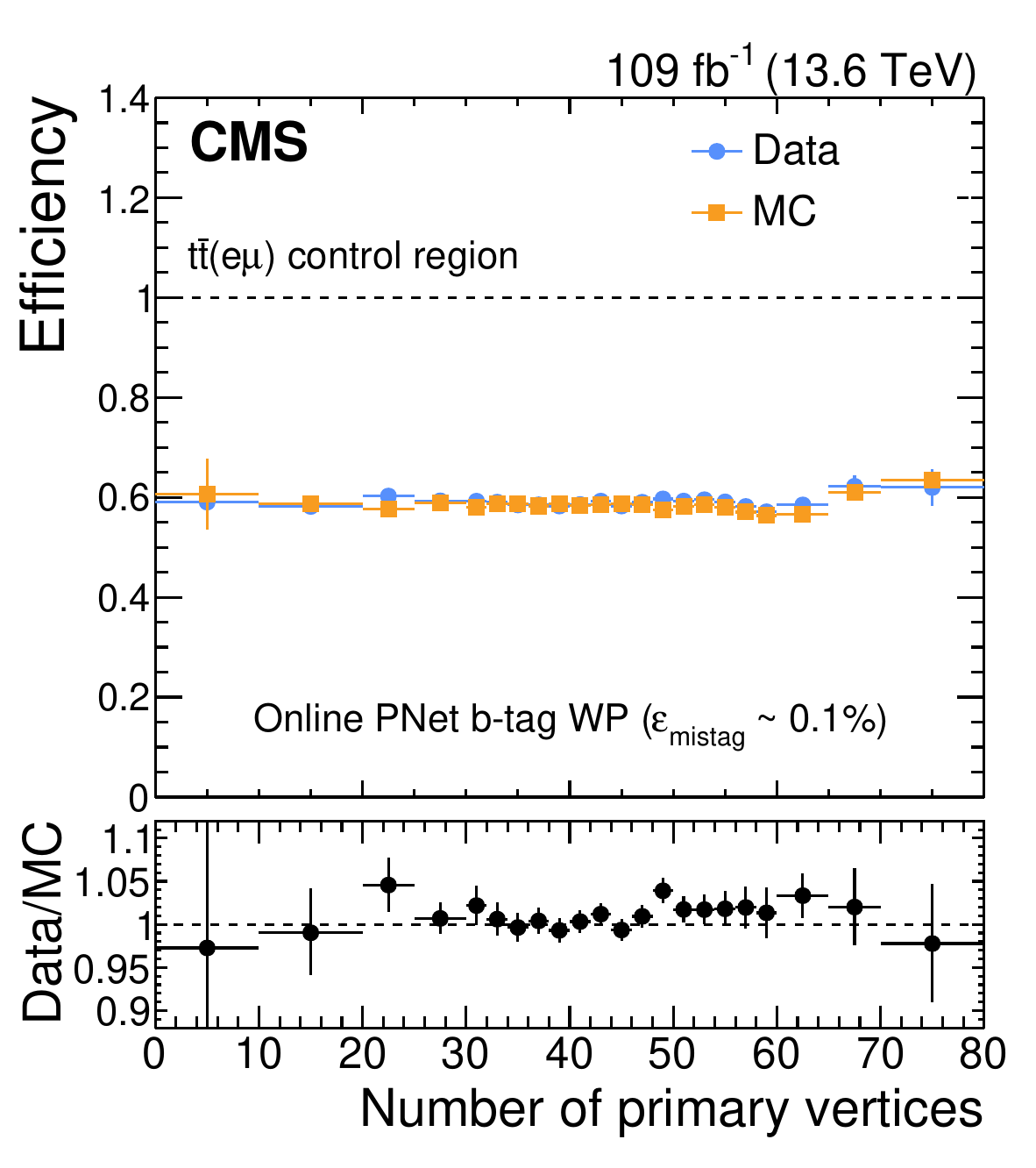}
    \caption{Efficiency of the online \pnet \PQb tagging algorithm as a function of the number of reconstructed primary vertices in the $\ttbar(\Pe\PGm)$ region, measured in data (blue) and simulation (orange). The panel definitions and styling follow Fig.~\ref{fig:btag_efficiency_hlt_pt}.}
    \label{fig:btag_efficiency_hlt_npv}
\end{figure*}

\begin{figure*}[!htb]  
    \centering
    \includegraphics[width=0.32\textwidth]{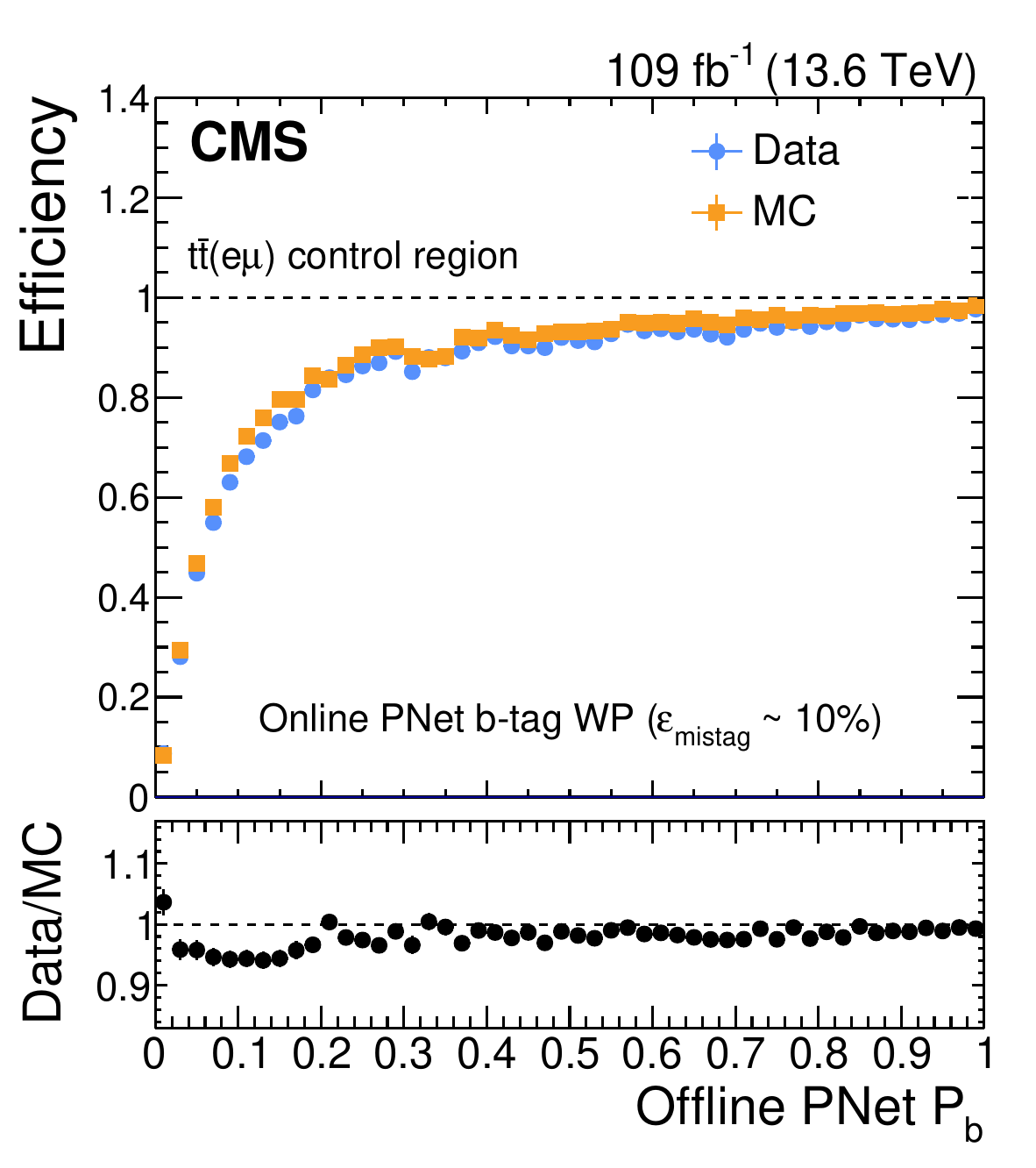}
    \includegraphics[width=0.32\textwidth]{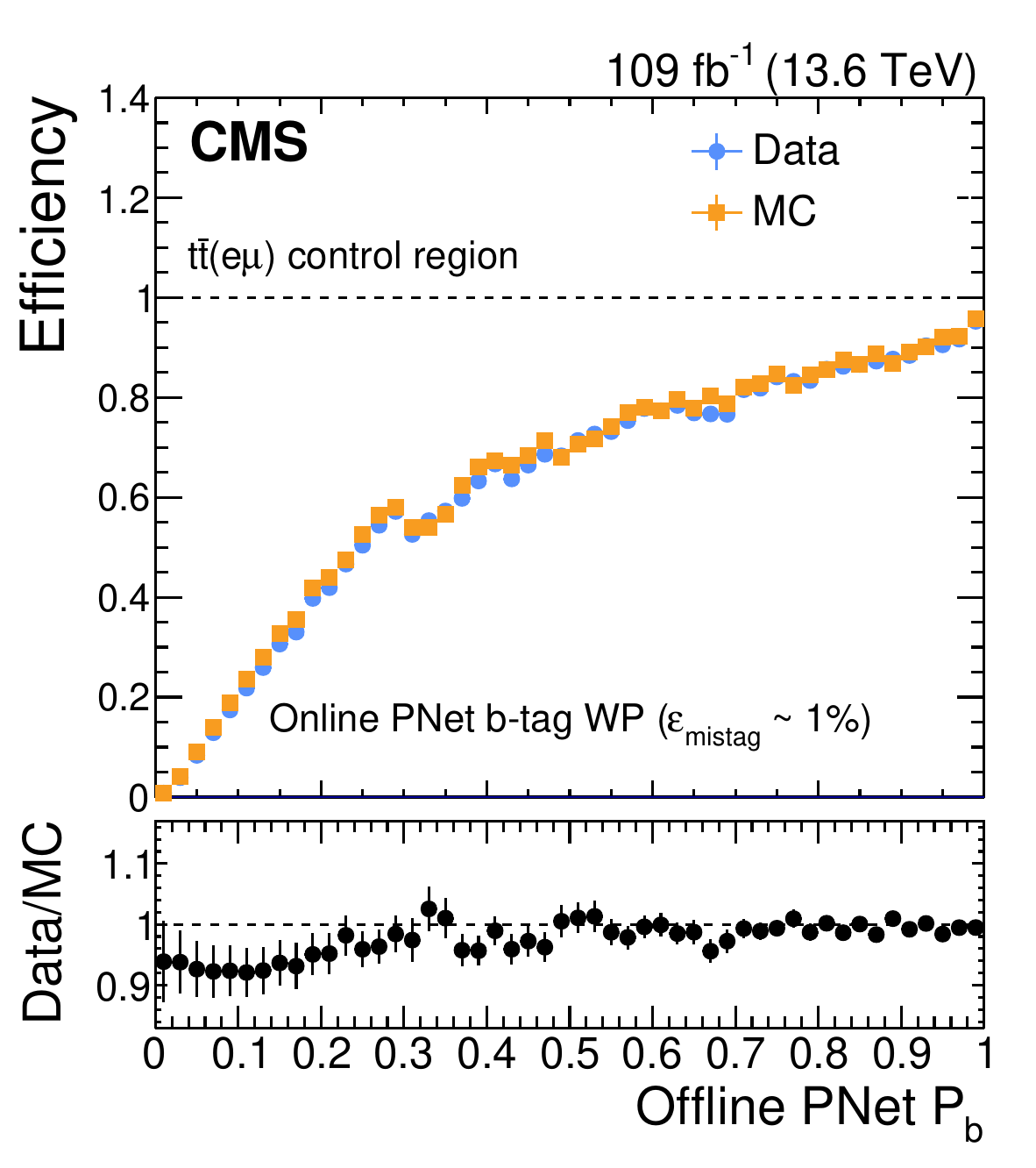}
    \includegraphics[width=0.32\textwidth]{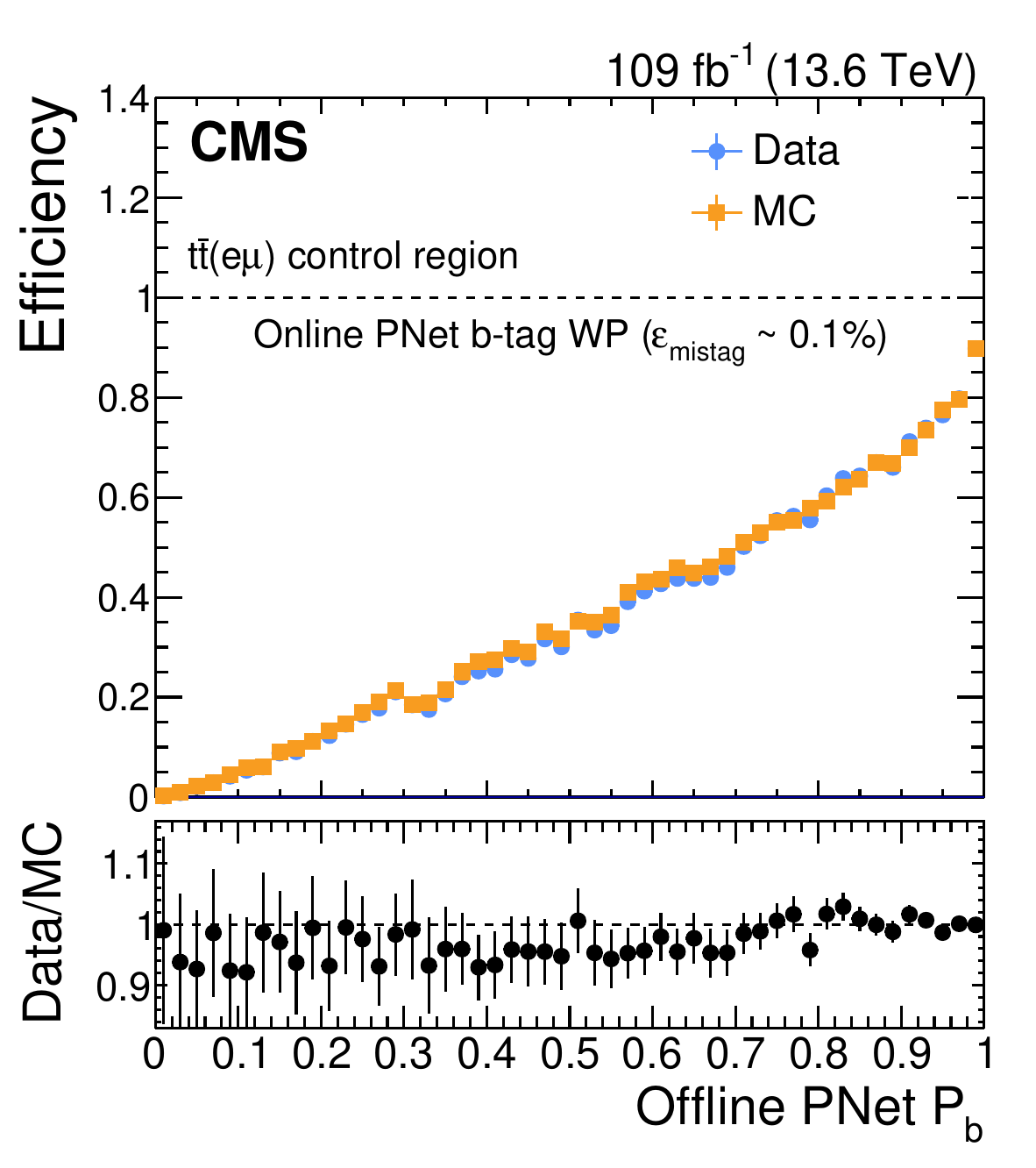}
    \caption{Efficiency of the online \pnet \PQb tagging algorithm as a function of the offline \pnet \probb score in the $\ttbar(\Pe\PGm)$ region, for data (blue) and simulation (orange). The panel definitions and styling follow Fig.~\ref{fig:btag_efficiency_hlt_pt}. For this measurement, no requirements are applied to the offline \probb score.}
    \label{fig:btag_efficiency_hlt_pbb}
\end{figure*}

\begin{figure*}[!htb]  
    \centering
    \includegraphics[width=0.32\textwidth]{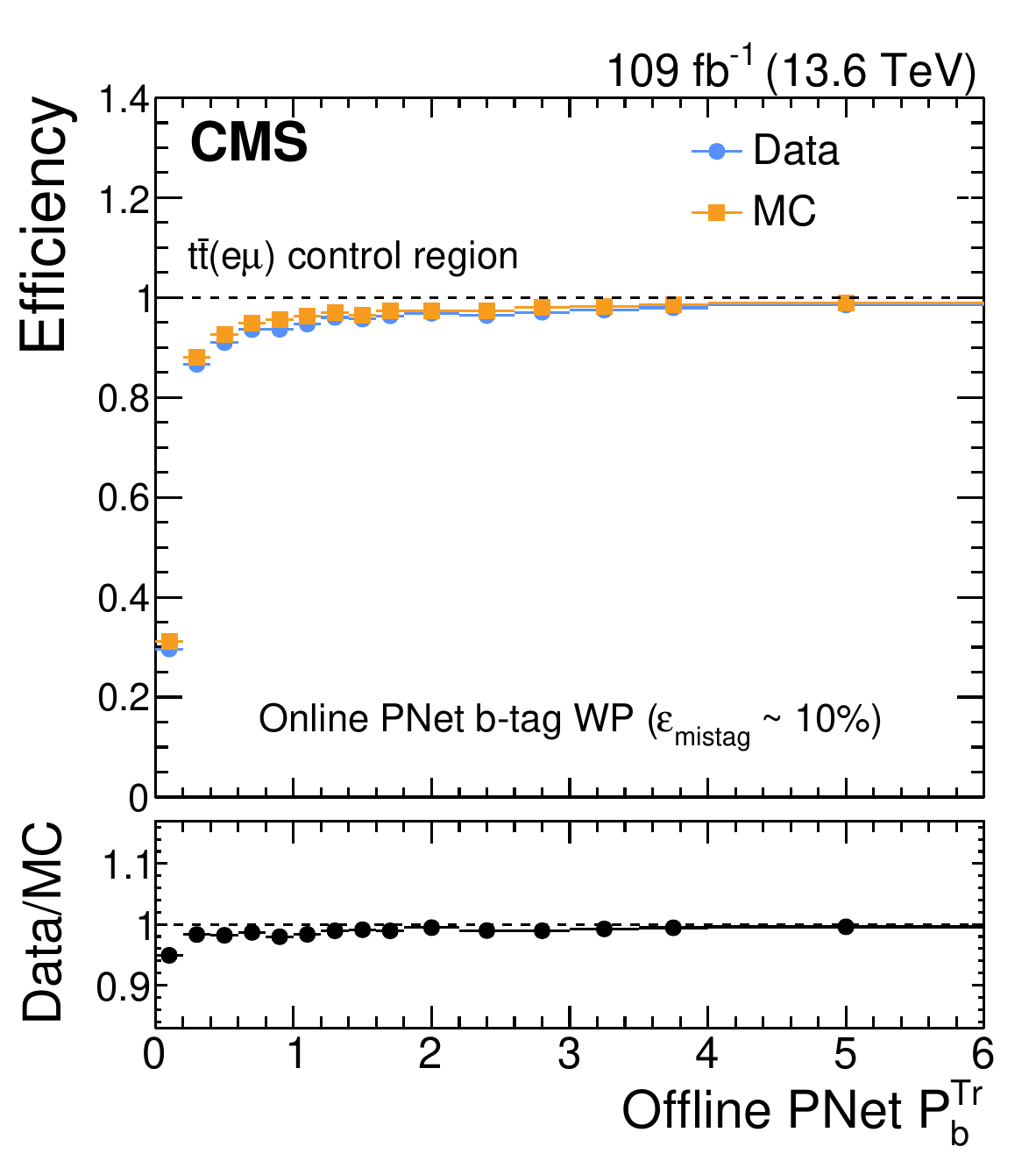}
    \includegraphics[width=0.32\textwidth]{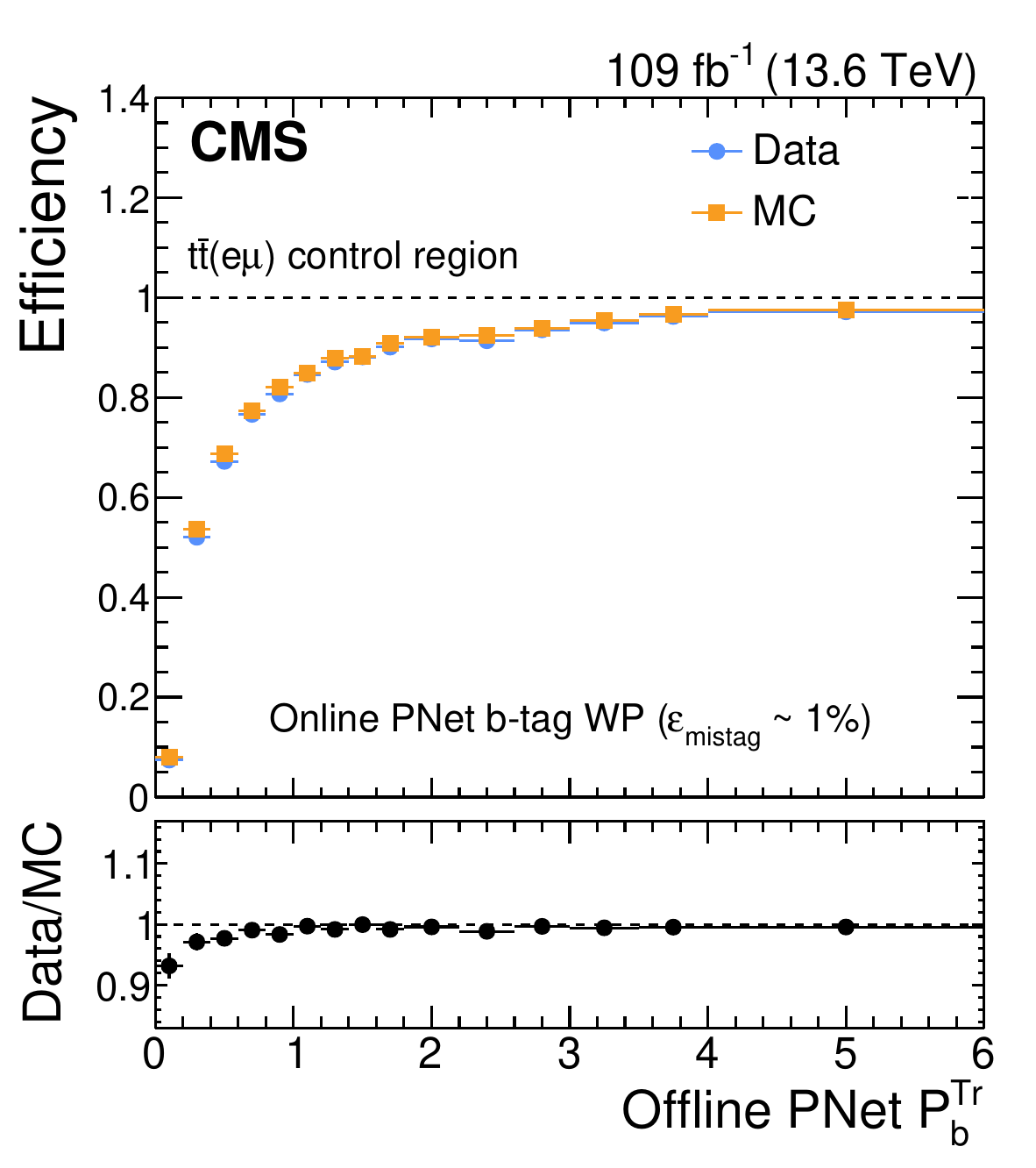}
    \includegraphics[width=0.32\textwidth]{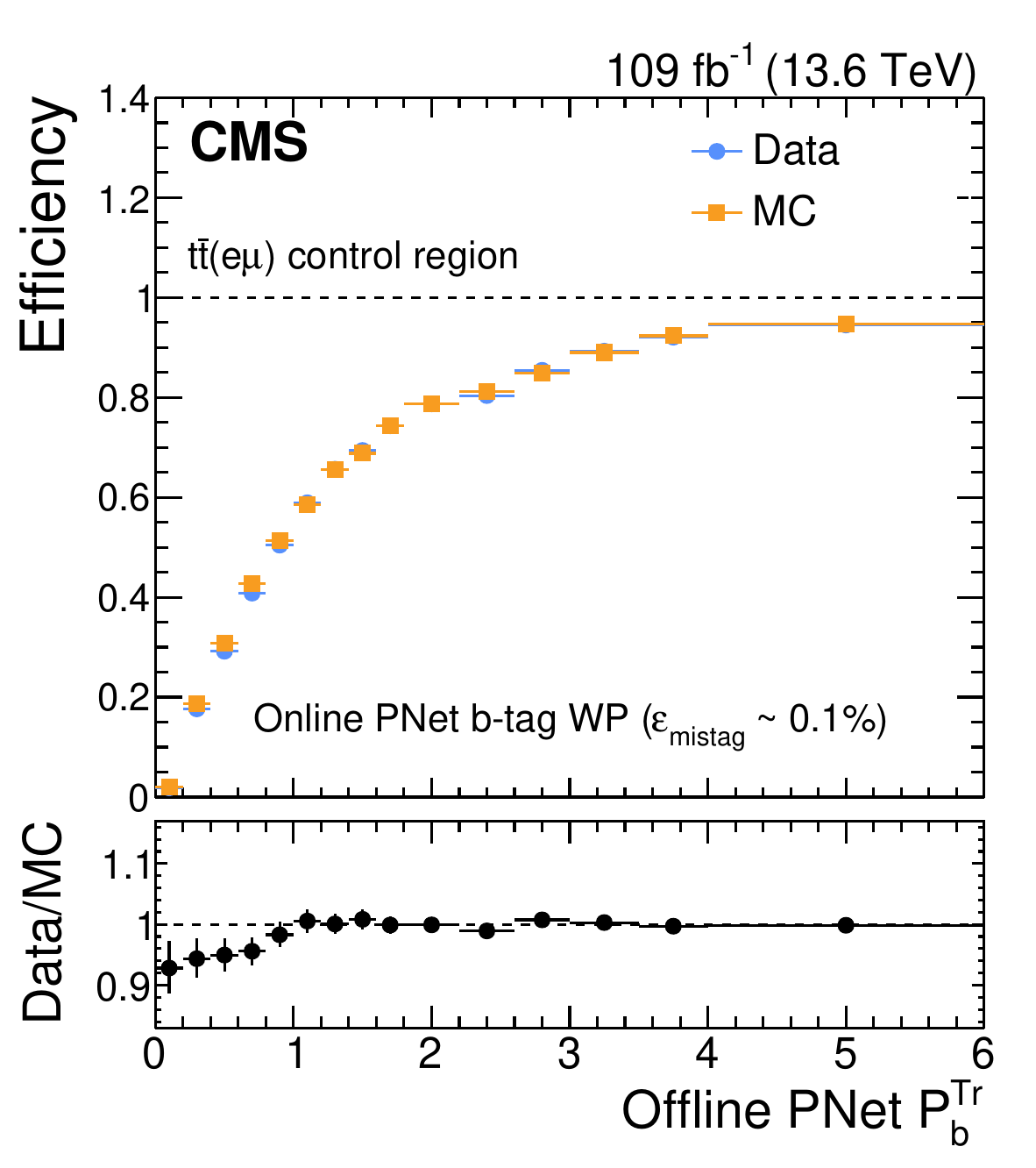}
    \caption{Efficiency of the online \pnet \PQb tagging algorithm as a function of the offline \pnet \probbtr score in the $\ttbar(\Pe\PGm)$ region, for data (blue) and simulation (orange). The panel definitions and styling follow Fig.~\ref{fig:btag_efficiency_hlt_pt}. For this measurement, no requirements are applied to the offline \probb score.}
    \label{fig:btag_efficiency_hlt_pbb_transf}
\end{figure*}

As described above, the online \pnet $\PQb$ tagging efficiency is measured in data and simulation using the two AK4 jets with the highest \probb score in each event. Efficiencies are evaluated for three online \probb working points (WPs), defined in simulated \ttbar events to target misidentification rates of 10\% (loose), 1\% (medium), and 0.1\% (tight) for background jets (\PQc, uds, and gluon). These definitions follow the convention used for offline \PQb tagging algorithms~\cite{Bols:2020bkb} and correspond to average online \PQb tagging efficiencies of 90, 75, and 60\%, respectively. The upper panels of Figs.~\ref{fig:btag_efficiency_hlt_pt}--\ref{fig:btag_efficiency_hlt_pbb_transf} show the online \PQb tagging efficiencies in data and simulation for the loose (\cmsLeft), medium (middle), and tight (\cmsRight) WPs as functions of the offline jet \pt, number of PVs, and the \probb score from the \pnet model trained on offline jets~\cite{CMS-DP-2024-066}. The lower panels display the ratio between the efficiencies in data and simulation. The online \pnet performance depends on jet \pt and $\eta$ for all WPs, with the highest efficiency for jets with ${\pt>60\GeV}$ and ${\abs{\eta}<1.5}$, while remaining stable as a function of pileup. Note that the loose requirement on the offline \probb score is omitted when measuring the trigger efficiency as a function of the \probb score in Figs.~\ref{fig:btag_efficiency_hlt_pbb}--\ref{fig:btag_efficiency_hlt_pbb_transf}.

Overall, the online \PQb tagging performance is well described by simulation, and the data-to-simulation efficiency ratios are close to unity with a spread of only a few percent. The largest discrepancies are observed for jets with ${\pt<50\GeV}$ (Fig.~\ref{fig:btag_efficiency_hlt_pt}) and for jets with small offline \probb values (Fig.~\ref{fig:btag_efficiency_hlt_pbb}). In most analyses, such jets passing the online \PQb tagging requirements but assigned a small offline \probb score are either removed by offline selections or do not contribute significantly to high signal-to-background regions. Consequently, these discrepancies have a limited impact and are accounted for by applying data-to-simulation correction factors to simulated events. The uncertainties associated with these correction factors are derived from the statistical uncertainties in both data and simulation.

\begin{figure*}[!htb]  
    \centering
    \includegraphics[width=0.32\textwidth]{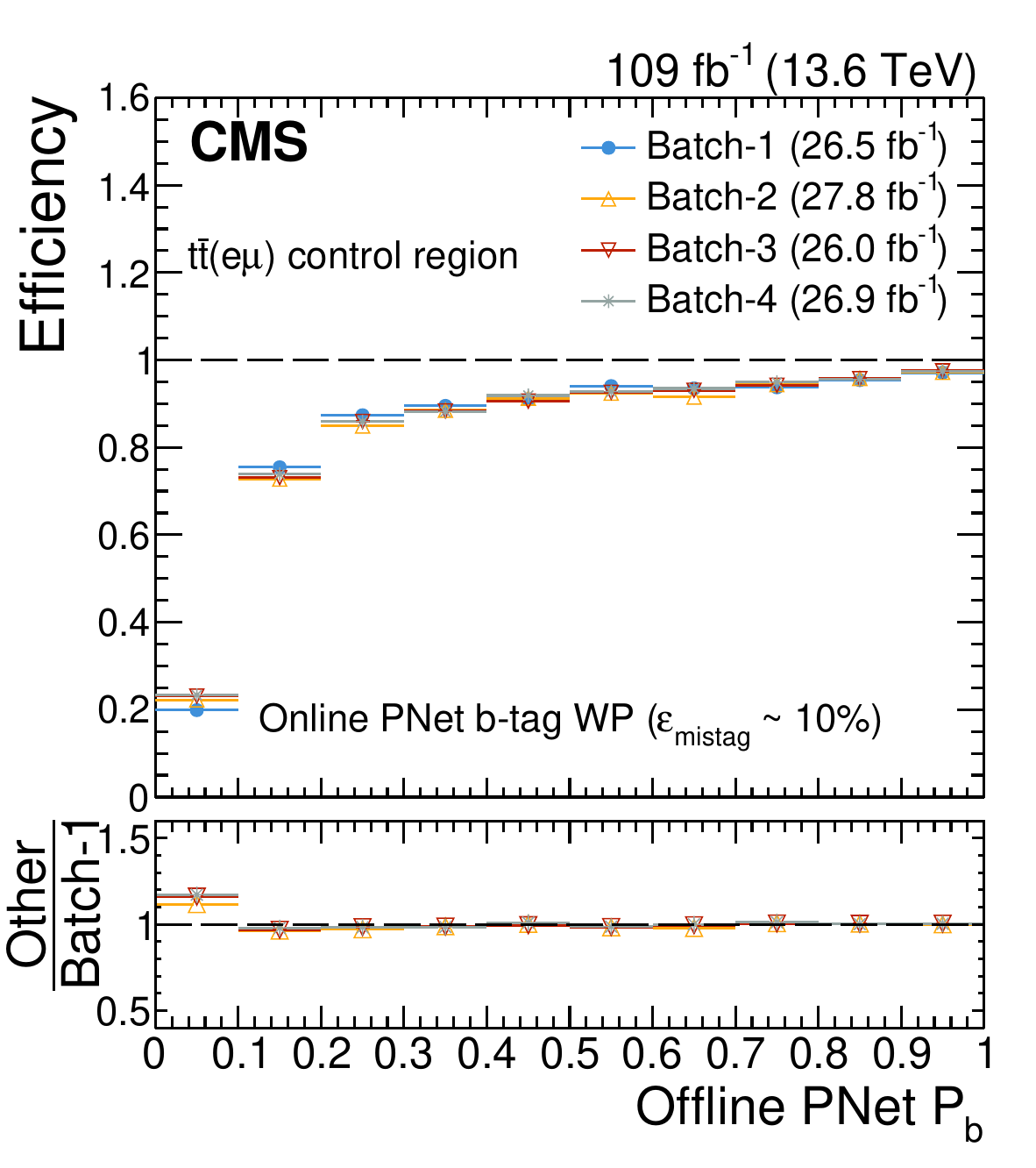}
    \includegraphics[width=0.32\textwidth]{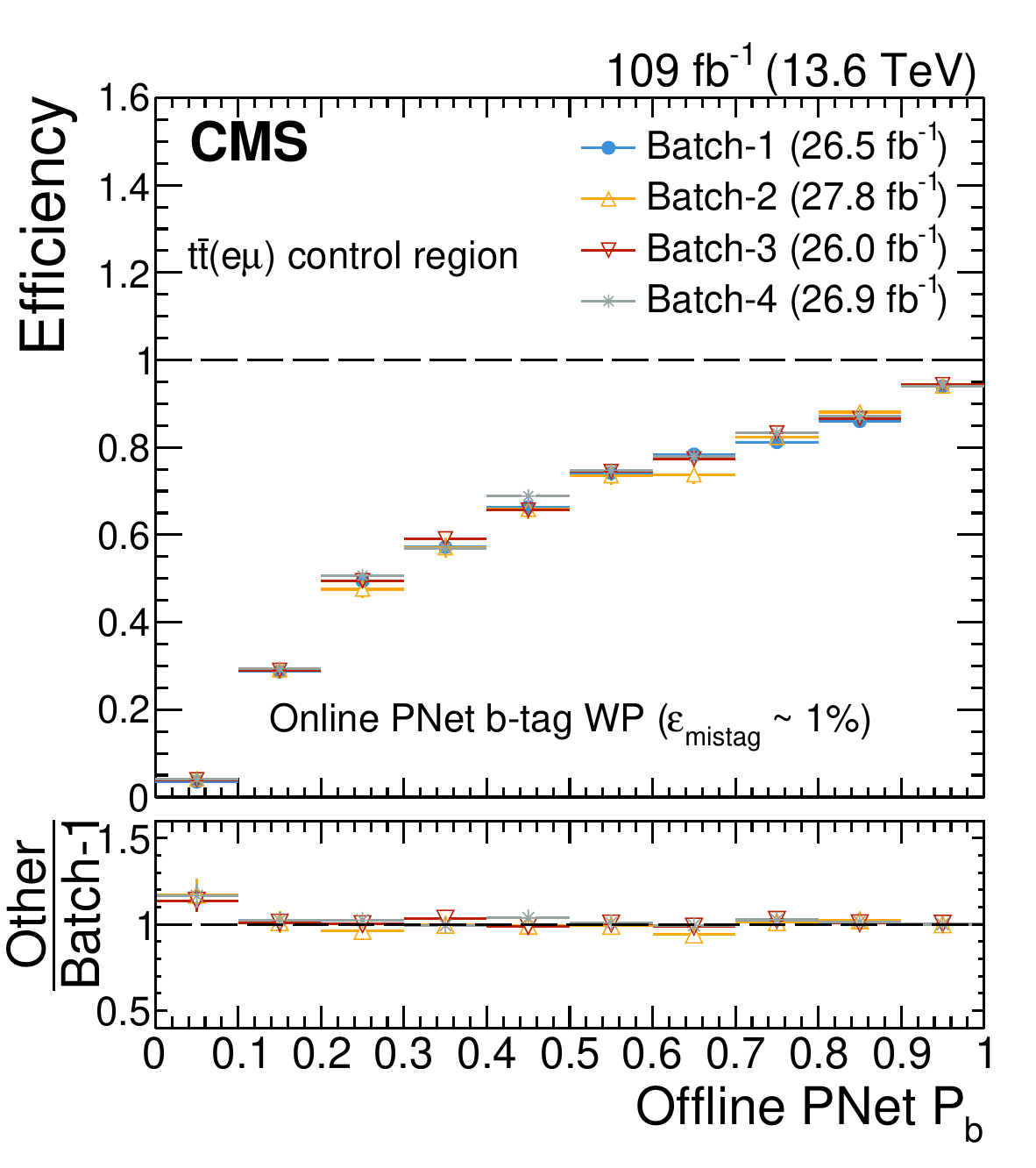}
    \includegraphics[width=0.32\textwidth]{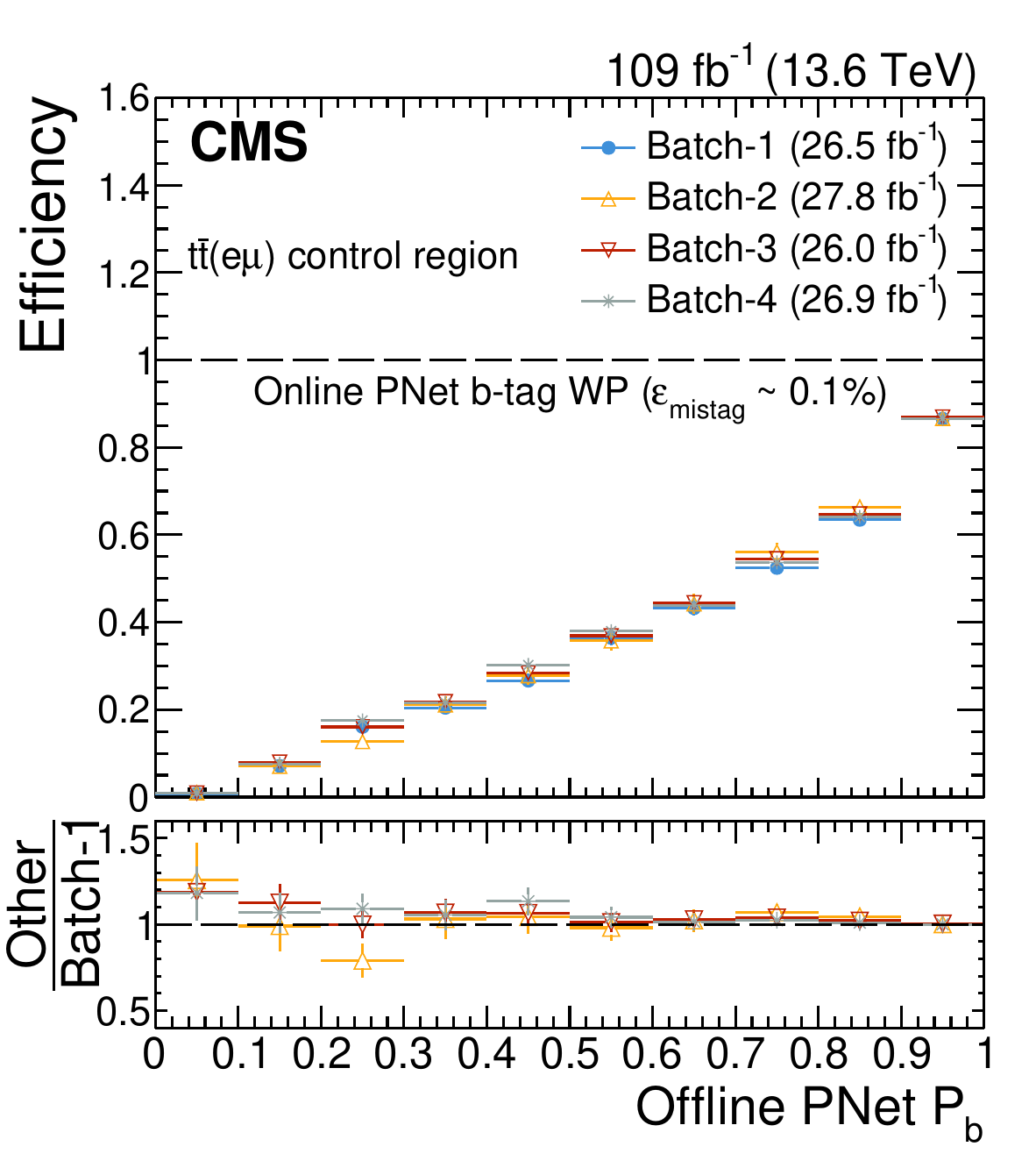}
    \caption{Efficiency of the online \pnet \PQb tagging algorithm as a function of the offline \pnet \probb score in the ${\ttbar(\Pe\PGm)}$ region, measured in data for four consecutive time periods (batch-1 to batch-4) during 2024. The lower panel shows the ratio of efficiencies in batch-$N$ ($N=2,\ldots,4$) to batch-1.}
    \label{fig:btag_efficiency_hlt_pbb_vs_time}
\end{figure*}

\begin{figure*}[!htb]  
    \centering
    \includegraphics[width=0.32\textwidth]{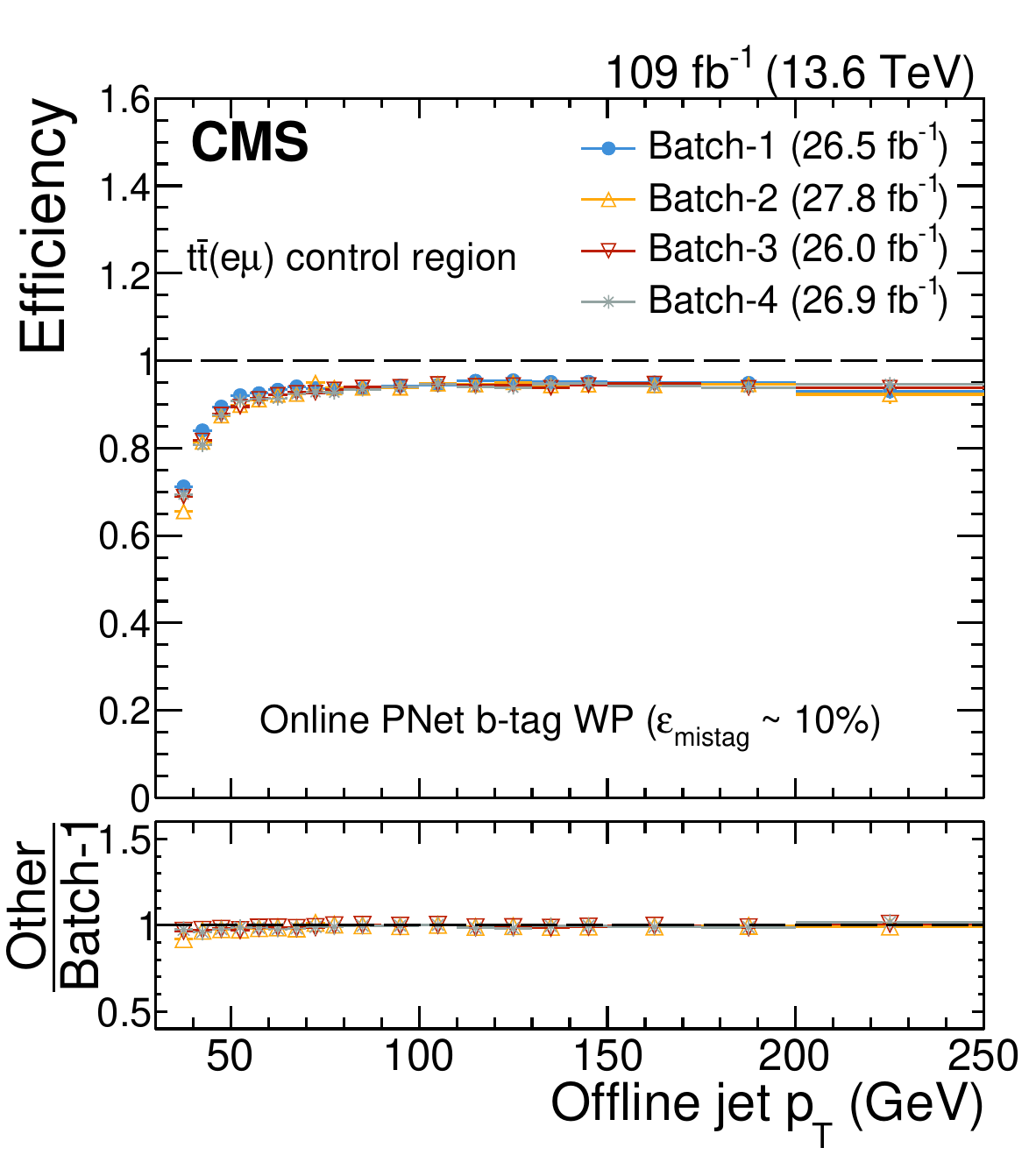}
    \includegraphics[width=0.32\textwidth]{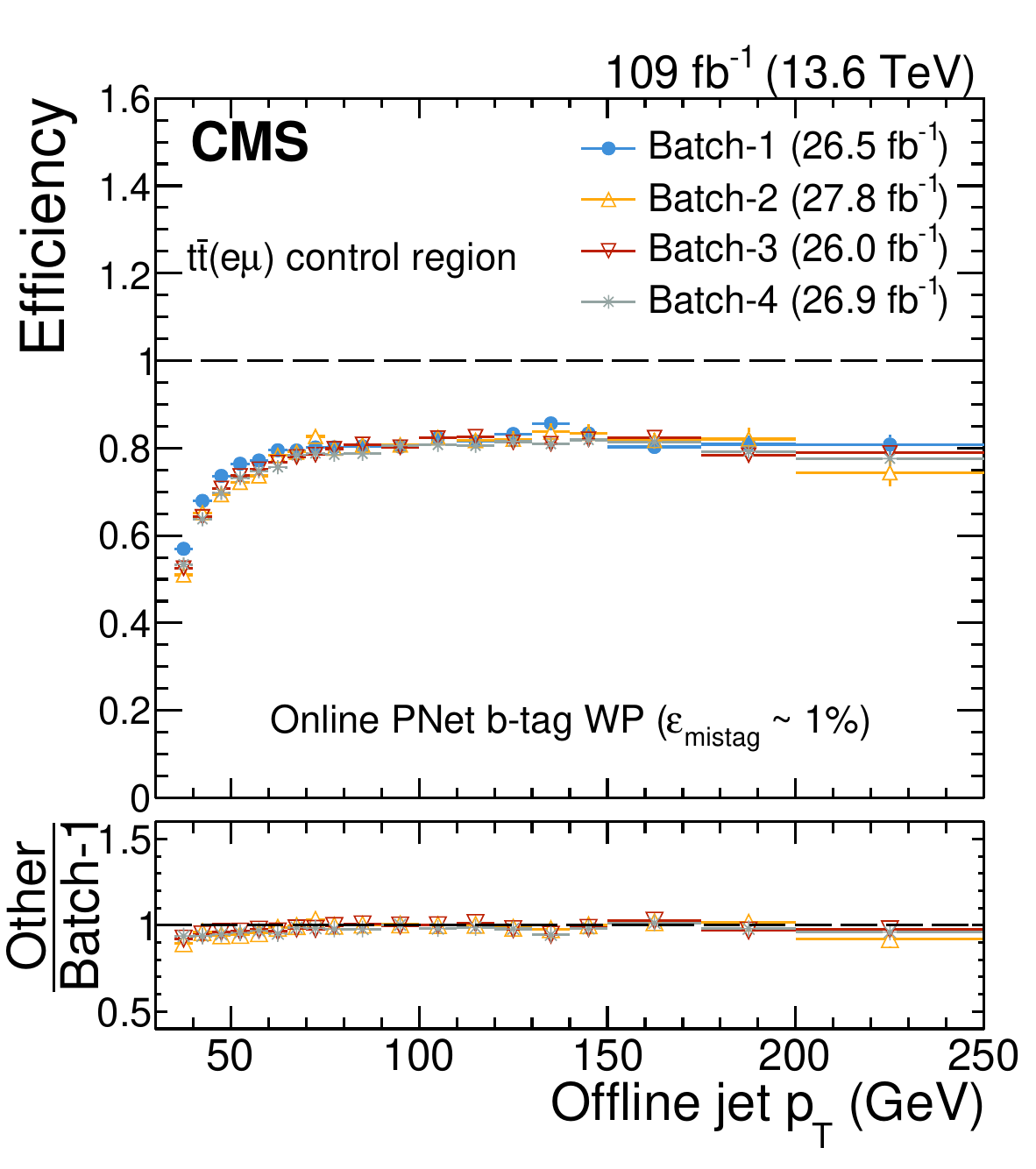}
    \includegraphics[width=0.32\textwidth]{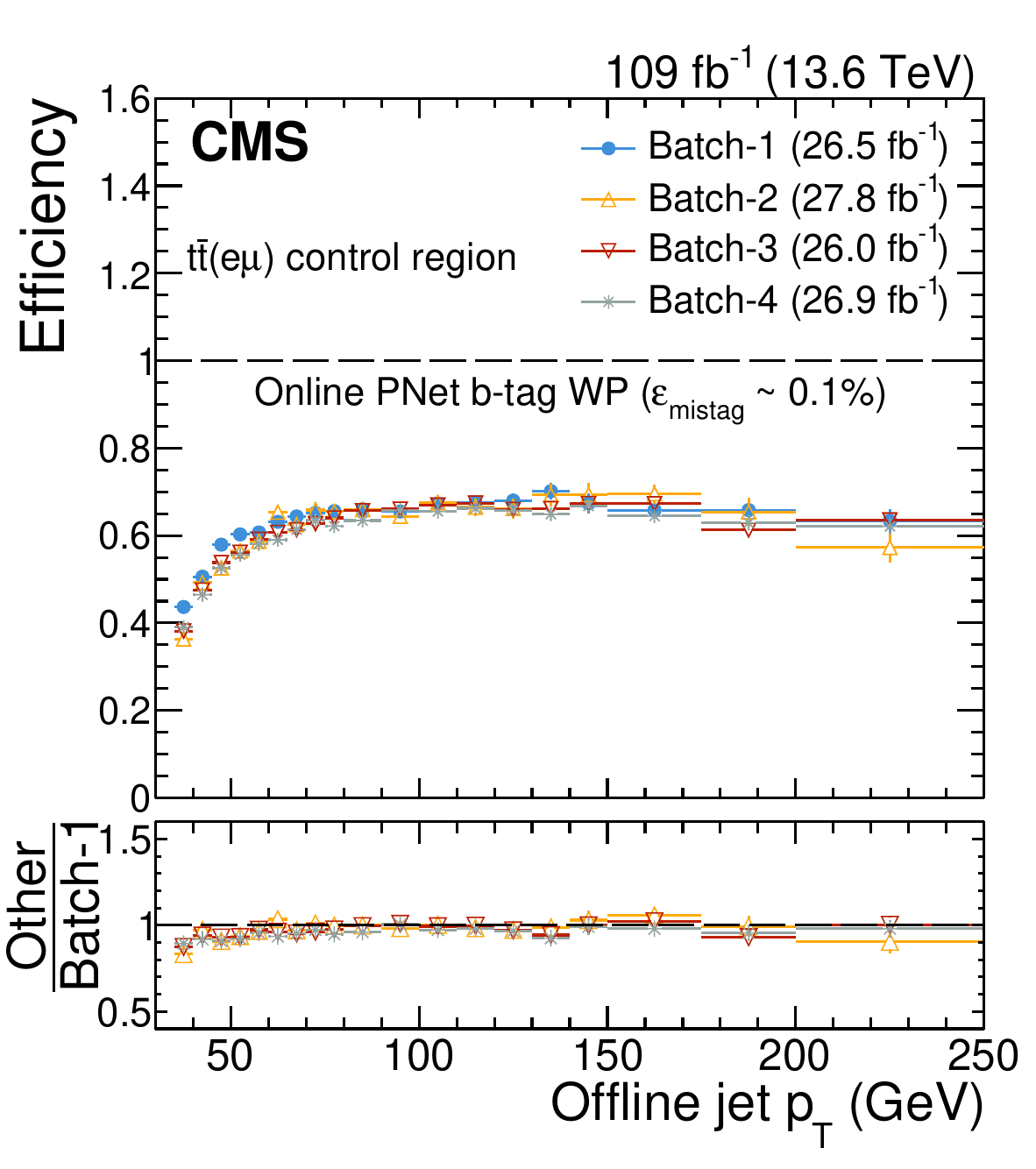}
    \caption{Efficiency of the online \pnet \PQb tagging algorithm as a function of the offline jet \pt in the ${\ttbar(\Pe\PGm)}$ region. The panel definition and ratios follow Fig.~\ref{fig:btag_efficiency_hlt_pbb_vs_time}.}
    \label{fig:btag_efficiency_hlt_pt_vs_time}
\end{figure*}

\begin{figure*}[!htb]  
    \centering
    \includegraphics[width=0.32\textwidth]{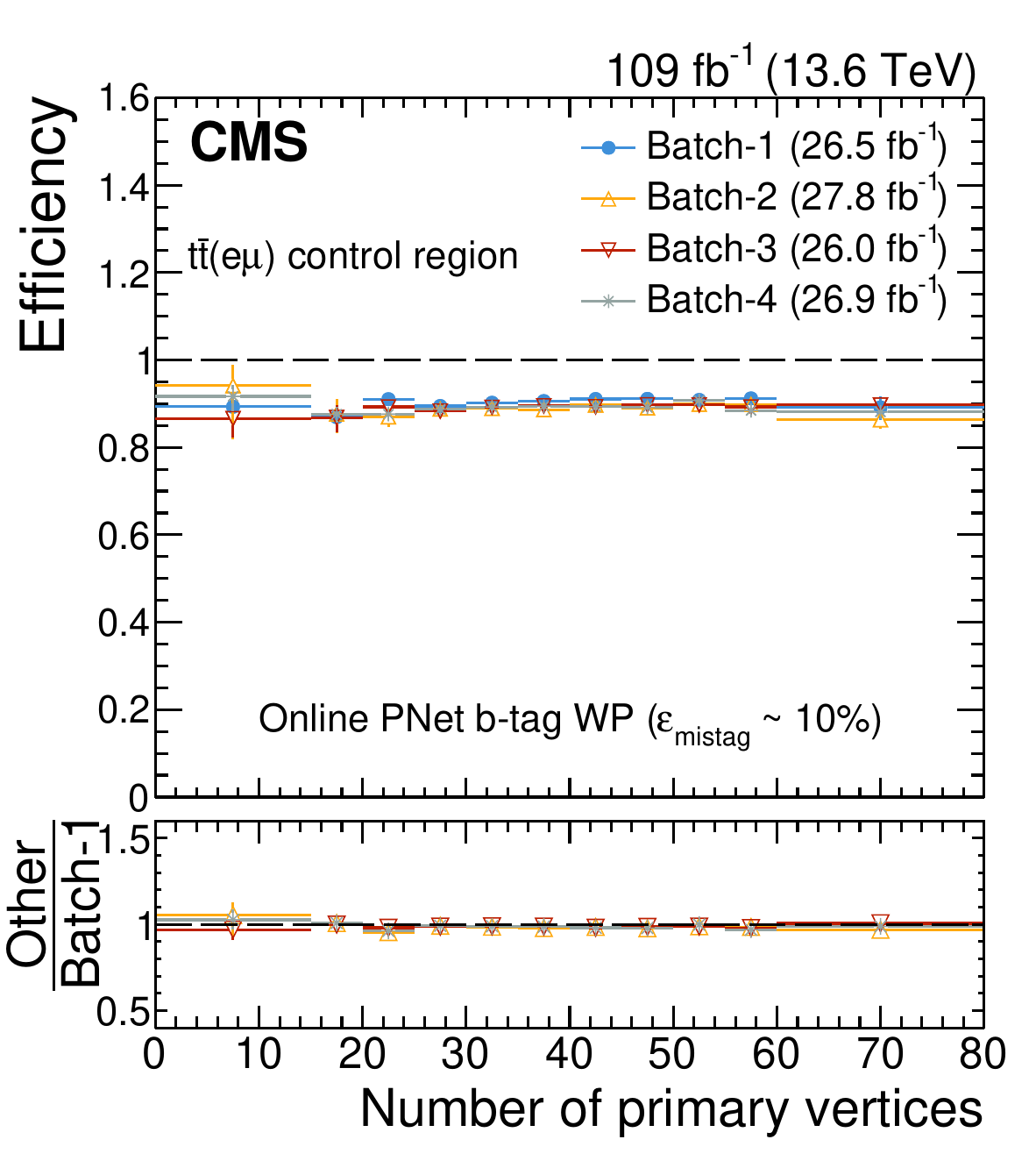}
    \includegraphics[width=0.32\textwidth]{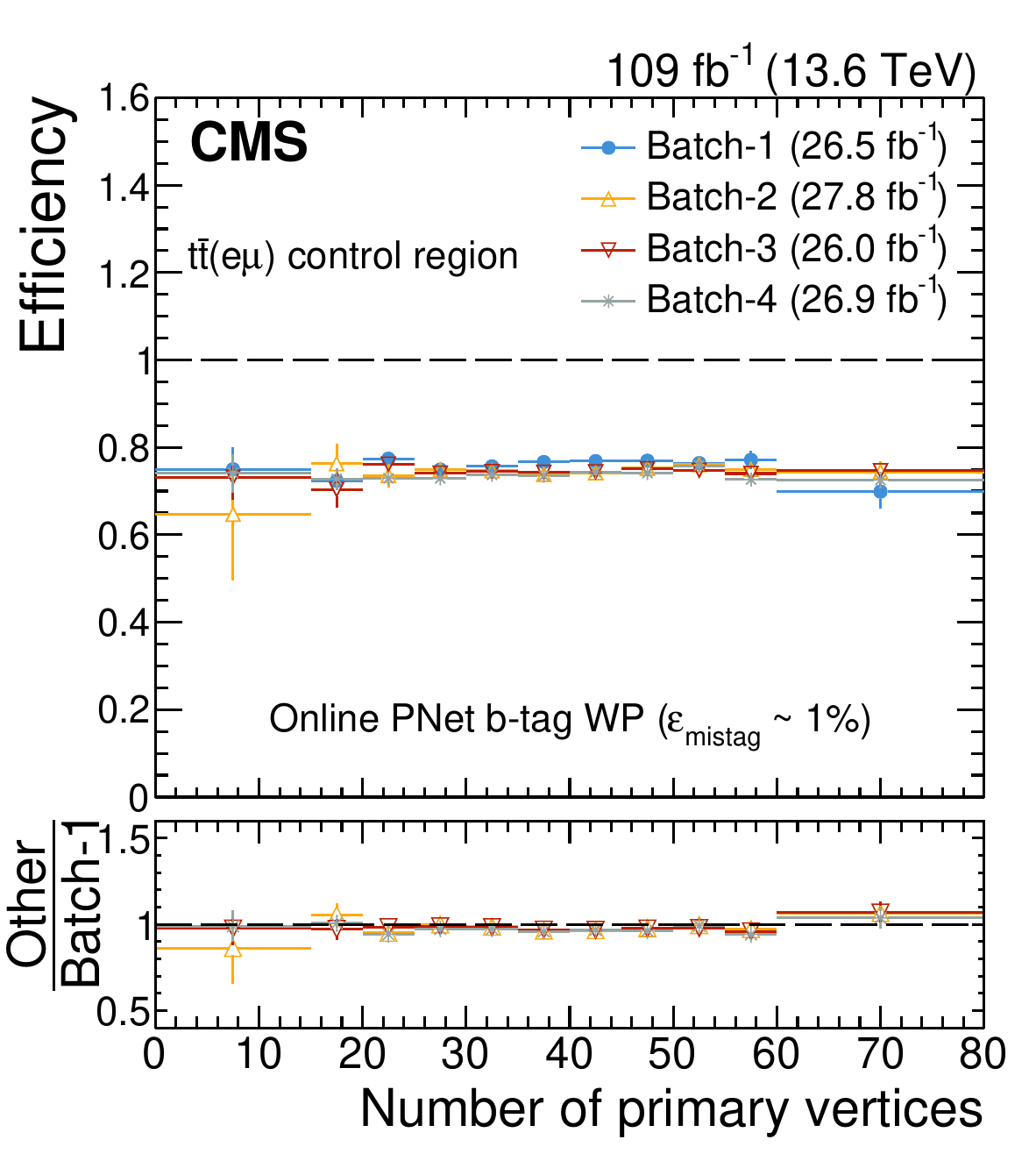}
    \includegraphics[width=0.32\textwidth]{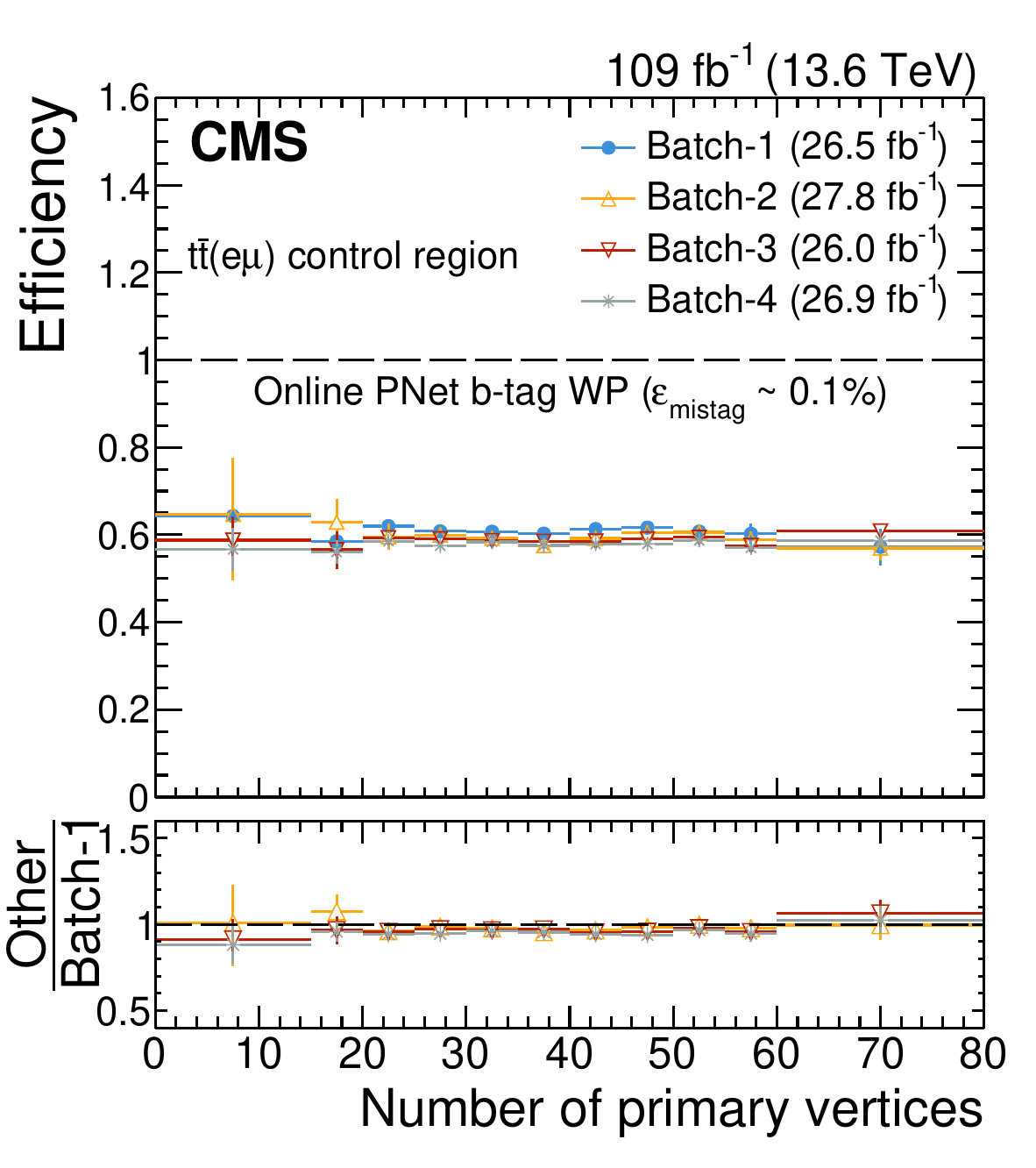}
    \caption{Efficiency of the online \pnet \PQb tagging algorithm as a function of the number of reconstructed primary vertices in the ${\ttbar(\Pe\PGm)}$ region. The panel definition and ratios follow Fig.~\ref{fig:btag_efficiency_hlt_pbb_vs_time}.}
    \label{fig:btag_efficiency_hlt_npv_vs_time}
\end{figure*}

The time dependence of the online \pnet \PQb tagging performance is studied by evaluating the efficiency in four consecutive data-taking periods, from batch-1 (start of the 2024 $\Pp\Pp$ run) to batch-4 (end of the 2024 $\Pp\Pp$ run). These batches reflect updates in detector conditions and calibrations, while maintaining similar integrated luminosity. Figure~\ref{fig:btag_efficiency_hlt_pbb_vs_time} shows the online \pnet \PQb tagging efficiency as a function of the offline \pnet \probb score for the three working points: loose (\cmsLeft), medium (middle), and tight (\cmsRight). The lower panels display the ratio of the efficiencies in batch-$N$ ($N=2,\ldots,4$) to that in batch-1, taken as reference. Across all WPs, and especially in the high-\probb region where most \PQb jets are found, the performance is stable over time. Variations up to about 5--10\% are observed in the mid-to-high \probb range, but remain compatible within statistical uncertainties. Figures~\ref{fig:btag_efficiency_hlt_pt_vs_time} and~\ref{fig:btag_efficiency_hlt_npv_vs_time} show the efficiency as functions of jet \pt and the number of reconstructed PVs, respectively. No significant time-dependent trends are observed for any of the tested \PQb tagging WPs.

\subsection{Measurement of online \texorpdfstring{\PQc}{c} tagging efficiency}\label{sec:monitor_ctag}

Measuring the online \pnet \PQc tagging efficiency is challenging due to the lack of high-purity \PQc jet control samples in data. Although ${\PW+\PQc}$ events provide the cleanest source~\cite{CMS:2017wtu}, their limited yield is insufficient for detailed \PQc tagging monitoring studies. Therefore, the efficiency is evaluated in QCD multijet events, in a phase space similar to that targeted by the VBF \hcc trigger described in Section~\ref{sec:trigger_menu}, selected with a dedicated ``control trigger''.

\begin{figure*}[!htb]
    \centering
    \includegraphics[width=0.5\textwidth]{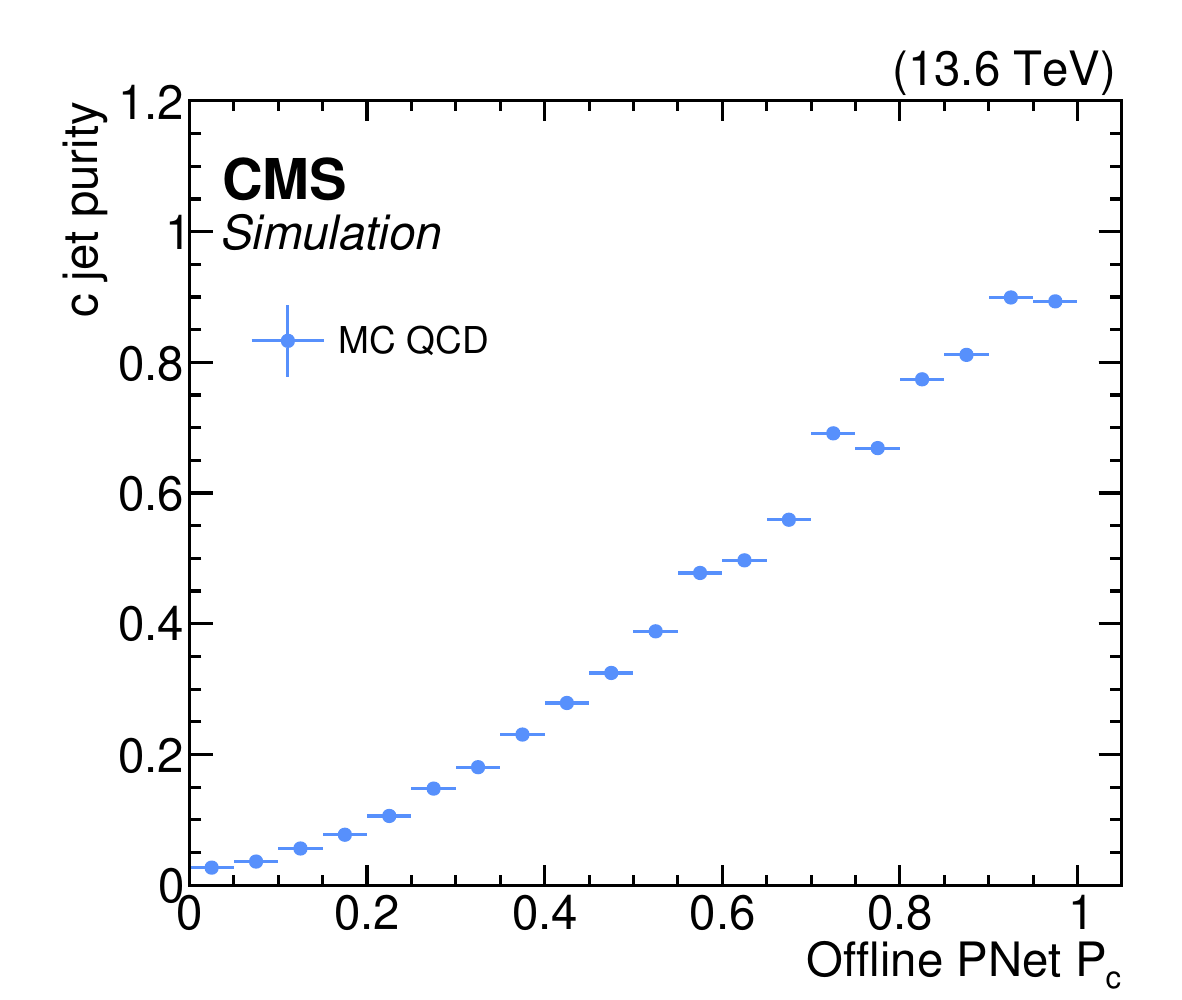}
    \caption{Purity of the leading \PQc tagged jet as a function of the offline \pnet \probc score in simulated QCD multijet events. The purity is the fraction of jets matched to a charm quark. The vertical bars represent the statistical uncertainty affecting the purity measurement. Selections correspond to the VBF \hcc phase space described in the text.}
    \label{fig:ctag_purity}
\end{figure*}

At the L1T, this control trigger selects events by requiring the scalar sum of \pt of all jets (\HT) to be larger than 280\GeV, computed from AK4 jets with ${\pt>30\GeV}$ and ${\abs{\eta}<2.5}$. The trigger acceptance is further increased by including additional hadronic triggers requiring two central (${\abs{\eta}<2.5}$) and one forward (${\abs{\eta}>2.5}$) high-\pt jet with ${\pt>95\GeV}$. At the HLT, at least four AK4 jets with ${\abs{\eta}<2.5}$ and ${\pt>100}$, 88, 70, and 30\GeV are required. This trigger is prescaled~\cite{CMS:2016ngn} and therefore the data recorded with it correspond to an integrated luminosity of about 0.4\fbinv out of the 109\fbinv collected in 2024. Offline, events are required to contain at least four AK4 jets with ${\pt>120}$, 105, 90, 40\GeV, respectively, and at least one pair (VBF jet pair) with ${\mjj>500\GeV}$, and ${\abs{\detajj}>4.0}$. In this control region, the \PQc tagging efficiency is measured using the central jet, not forming the VBF jet pair, with the highest \probc score. Figure~\ref{fig:ctag_purity} shows the probability (purity) for the chosen jet to originate from a \PQc quark in  simulated QCD multijet events. The purity increases from about 10\% at ${\probc \approx 0.2}$ to more than 90\% at ${\probc > 0.9}$.
 
\begin{figure*}[!htb]
    \centering
    \includegraphics[width=0.45\textwidth]{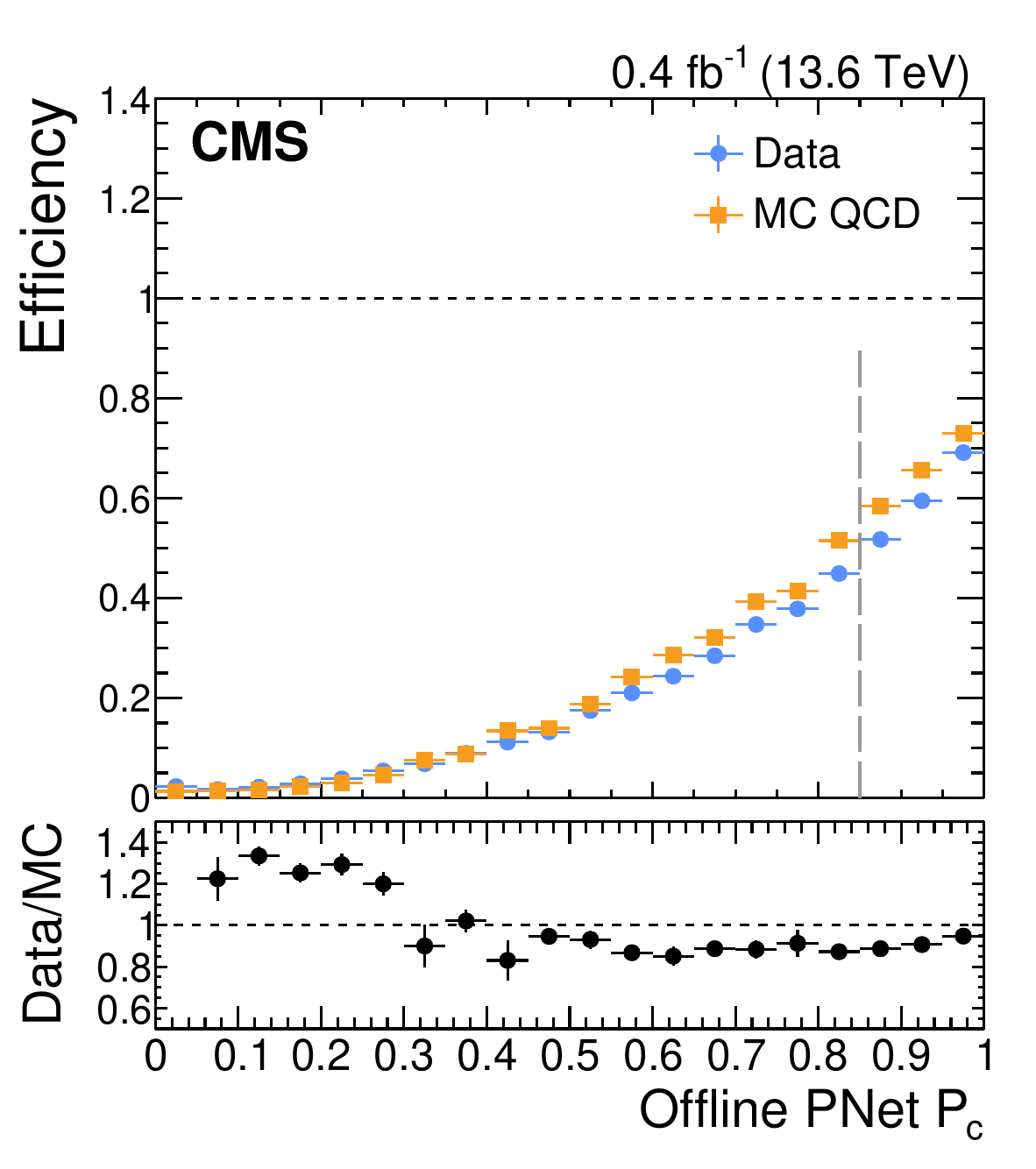}
    \includegraphics[width=0.45\textwidth]{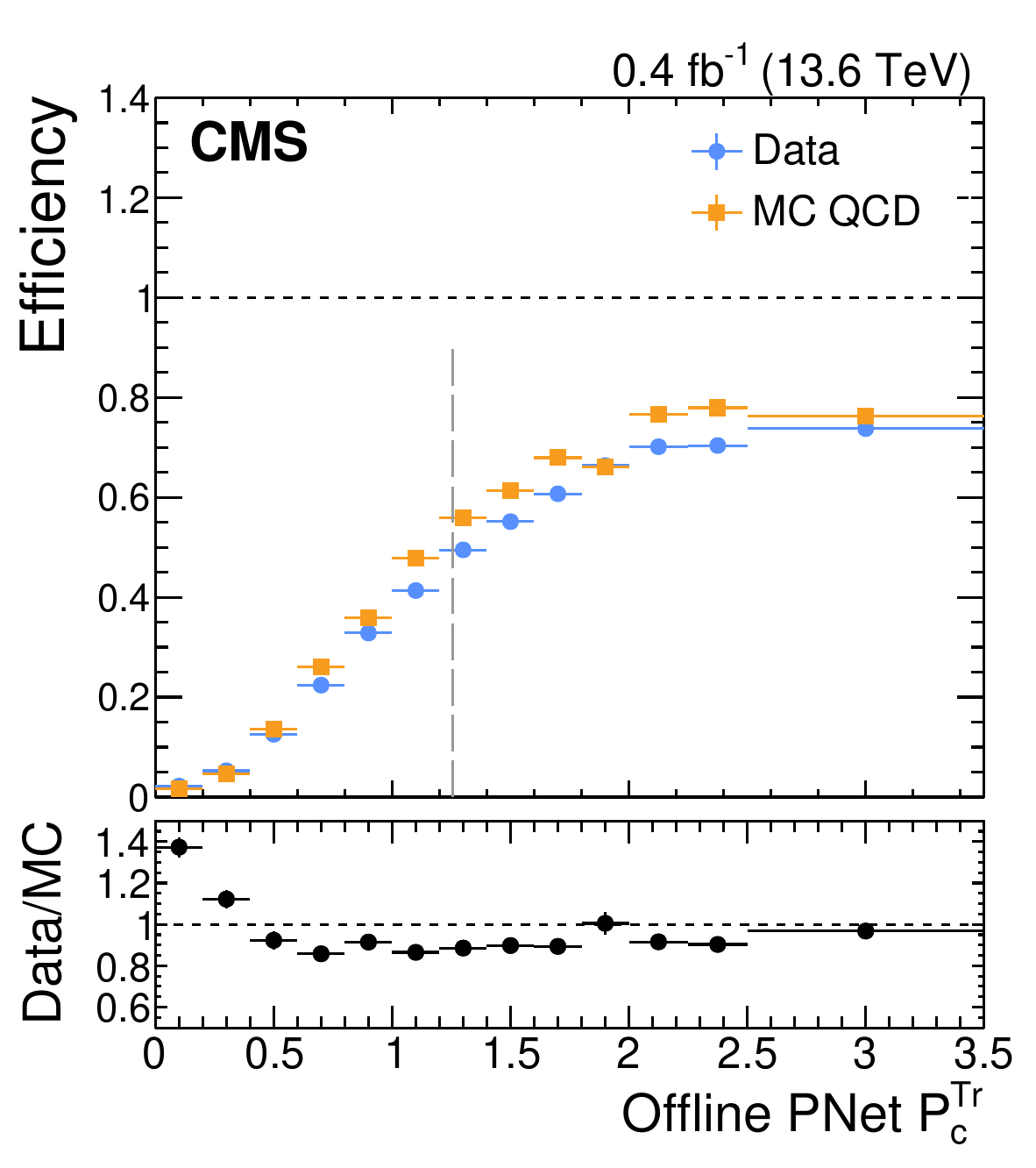}
    \caption{Online \pnet \PQc tagging efficiency in the selected VBF-like QCD multijet events as a function of offline \probc (\cmsLeft) and \probctr (\cmsRight) scores, as measured in data (blue) and simulation (orange) in a QCD-enriched region. The vertical bars on the data and simulation points represent the statistical uncertainty. Only the highest-\probc AK4 jet in the event is considered. The vertical dashed line denotes the reference value $\probc=0.85$, corresponding to a \PQc tagging purity above 80\%.}
    \label{fig:ctag_efficiency_hlt_pcc}
\end{figure*}

\begin{figure*}[!htb]
    \centering
    \includegraphics[width=0.45\textwidth]{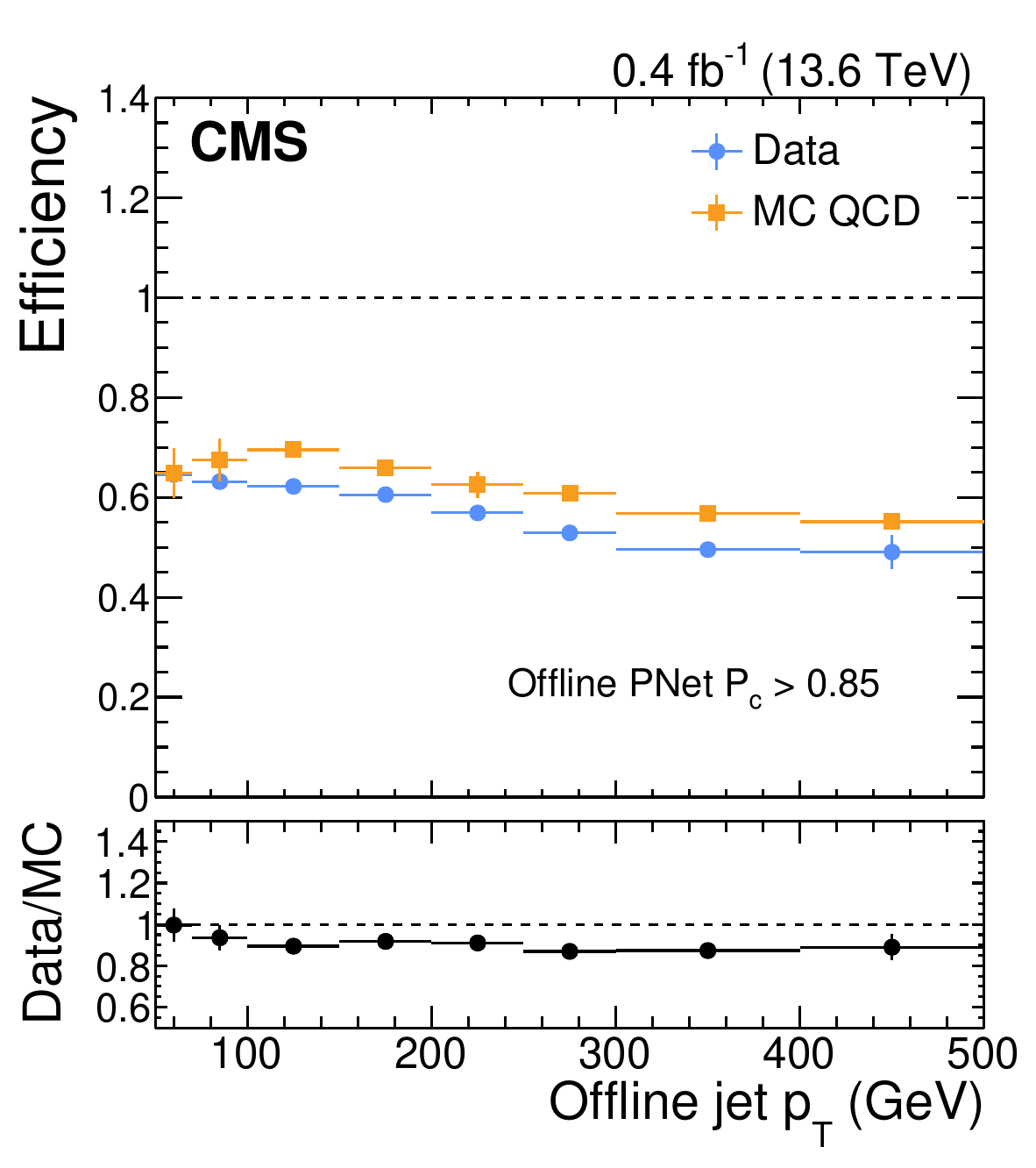}
    \includegraphics[width=0.45\textwidth]{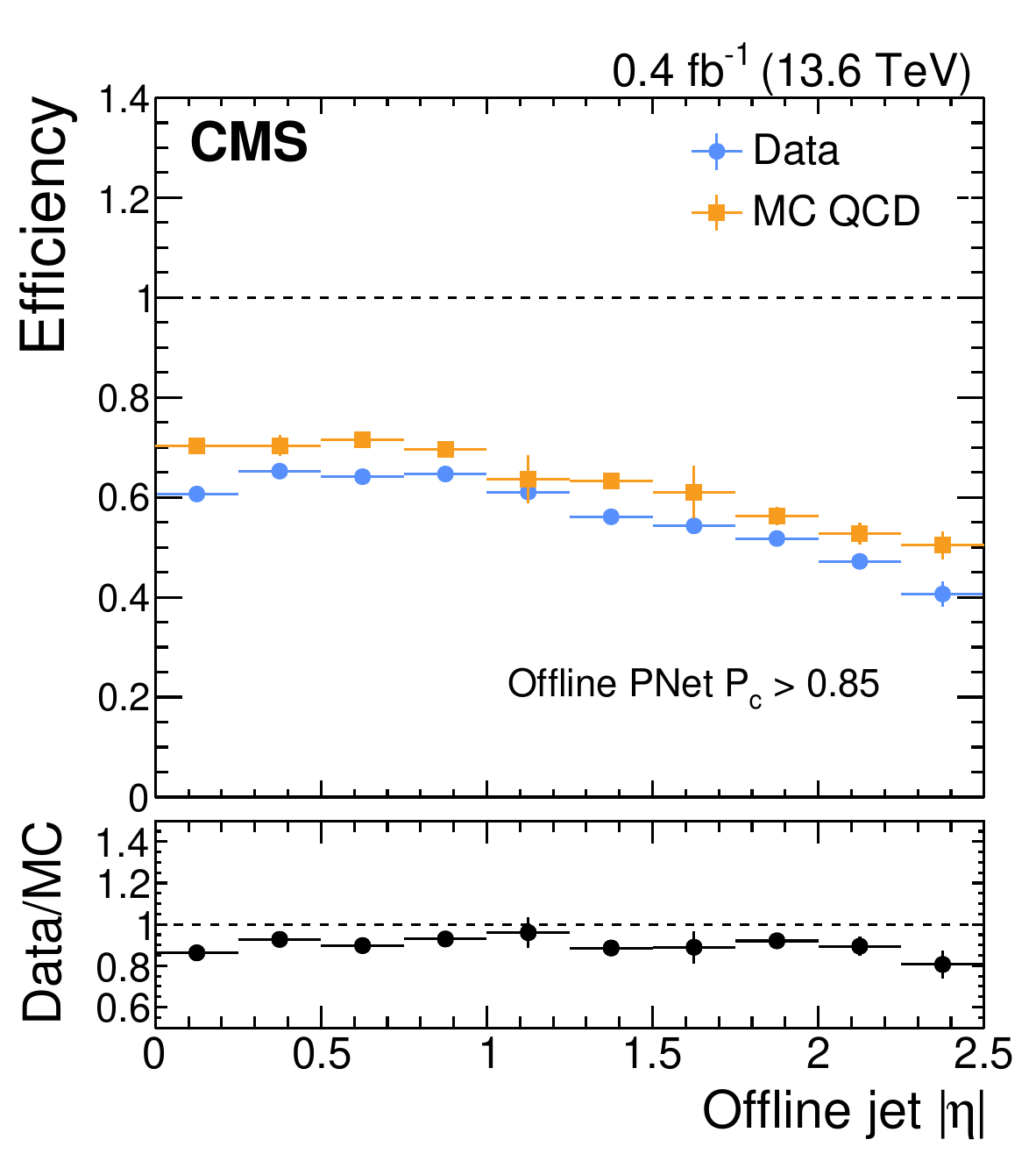}
    \caption{Online \pnet \PQc tagging efficiency in the selected VBF-like QCD multijet events as a function of offline \PQc tagged jet \pt (\cmsLeft) and $\eta$ (\cmsRight), as measured in data (blue) and simulation (orange) in a QCD-enriched region. The vertical bars on the data and simulation points represent the statistical uncertainty.}
    \label{fig:ctag_efficiency_hlt_kin}
\end{figure*}

Figure~\ref{fig:ctag_efficiency_hlt_pcc} shows the online \PQc tagging efficiency as a function of the offline \probc (\cmsLeft) and transformed \probctr (\cmsRight) scores. The efficiency is defined using events passing the previously described selection and corresponds to an online \probc requirement targeting a 1\% misidentification rate for light-flavour jets in simulated \ttbar events (see Fig.~\ref{fig:roc_performance_ctag}). This working point corresponds to a average online \PQc-jet efficiency of 30\% and a \PQb-jet misidentification rate of 7\%. Overall, the QCD multijet simulation reproduces the data well, with data-to-simulation ratio close to unity in the high-\probc region. Larger deviations are observed at intermediate \probc values, where contamination from non-\PQc jets is more significant, with data-to-simulation scale factors ranging between 0.85 and 0.95. Given the low \PQc jet purity in this \probc score region, the efficiency of the online \PQc tagging has also been measured as a function of the \PQc tagged jet \pt and $\eta$ for offline ${\probc>0.85}$, as shown in Fig.~\ref{fig:ctag_efficiency_hlt_kin}.

\begin{figure*}[!htb]
    \centering
    \includegraphics[width=0.45\textwidth]{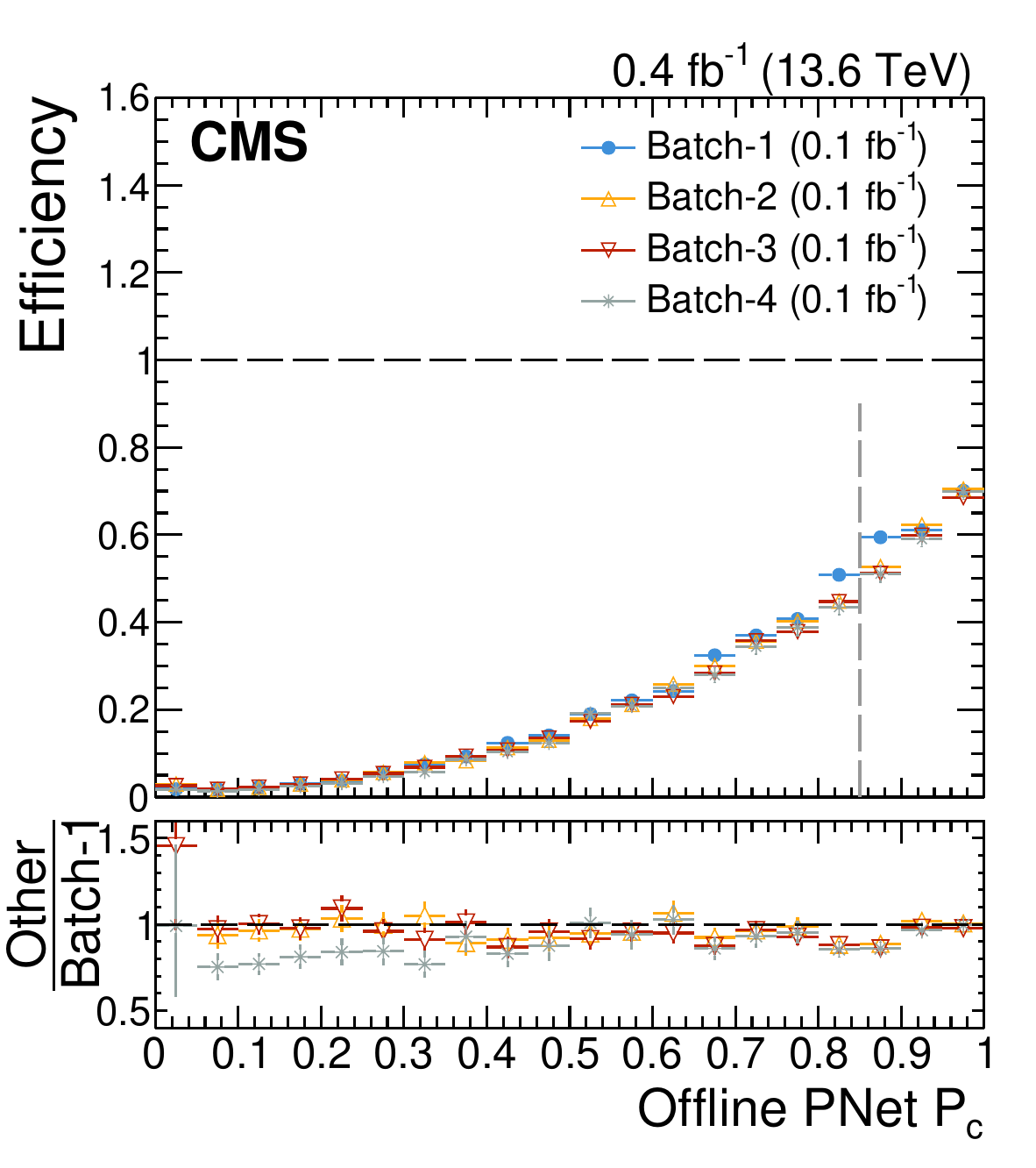}
    \includegraphics[width=0.45\textwidth]{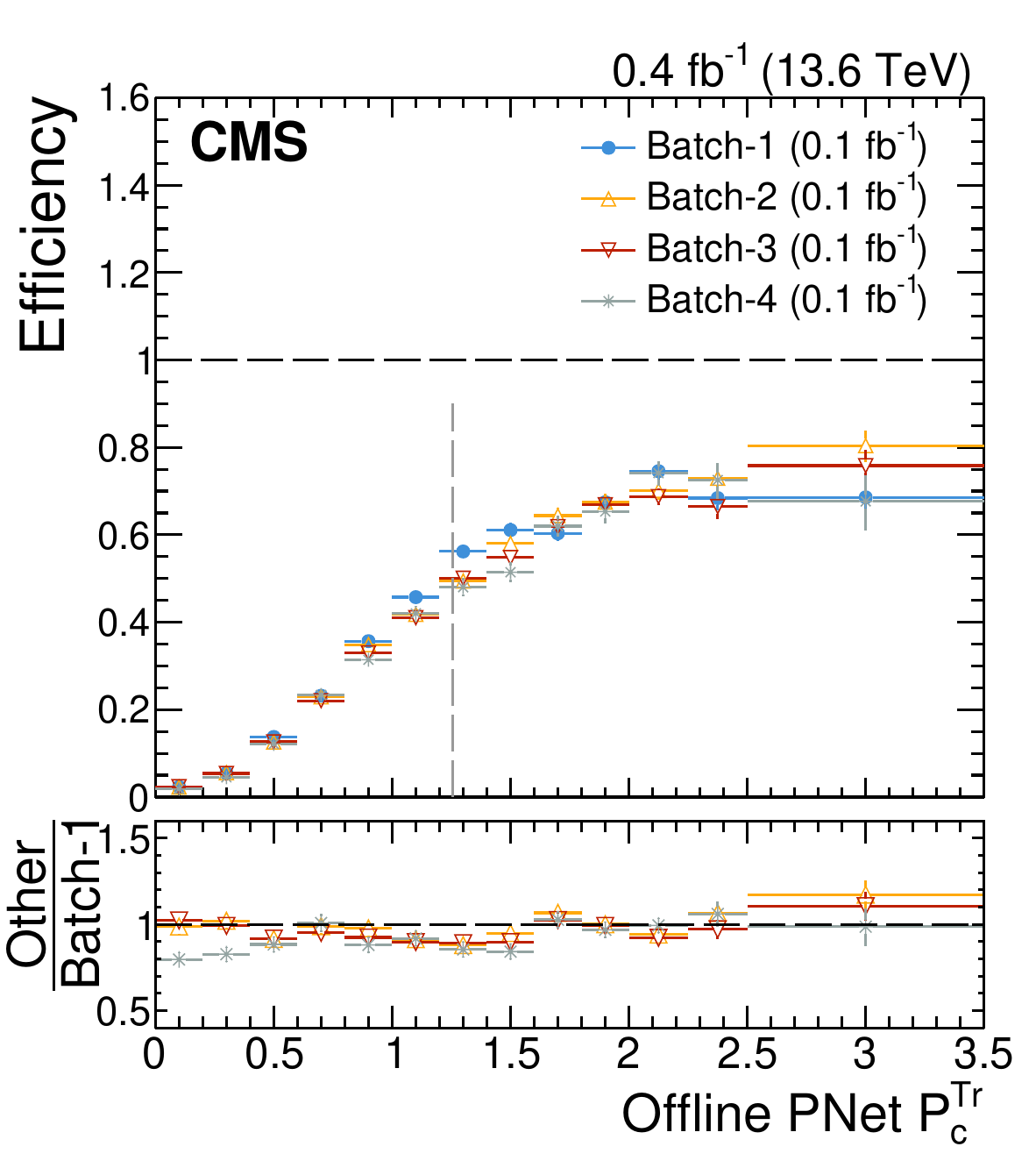}
    \caption{Efficiency of the online \pnet \PQc tagging algorithm in the selected VBF-like QCD multijet events as a function of the offline \pnet \probc (\cmsLeft) and \probctr (\cmsRight) scores, measured in data for four consecutive time periods (batch-1 to batch-4) during 2024. The vertical bars on the points represent the statistical uncertainty. The lower panels show the ratio of the efficiencies in batch-$N$ ($N=2,\ldots,4$) relative to batch-1. The vertical dashed line denotes the reference value $\probc=0.85$, corresponding to a \PQc tagging purity above 80\%.}
    \label{fig:ctag_efficiency_hlt_pcc_vs_time}
\end{figure*}

Finally, the time dependence of the online \pnet \PQc tagging performance is evaluated in four data-taking periods, following the same procedure as for \PQb tagging in Section~\ref{sec:monitor_btag}. Figure~\ref{fig:ctag_efficiency_hlt_pcc_vs_time} shows the online \PQc tagging efficiency for the chosen WP as a function of \probc (\cmsLeft) and \probctr (\cmsRight) in each of the four data-taking batches. The lower panels display the ratio of efficiencies in later batches relative to batch-1. The performance is stable over time, particularly in the high-\probc region. The larger deviations observed for batch-4 at low \probc values are concentrated in a region with low \PQc purity.

\subsection{Measurement of online \texorpdfstring{\hbb}{hhbb} tagging efficiency}

The efficiency of the \pnet algorithm for identifying high-\pt \xbb candidates at the HLT, reconstructed as single AK8 jets, should be measured using pure \hbb or \zbb samples in data. However, this is not feasible at the LHC due to overwhelming backgrounds. Although relatively pure high-\pt \zbb samples can be selected in $\Pp\Pp$ collisions~\cite{CMS:2025kje}, they require tight \probXbb selections that are not representative of those applied at the trigger level. Therefore, the best high-event-count proxy of merged \xbb decays is provided by AK8 jets, with ${\pt>300\GeV}$ and ${\mSD>50\GeV}$, selected in \ttbar lepton + jets enriched events where one $\PW$ boson decays hadronically. These selected jets may correspond to the decay products of a $\PW$ boson (merged $\PW$ boson), a bottom quark together with one quark from the $\PW$ (partially merged top quarks), or fully merged top quarks, with the last becoming significant only for $\pt>450\GeV$~\cite{CMS:2020poo}. While fully merged top quark jets contribute mainly at low \probXbb values, the high-\probXbb region is dominated by merged $\PW$ and partially merged top quark jets. The validity of these proxies for \hbb signal jets is verified using simulated \hhbb events, where the selected AK8 jets are matched to genuine \hbb candidates at the generator level. The measured efficiency is found to be consistent with that obtained in simulated \ttbar lepton + jets enriched events, particularly in the high-\probXbb region relevant for physics analyses.

Events are collected with a single-muon trigger requiring ${\pt> 22 \GeV}$ at the L1T and ${\pt > 50\GeV}$ at the HLT. Offline, the muon must satisfy ${\pt>55\GeV}$, ${\abs{\eta}<2.4}$, and tight identification and isolation criteria. Events with additional loosely identified leptons with ${\pt>10\GeV}$ and ${\abs{\eta}<2.5}$ are vetoed. At least two AK4 jets with ${\pt>25\GeV}$ and ${\abs{\eta}<2.5}$, separated from the muon by ${\Delta R>0.4}$, are required and must pass a loose offline \pnet \probb selection, yielding a \ttbar purity of about 90\%. To enrich the sample in high-\pt ${\PW\to\mathrm{qq'}}$ decays, at least one AK8 jet with ${\pt>300\GeV}$ and ${\mSD>60\GeV}$ is required. This jet (J) must recoil against the muon and its nearest \PQb tagged jet ($j_{\PQb}$), such that $\Delta\phi(J,\PGm)$ and $\Delta\phi(J,j_{\PQb})$ are both larger than $\pi/2$.

\begin{figure*}[!htb]
    \centering
    \includegraphics[width=0.45\textwidth]{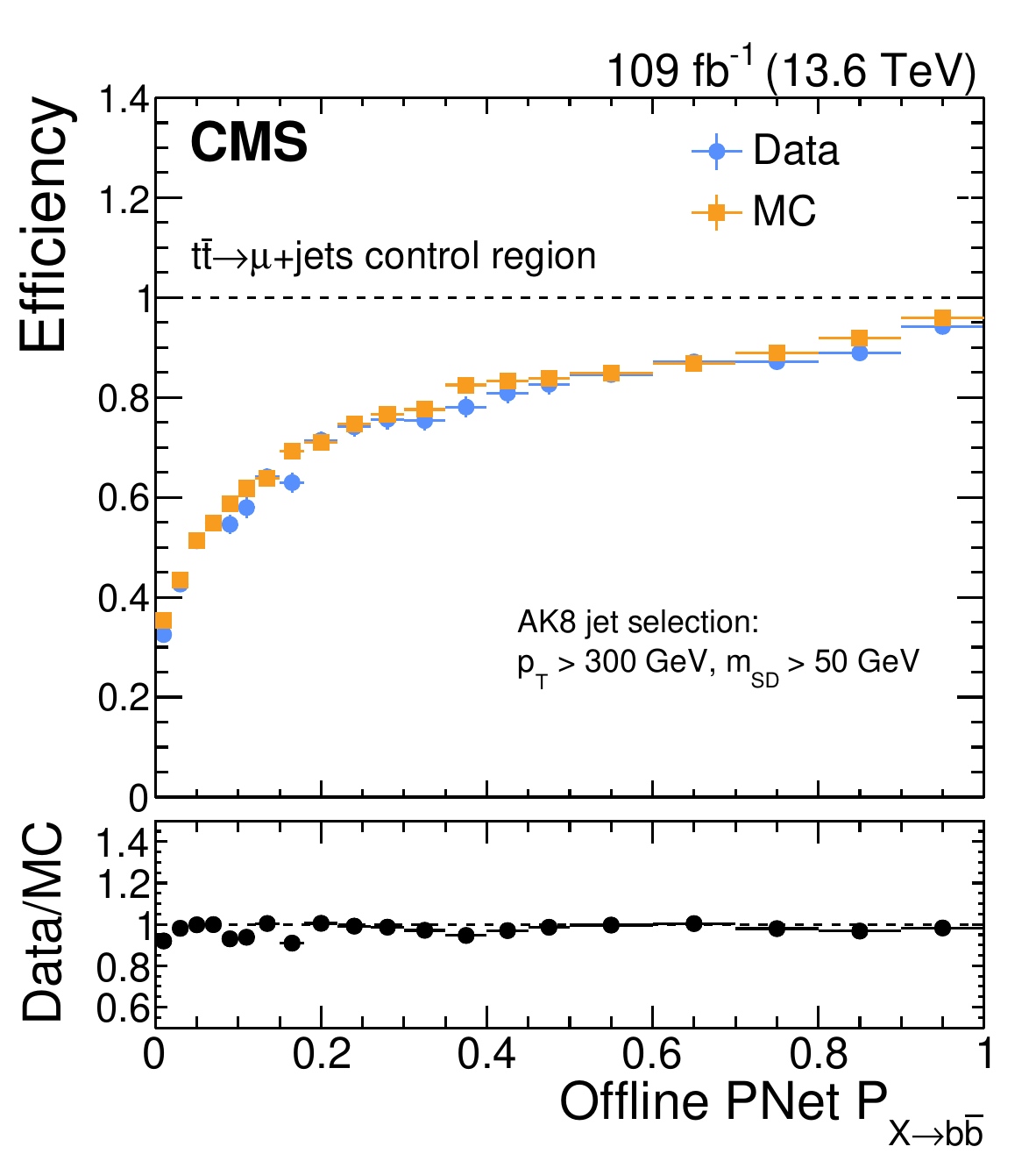}
    \includegraphics[width=0.45\textwidth]{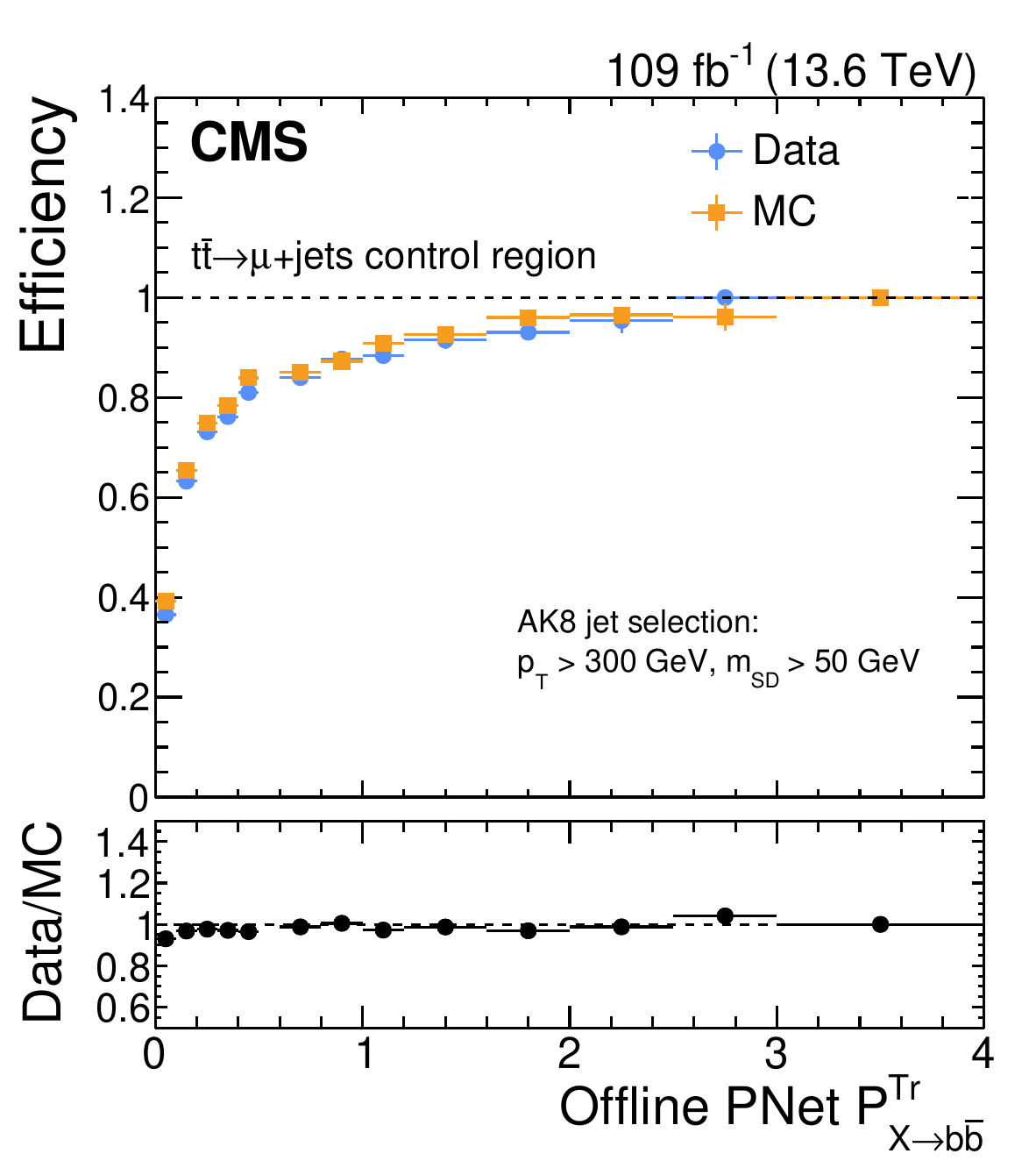}
    \caption{Online \pnet \xbb tagging efficiency as a function of the offline \probXbb (\cmsLeft) and \probXbbtr (\cmsRight) scores for AK8 jets with ${\pt>300\GeV}$ and ${\mSD>50\GeV}$ in a \ttbar lepton + jets enriched region, measured in data (blue) and simulation (orange). The vertical bars on the data and simulation points represent the statistical uncertainty.}
    \label{fig:xbb_efficiency_xbb_boosted}
\end{figure*}

The online \pnet \xbb tagging efficiency is measured for a loose \probXbb WP, matching that used in triggers targeting boosted \zbb and \hbb signatures as described in Section~\ref{sec:trigger_menu}. Figure~\ref{fig:xbb_efficiency_xbb_boosted} shows the efficiency in data and simulation, as well as their ratio, as a function of the offline \probXbb (\cmsLeft) and \probXbbtr (\cmsRight) scores. Good agreement is observed between observed and expected efficiencies, with a ratio close to unity with a few-percent spread across the whole \probXbb range. The ratios of data to simulation measured in the \ttbar lepton + jets enriched sample are expected to be applicable to genuine \xbb candidates, since they probe jets originating from real \PQb quarks reconstructed with similar kinematic properties and tagging features in the relevant analysis phase space.

\section{New heavy-flavour jet triggers in Run~3}\label{sec:trigger_menu}

Several triggers targeting fully hadronic final states with heavy-flavour jets or high-\pt \xbb candidates have been implemented in the HLT since the start of Run~3. The kinematic and \pnet-based jet flavour selections (as discussed in Sections~\ref{sec:jet_tagging} and~\ref{sec:monitoring}) were optimized to maximize, within the available trigger bandwidth, the acceptance for key benchmark processes, including \hhbb, ${\ttbar\PH(\hbb)}$, ${\ttbar\PH(\hcc)}$, VBF \hcc, and measurements of high-\pt \hbb production. Trigger thresholds and tagging requirements evolved during data taking to accommodate increasing instantaneous luminosity, updated detector conditions, and changes in the bandwidth allocated to heavy-flavour HLT triggers and their seeding L1 triggers. Since 2024, the configurations have remained stable, with a single version collecting most of the Run~3 data. Their impact on benchmark signals is discussed in Section~\ref{sec:physics_impact}.

Events are first selected at the L1T using either an \HT threshold or the presence of one jet with ${\pt>180\GeV}$ and ${\abs{\eta}<2.4}$. The jet \pt is calibrated in situ and the average pileup contributions are subtracted. The \HT threshold of the lowest unprescaled trigger decreased from ${\HT>360\GeV}$ in 2022 to ${\HT>280\GeV}$ in 2023--24. At peak instantaneous luminosity, the corresponding input rate to heavy-flavour triggers increased from about 3\unit{kHz} (2022) to 8\unit{kHz} (2023--24), while the average HLT output rate was increased from 150 to 160\unit{Hz} and 280\unit{Hz} in 2022, 2023, and 2024, respectively.

A subset of representative triggers is summarized in Table~\ref{tab:hlt_trigger_rate_run3}. In 2022, most heavy-flavour triggers relied on \deepjet for \PQb tagging. After validation with 2022 data, \pnet was adopted as the default tagger from 2024, providing significantly improved background rejection. This enabled lower \pt thresholds without increasing the overall output rate. In 2024, the available bandwidth was further expanded with additional and more relaxed trigger configurations.

\begin{table*}[!htb]
    \centering
    \caption{Summary of representative Run~3 HLT triggers using \pnet-based heavy-flavour tagging. All triggers are seeded by common L1 \HT requirements, ${\HT>360\GeV}$ in 2022 to ${\HT>280\GeV}$ in 2023--24, with AK8 triggers also accepting events with at least one jet with ${\pt>180\GeV}$ and ${\abs{\eta}<2.5}$. Rates correspond to at an instantaneous luminosity of ${{\approx} 2\times10^{34}\percms}$. The significant reduction in the rate of the AK8 \xbb trigger in 2023--24 compared to 2022 was made possible by a retrained \pnet model with improved discrimination power.}
    \cmsTable{
    \begin{tabular}{c c c c c} 
    HLT selection  & Rate (\unit{Hz}) & Year  & Target signal & Acronym \\ 
    \hline 
    \\[-1ex]
    Four AK4 jets $\pt>70,50,40,35\GeV$, 2 \PQb-jets &  60 & 2022 & \multirow{3}{*}{\hhbb, $\ttbar\PH(\hbb)$} & \multirow{3}{*}{AK4: 4j2b}\\
    Four AK4 jets $\pt>30\GeV$, $\HT > 280\GeV$, 2 \PQb-jets & 115 & 2023 & & \\
    Four AK4 jets $\pt>25\GeV$, $\HT > 250\GeV$, 2 \PQb-jets & 170 & 2024 & & \\[1ex]
    Four AK4 jets $\pt>100,88,70,30\GeV$, 1 \PQc-jet  & \multirow{2}{*}{15} & \multirow{2}{*}{2023--24} & \multirow{2}{*}{VBF \hcc} &  \multirow{2}{*}{AK4: VBF-4j1c} \\
    $\mjj>450\GeV$, $\abs{\detajj}>3.5$ & & & \\[1ex]
    One AK8 jet $\pt>250\GeV$, $\mSD>40\GeV$, 1 \xbb jet & 47 & 2022 &  \multirow{2}{*}{Merged \hbb} & \multirow{2}{*}{AK8: 1jX$_{\PGb\PQb}$}\\
    One AK8 jet $\pt>230\GeV$, $\mSD>40\GeV$, 1 \xbb jet & 17 & 2023--24 & & \\[1ex]
    \end{tabular}
    }
    \label{tab:hlt_trigger_rate_run3}
\end{table*}

\begin{figure*}[!htb]
    \centering
    \includegraphics[width=0.45\textwidth]{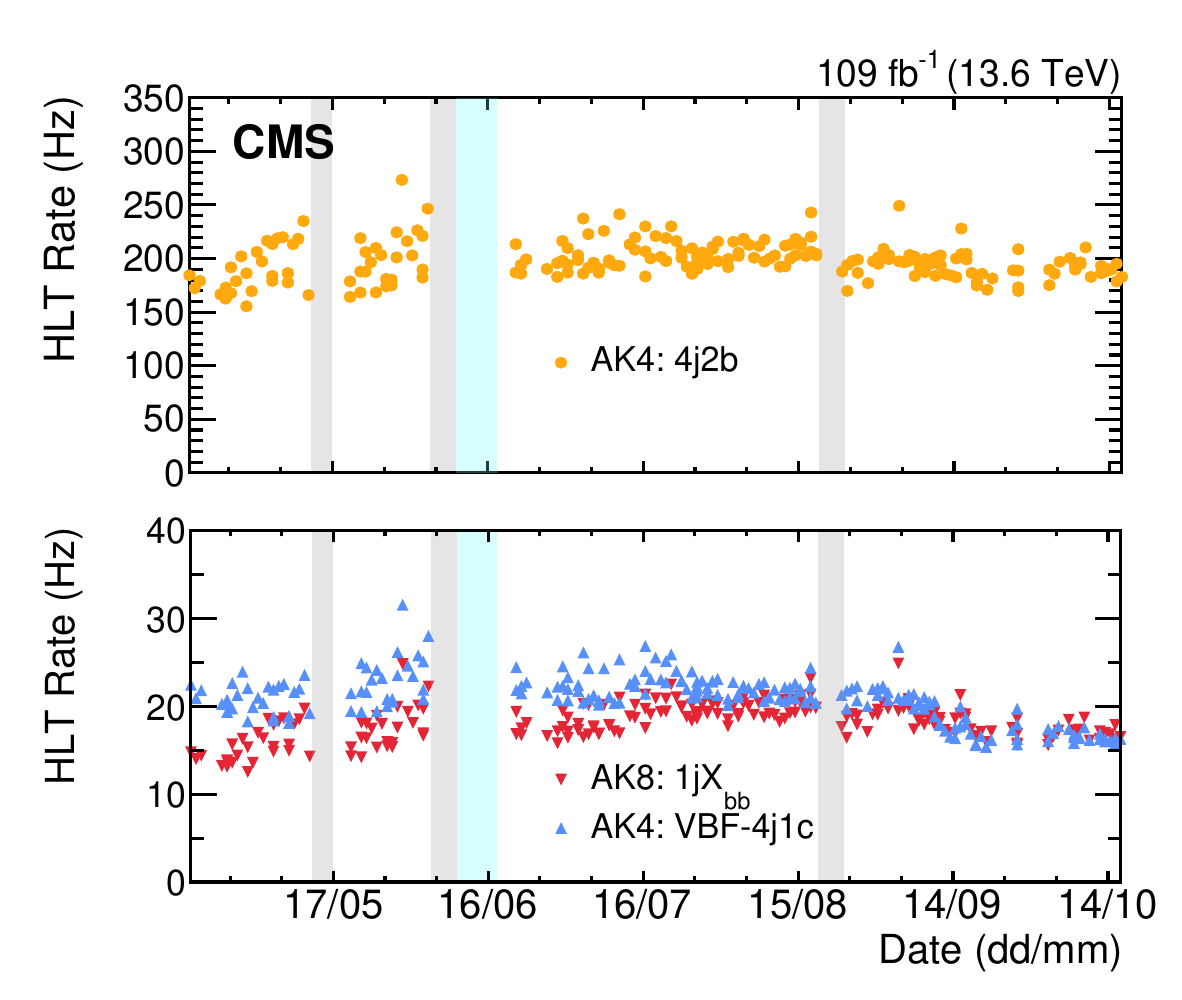}
    \includegraphics[width=0.45\textwidth]{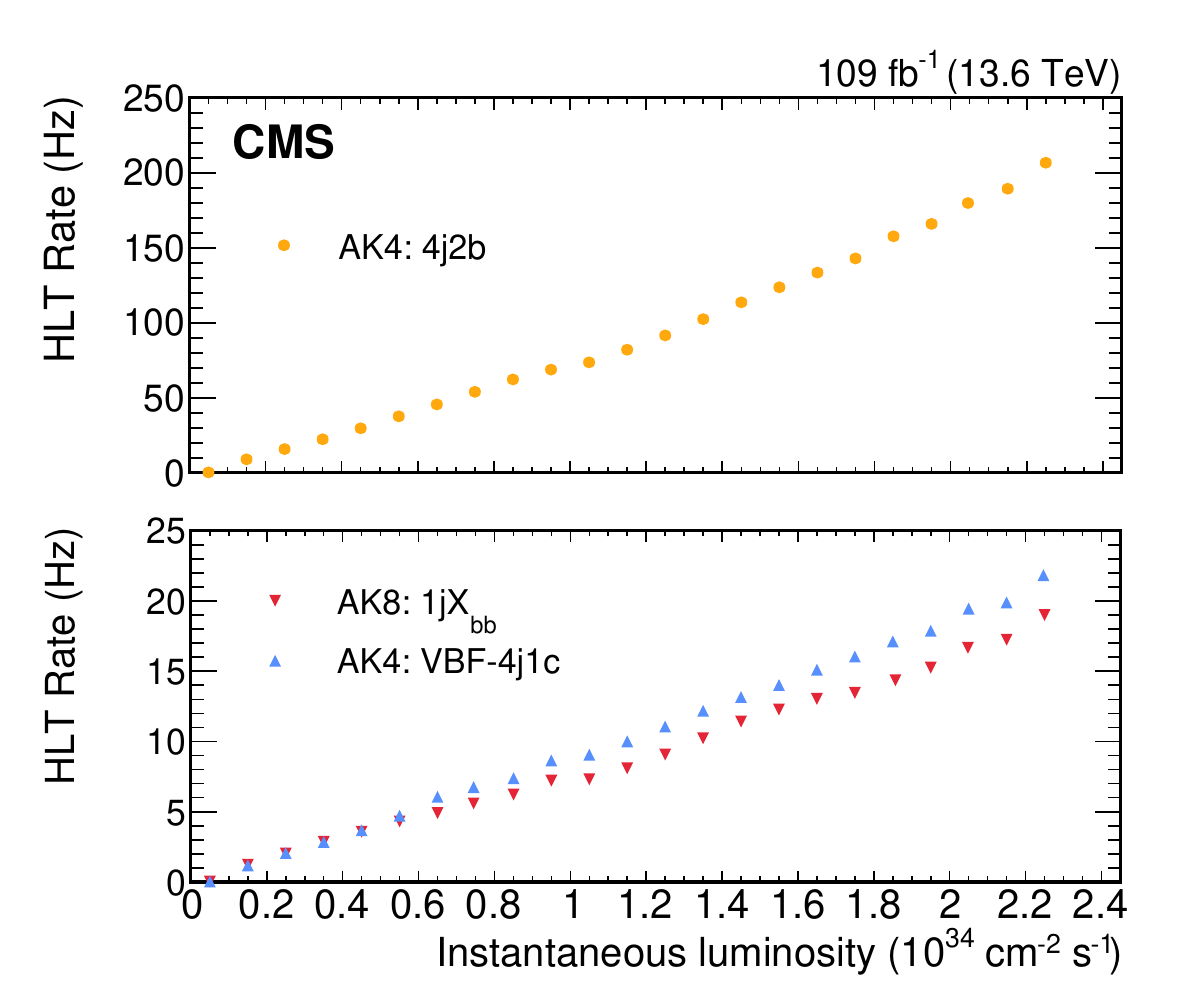}
    \caption{Average output rates correspond to an instantaneous luminosity of ${{\approx}2\times10^{34}\percms}$ for the selected heavy-flavour triggers reported in Table~\ref{tab:hlt_trigger_rate_run3} during 2024 as a function of time (\cmsLeft) and instantaneous luminosity (\cmsRight).} 
    \label{fig:hlt_trigger_rate_run3}
\end{figure*}

Figure~\ref{fig:hlt_trigger_rate_run3} (\cmsLeft) shows the evolution of the trigger rates corresponding to an instantaneous luminosity of ${{\approx}2\times10^{34}\percms}$ in 2024. Rates are evaluated for LHC $\Pp\Pp$ fills with the highest number of colliding bunches, averaging over the time period in which the instantaneous luminosity is at its peak value. The lower panel shows the corresponding average instantaneous luminosity for the selected periods, demonstrating that the rates are compared under similar luminosity conditions throughout the year. The grey and azure bands indicate periods without data taking due to machine development or technical stops. Variations in detector conditions are often correlated with discontinuities in the rates. Overall, the trigger rates remain stable, with fluctuations of 10--15\%. Figure~\ref{fig:hlt_trigger_rate_run3} (\cmsRight) shows the average rate for each trigger reported in Table~\ref{tab:hlt_trigger_rate_run3} as a function of instantaneous luminosity, exhibiting a linear dependence.

\section{Impact on the CMS physics programme}\label{sec:physics_impact}

The online \pnet-based flavour tagging for AK4 and AK8 jets significantly improved the discrimination of heavy-flavour and high-\pt \xbb jets from background. The resulting reduction in misidentification rates, reported in Figs.~\ref{fig:roc_performance_btag}--\ref{fig:trigger_merged_xbb}, combined with the increased HLT bandwidth in Run~3, enabled the development of new triggers targeting fully hadronic final states with substantially lower kinematic thresholds. This section highlights benchmark processes for which the triggers listed in Table~\ref{tab:hlt_trigger_rate_run3} improve acceptance, allowing access to the previously unexplored regions of phase space.

A key goal of the CMS physics programme is the measurement of the Higgs boson self-coupling via non-resonant $\PH\PH$ production. Based on results of previous searches for this process by the ATLAS and CMS Collaborations~\cite{CMS:2022dwd,ATLAS:2024ish}, the \hhbb final state is among the most sensitive channels. In Run~2, the resolved \hhbb analysis---where the jets originating from the hadronization of each \PQb quark are reconstructed as AK4 jets---relied on triggers requiring at least four AK4 jets, with at least three \PQb tagged. The combination of tight requirements on the jet \pt and the less performant online \PQb tagging algorithms severely limited the overall trigger acceptance for signal events. In addition, these requirements reduced the ability to collect statistically enriched control regions for background modelling.

\begin{figure*}[!htb]
    \centering
    \includegraphics[width=0.55\textwidth]{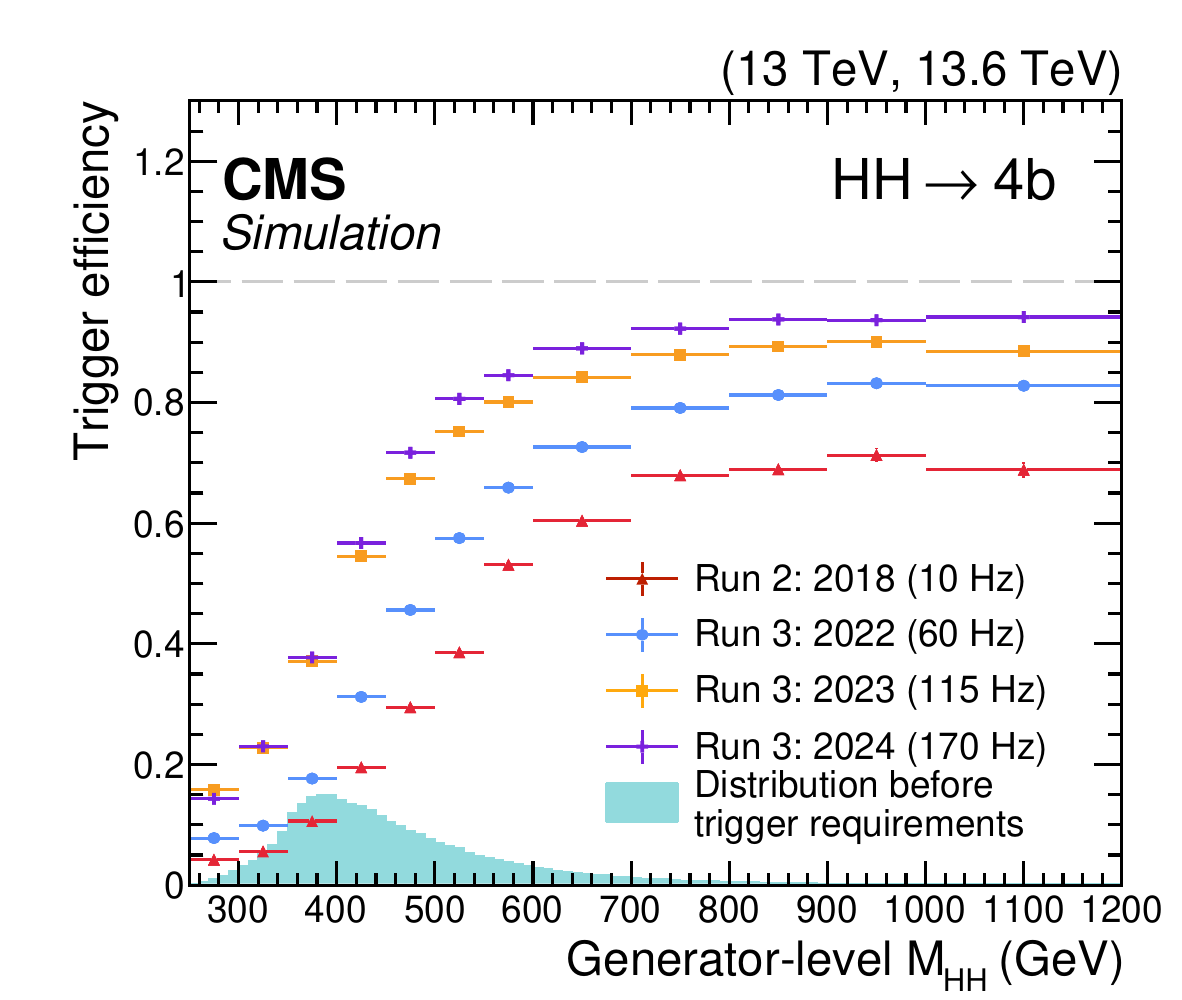}
    \caption{Efficiency of the Run~3 4j2b trigger in simulated \hhbb events for 2022 (azure), 2023 (orange), and 2024 (purple) configurations as a function of generator-level \mhh. The vertical error bars indicate the statistical uncertainty in the measured trigger efficiency. Four generator-level jets are required in the selected events with ${\pt>25\GeV}$ and ${\abs{\eta}<2.5}$. Results are compared to Run~2 triggers (red). The corresponding trigger rates are 10, 60, 115, and 170\unit{Hz} for 2018, 2022, 2023, and 2024, respectively, at an instantaneous luminosity of ${{\approx}2\times10^{34}\percms}$. The increase in trigger rate across Run~3 configurations reflects the progressive relaxation and optimisation of the trigger selection in later years. The teal histogram shows the expected distribution of the simulated events at ${\sqrt{s}=13.6\TeV}$ before trigger requirements, scaled by an arbitrary factor for visibility.}    
    \label{fig:trigger_efficiency_hh4b_gen_phys}
\end{figure*}

In Run~3, events in the resolved topology are selected with the 4j2b trigger (see Tab.~\ref{tab:hlt_trigger_rate_run3}) by requiring at least four AK4 jets with ${\abs{\eta}<2.5}$, of which at least two are \PQb tagged using \pnet. In 2022, this trigger operated at an output rate of ${\sim}60\unit{Hz}$, with jet \pt thresholds of 70, 50, 40, and 35\GeV. In 2023 (2024), the available bandwidth increased to 115~(170)\unit{Hz} thanks to the development of a dedicated data parking stream, referred to as ``ParkingHH''~\cite{CMS:2024zhe}, allowing progressively looser selections at the cost of a potentially delayed offline event reconstruction. The \HT threshold of the lowest unprescaled L1T trigger was lowered to 280\GeV, and the jet \pt thresholds at HLT were reduced to 30\GeV in 2023 and 25\GeV in 2024, with \PQb tagging applied only on jets with ${\pt>30\GeV}$. Figure~\ref{fig:trigger_efficiency_hh4b_gen_phys} shows the efficiency of the 4j2b trigger in 2022 (azure), 2023 (orange), and 2024 (purple) as a function of the generator-level \mhh for the SM $\PH\PH$ production via the gluon fusion mechanism, requiring at least four generator-level jets in the event with ${\pt>25\GeV}$ and ${\abs{\eta}<2.5}$. The efficiency gain relative to Run~2 ranges from a factor of two (four) at ${\mhh<300\GeV}$ to 25\% (35\%) for ${\mhh>800\GeV}$ in 2022~(2024).

\begin{figure*}[!htb]
    \centering
    \includegraphics[width=0.45\textwidth]{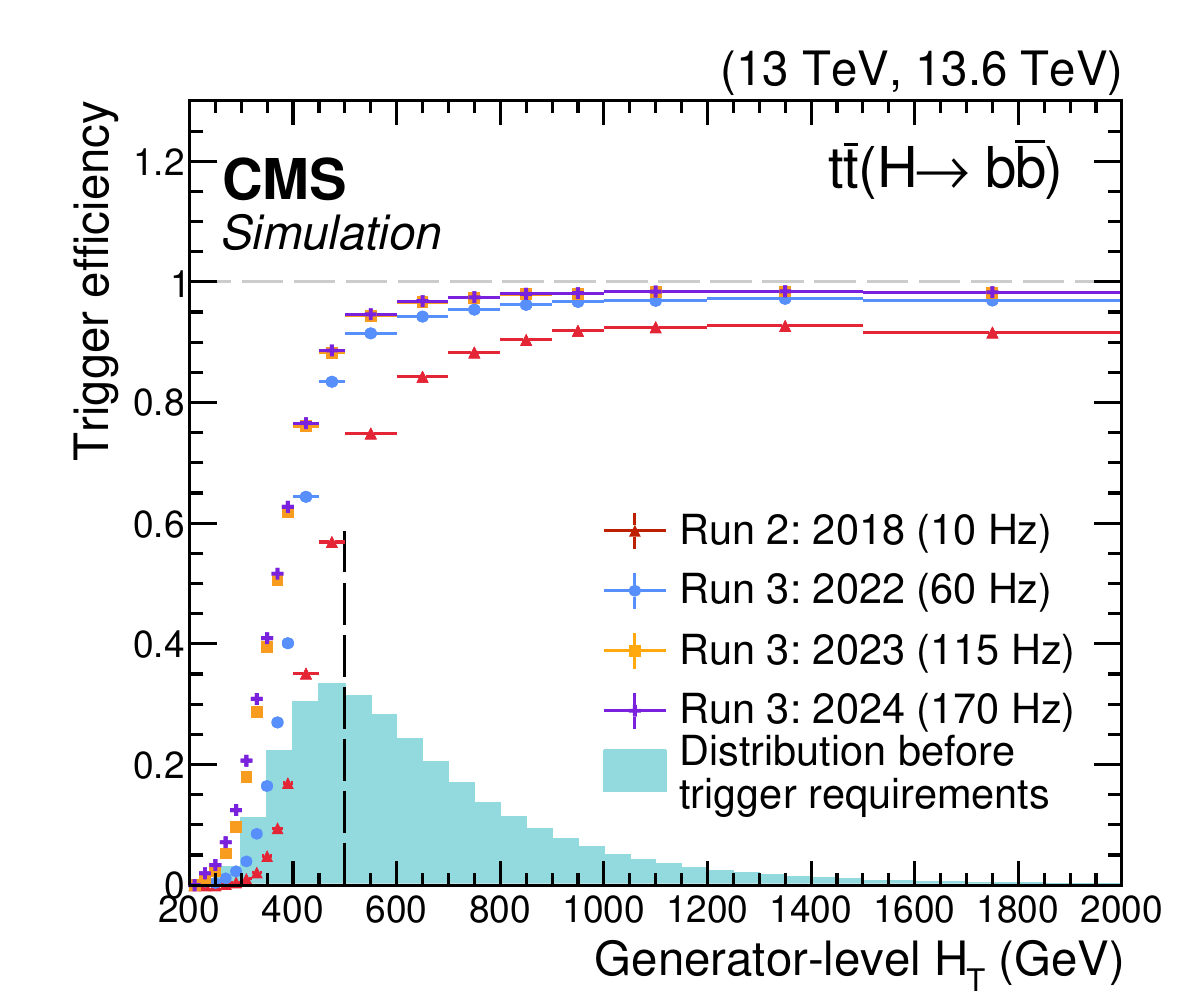}
    \includegraphics[width=0.45\textwidth]{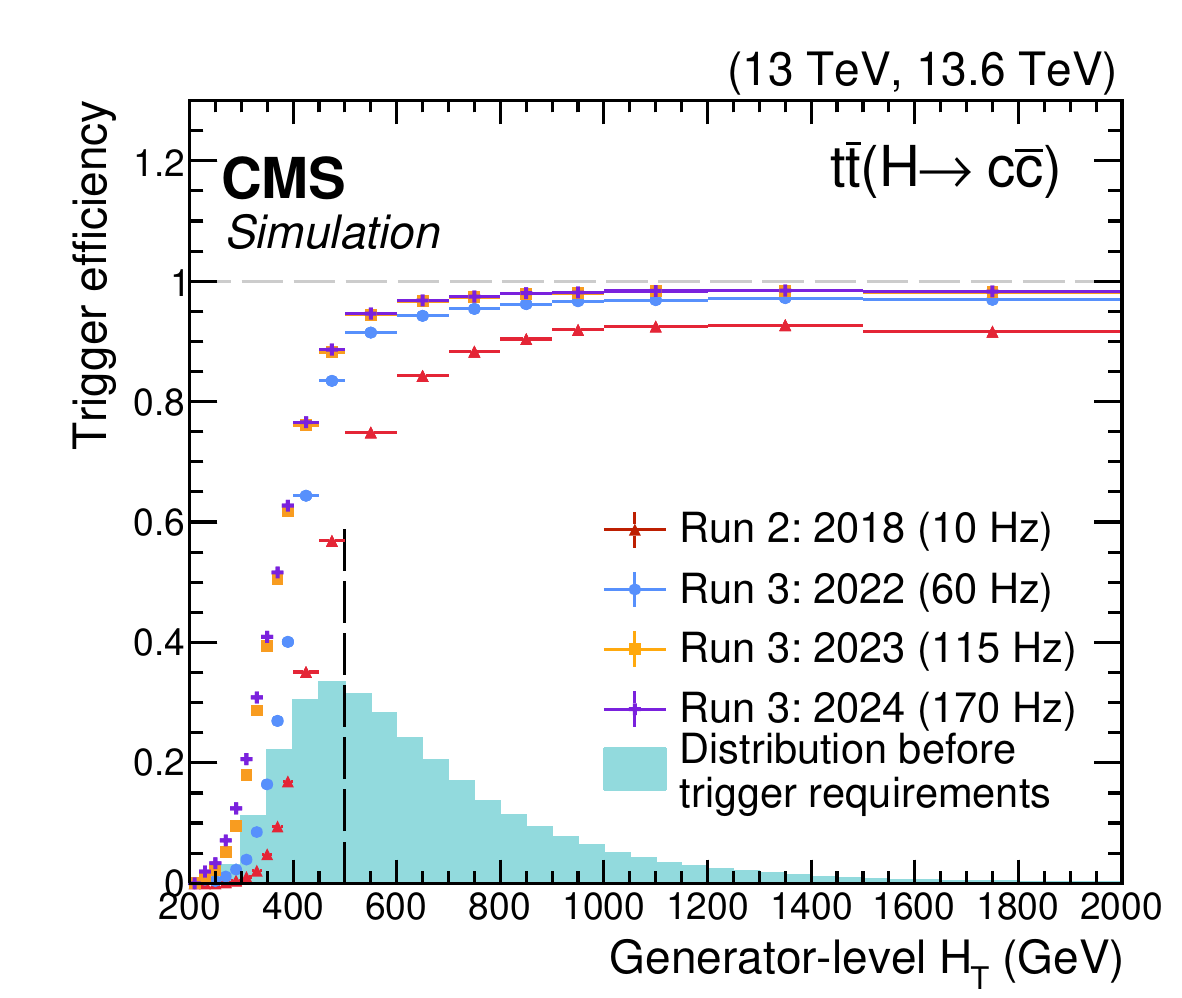}
    \caption{Efficiency of the Run~3 4j2b  trigger in simulated \ttHbb (\cmsLeft) and \ttHcc (\cmsRight) events, shown for 2022 (azure), 2023 (orange), and 2024 (purple) configurations as a function of generator-level \HT. The vertical error bars indicate the statistical uncertainty in the measured trigger efficiency. Six generator-level jets in the event are required to satisfy ${\pt>25\GeV}$ and ${\abs{\eta}<2.5}$. Results are compared to Run~2 triggers (red). The dashed line at ${\HT=500\GeV}$ indicates the Run~2 offline threshold. The corresponding trigger rates are 10, 60, 115, and 170\unit{Hz} for 2018, 2022, 2023, and 2024, respectively, at an instantaneous luminosity of ${{\approx}2\times10^{34}\percms}$. The teal histogram shows the expected distribution of the simulated events at ${\sqrt{s}=13.6\TeV}$ before trigger requirements, scaled by an arbitrary factor for visibility.}
    
    \label{fig:trigger_efficiency_ttH_gen_phys}
\end{figure*}

The same trigger also benefits other analyses with similar final states. In the measurement of \hbb and \hcc decays via $\ttbar\PH$ production, the fully hadronic final state achieves nearly the same sensitivity as the semileptonic channel, as reported in Refs.~\cite{CMS:2024fdo,CMS:2025dsh}. The Run~2 selection relied on multijet triggers requiring at least six AK4 jets, with at least one or two of them \PQb tagged. The Run~3 4j2b trigger applies looser requirements on the jet multiplicity, \PQb tagging, and \pt thresholds, leading to significant efficiency gains. As shown in Fig.~\ref{fig:trigger_efficiency_ttH_gen_phys}, the efficiency measured as a function of the generator-level \HT, computed in simulated $\ttbar\PH$ events with at least six generator-level jets with ${\pt>25\GeV}$ and ${\abs{\eta}<2.5}$, increases by 20\% (25\%), 28\% (36\%), and 30\% (37\%) for \ttHbb (\ttHcc) in 2022, 2023, and 2024, respectively, relative to the 2018 configuration. The gain is most pronounced in the turn-on region at ${\HT<700\GeV}$. The absolute efficiency is lower for \ttHcc (\cmsRight) than for \ttHbb (\cmsLeft) due to the reduced number of \PQb jets available for tagging.

Another important milestone in Higgs boson physics is the measurement of its couplings to second-generation fermions. While the most sensitive channel at the LHC is ${\PH\to\PGm\PGm}$~\cite{CMS:2020xwi}, recent searches for Higgs boson couplings to charm quarks in the $\PV\PH$~\cite{CMS:2022psv} and \ttH production~\cite{CMS:2025dsh} still leave open the possibility of observing \hcc decays at the high-luminosity LHC. In these channels, the presence of additional objects (leptons or multiple jets) in the final state is exploited to trigger the events. In contrast, searches in gluon-fusion ($\Pg\Pg\PH$) and VBF production are more challenging due to the lack of suitable triggers and a much higher contamination from hadronic backgrounds. As a result, no VBF \hcc search was performed with Run~2 data, and ${\Pg\Pg\hcc}$ studies were limited to the high-\pt regime~\cite{CMS:2022fxs}.

With the introduction of \pnet-based jet flavour tagging at the HLT in Run~3, a dedicated VBF \hcc trigger has been deployed since early 2023. It requires at least four AK4 jets with ${\pt>100}$, 88, 70, and 30\GeV and ${\abs{\eta}<4.7}$, including a jet pair exhibiting a large invariant mass (${\mjj>450\GeV}$) and a large pseudorapidity separation (${\abs{\detajj}>3.5}$), as expected for VBF production. In addition, at least one central jet ($\abs{\eta}<2.5$) must satisfy a \PQc tagging requirement corresponding to an average efficiency of about 40\%. The VBF-4j1c trigger operates within a rate of about $15\unit{Hz}$ at an instantaneous luminosity of ${{\approx 2}\times10^{34}\percms}$ and achieves an absolute efficiency of about 2\% on simulated VBF \hcc events, limited mainly by the jet \pt thresholds and the \probc selection.

\begin{figure*}[!htb]
    \centering
    \includegraphics[width=0.45\textwidth]{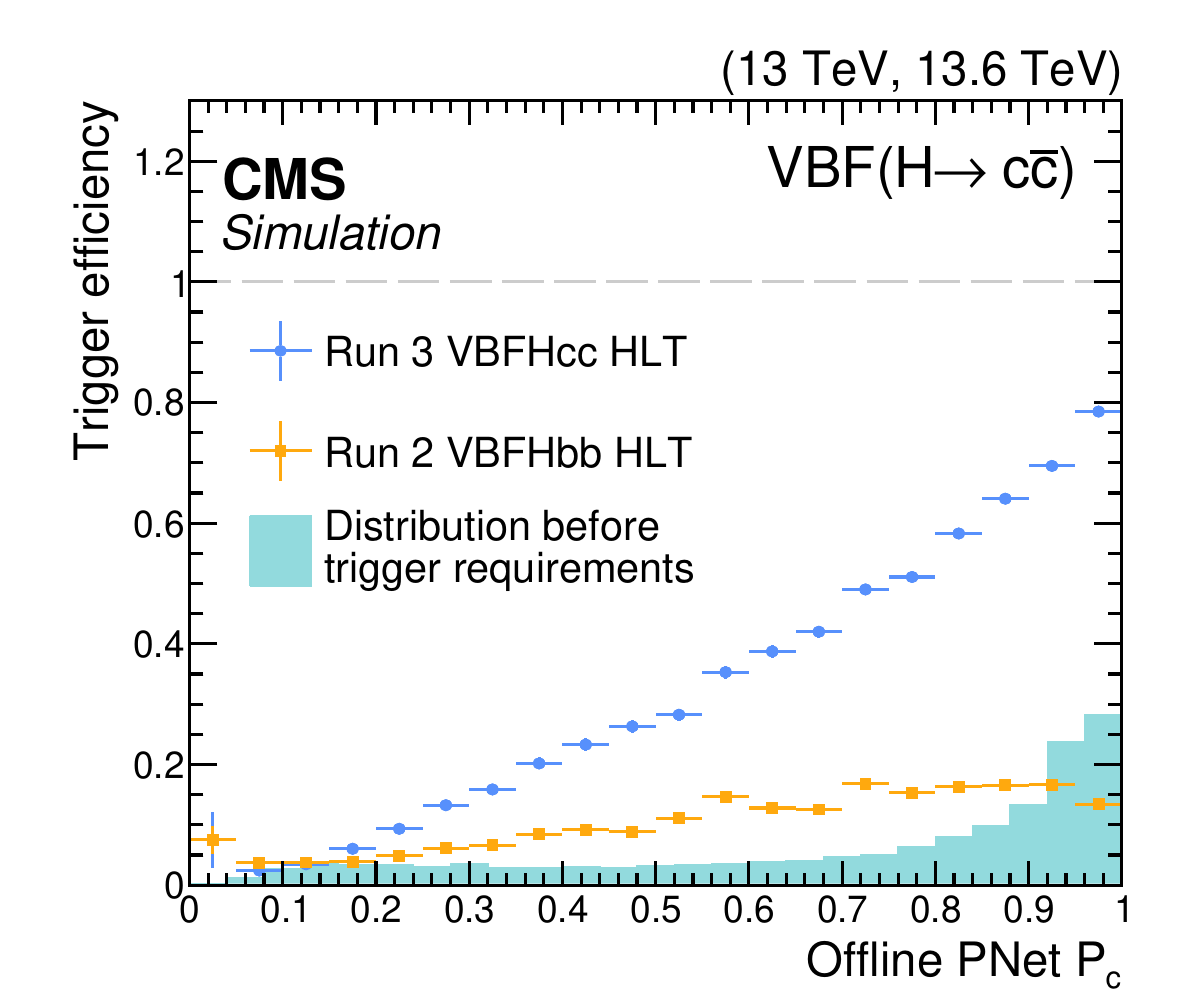}
    \includegraphics[width=0.45\textwidth]{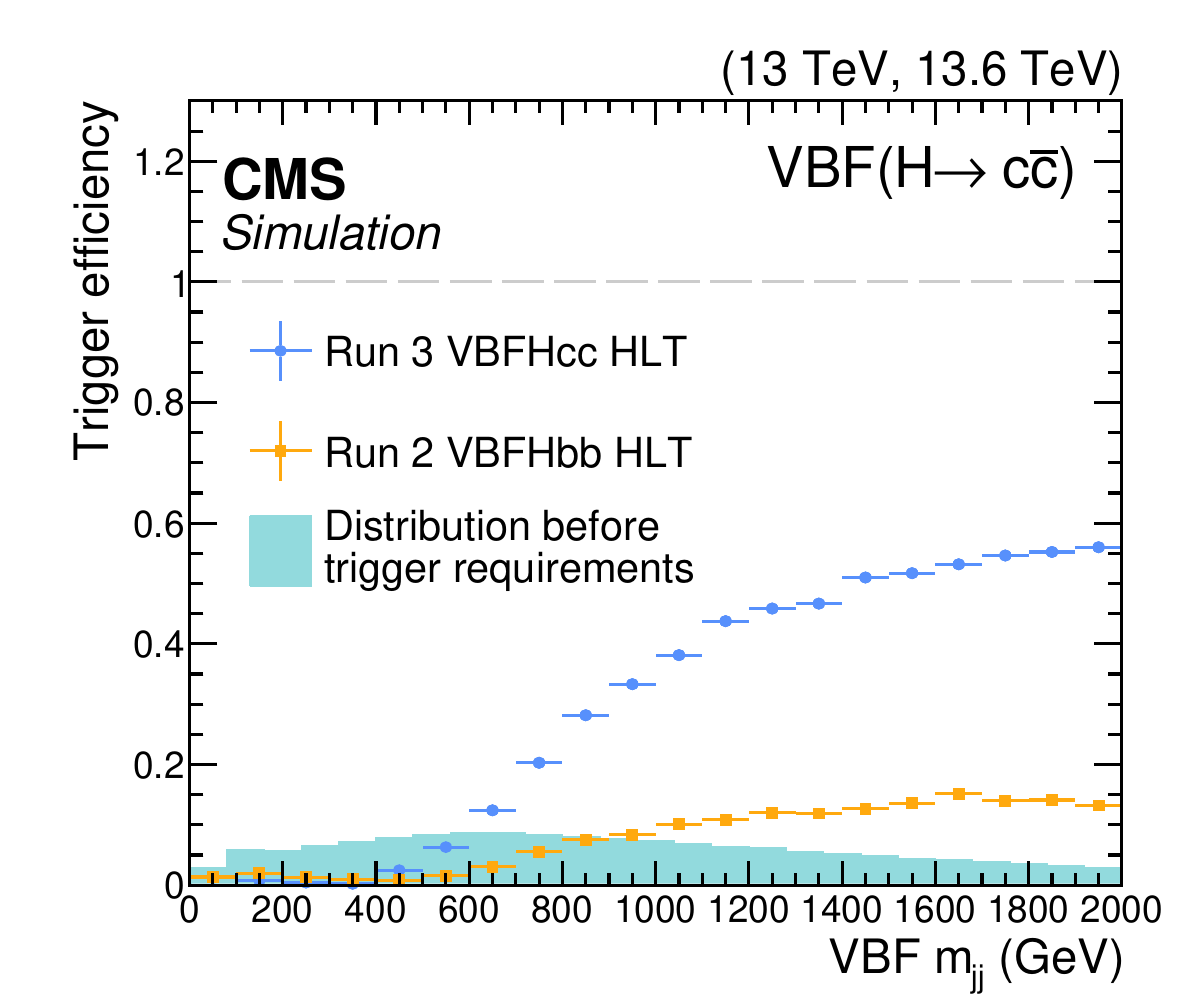}
    \caption{Efficiency of the VBF-4j1c trigger (blue) in simulated VBF \hcc events as a function of the offline \probc score of the leading \PQc tagged jet (\cmsLeft) and \mjj (\cmsRight). The vertical error bars indicate the statistical uncertainty in the measured trigger efficiency. Events require at least four AK4 jets with ${\abs{\eta}<4.7}$ and ${\pt>100}$, 88, 70, and 30\GeV. For the \cmsRight, at least one jet pair must satisfy ${\mjj>500\GeV}$ and ${\abs{\detajj}>4}$. Results are compared to the Run~2 VBF \hbb triggers~\cite{CMS:2023tfj} (orange). An improvement exceeding a factor of two is observed at large \mjj and high \probc. The teal histogram shows the expected distribution of the simulated events at ${\sqrt{s}=13.6\TeV}$ before trigger requirements, scaled by an arbitrary factor for visibility.}
    \label{fig:ctg_sf_signal}
\end{figure*}

Figure~\ref{fig:ctg_sf_signal} (\cmsLeft) shows the trigger efficiency as a function of \mjj in simulated VBF \hcc signal events, requiring at least four offline AK4 jets with ${\pt>110}$, 90, 65, and 30\GeV. In Run~2, such a search was not feasible because the available fully hadronic triggers targeting a VBF-like signature required at least one \PQb tagged jet~\cite{CMS:2023tfj}, significantly reducing the efficiency for VBF \hcc. Nevertheless, the Run~2 trigger efficiency obtained for the VBF \hcc events is also reported to showcase the substantial improvements achieved with the introduction of the VBF-4j1c trigger. The new trigger improves the efficiency by more than a factor of three for ${\mjj>1\TeV}$. Similarly, Fig.~\ref{fig:ctg_sf_signal} (\cmsRight) shows the efficiency as a function of the offline \probc score of the leading \PQc tagged jet, using the same selection employed before with the additional requirements of ${\mjj>500\GeV}$ and ${\abs{\detajj}>4}$. The efficiency increases with \probc, reaching about 80\% for ${\probc>0.95}$.
 
Several Higgs boson analyses exploit \hbb decays to measure differential cross sections at high \ptH, as the \hbb decay has the largest branching fraction. As \ptH increases, the angular separation of the two \PQb quarks decreases, and for ${\ptH > 350\GeV}$ they are typically reconstructed within a single AK8 jet. In the last year of Run~2 operations, triggers targeting high-\pt \hbb candidates relied on the \DoubleB tagger, whose limited performance required a jet \pt threshold of 330\GeV, reaching full efficiency only above 450\GeV. Consequently, this configuration significantly reduced the acceptance at the lower end of the merged \ptH regime. In Run~3, a new trigger was developed exploiting the improved performance of \pnet \xbb tagging described in Section~\ref{sec:ak8_jet_tagging}. In 2022, the trigger required at least one AK8 jet with ${\pt>250\GeV}$, ${\abs{\eta}<2.5}$, ${\mSD>40\GeV}$, and a selection on the \probXbb score. As reported in Table~\ref{tab:hlt_trigger_rate_run3}, the trigger operated at about 50\unit{Hz}, higher than initially expected. In 2023, a retrained \pnet model with improved discrimination enabled a significant reduction of the trigger rate. The trigger configuration was updated by lowering the jet \pt threshold to 230\GeV while tightening the \probXbb requirement, resulting in a small loss in efficiency at high \pt.

\begin{figure*}[!htb]
    \centering
    \includegraphics[width=0.45\textwidth]{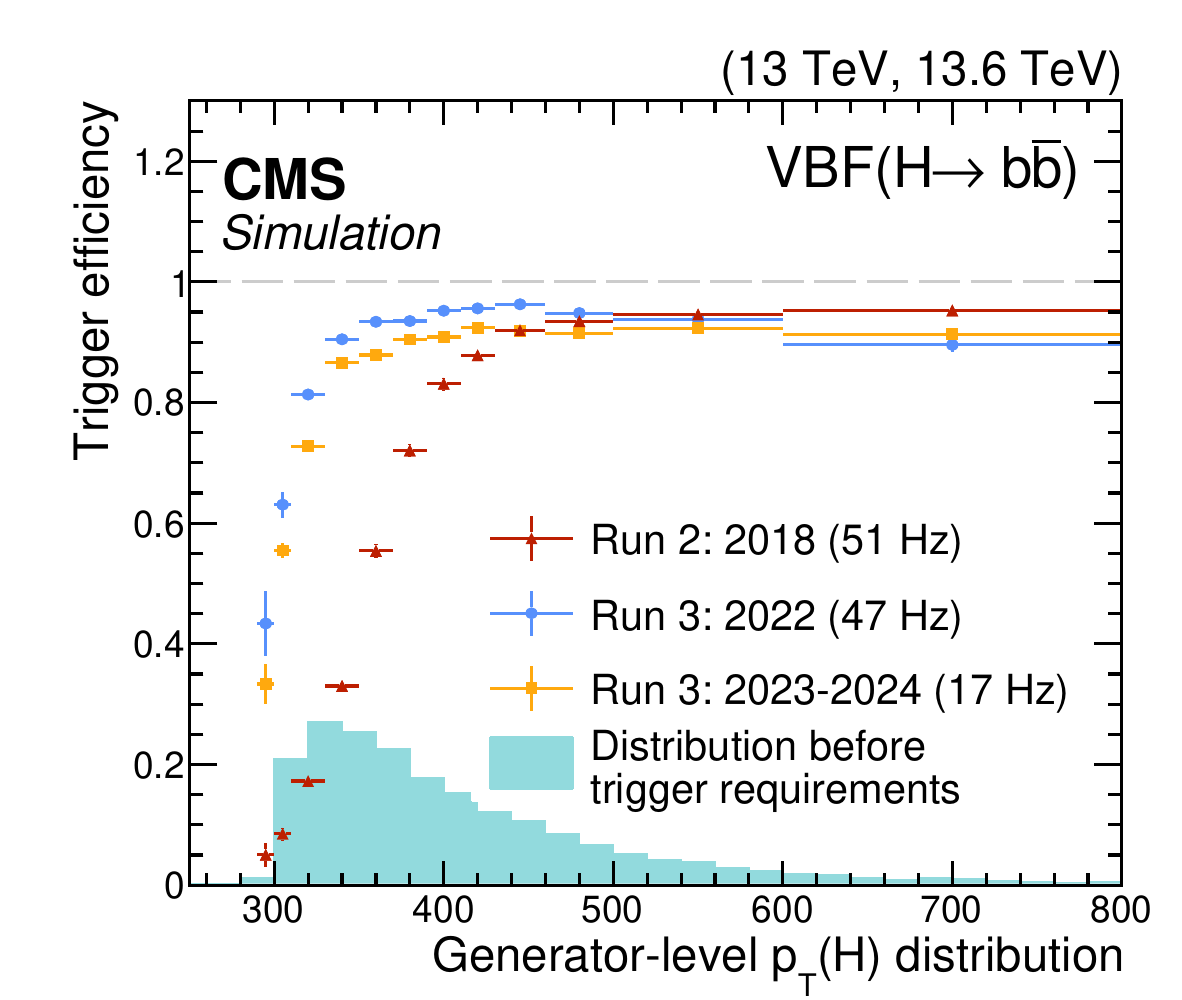}
    \includegraphics[width=0.45\textwidth]{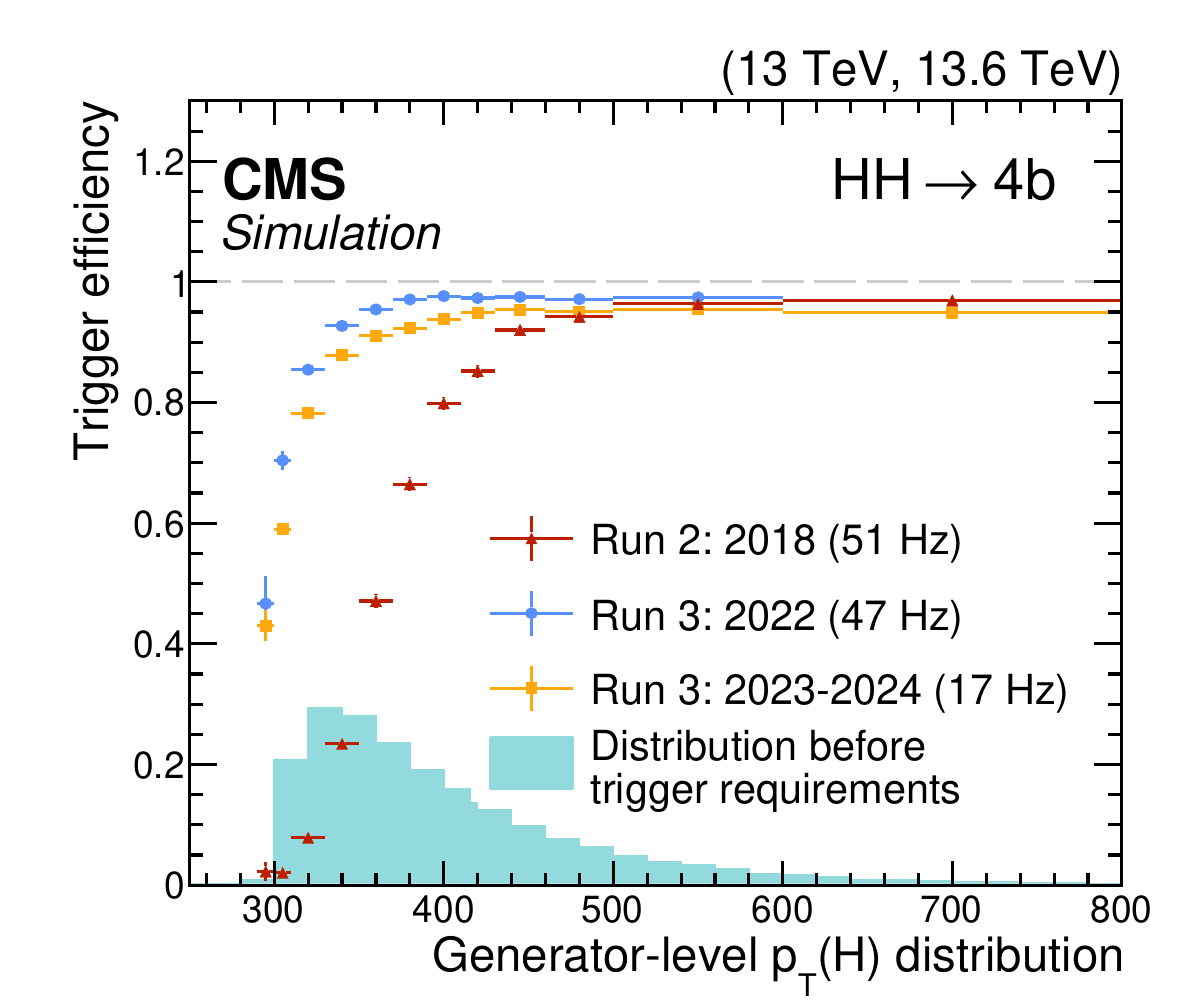}
    \caption{Efficiency of the Run~3 \xbb triggers in 2022 (azure) and 2023--24 (orange) in simulated VBF \hbb (\cmsLeft) and \hhbb (\cmsRight) events as a function of the generator-level Higgs boson \pt. The vertical error bars indicate the statistical uncertainty in the measured trigger efficiency. Only events in which at least one \hbb candidates has ${\Delta R(\PQb,\PAQb)<0.8}$ are considered. Results are compared to Run~2 (red). The corresponding trigger rates are 51, 47, and $17\unit{Hz}$ for 2018, 2022, and 2023, respectively, at an instantaneous luminosity of ${{\approx}2\times10^{34}\percms}$. The teal histogram shows the expected distribution of the simulated events at ${\sqrt{s}=13.6\TeV}$ before trigger requirements, scaled by an arbitrary factor for visibility.}
    \label{fig:trigger_efficiency_gen_phys_boosted}
\end{figure*}

Figure~\ref{fig:trigger_efficiency_gen_phys_boosted} compares the efficiencies of the 2022 (azure) and 2023--24 (orange) triggers with the Run~2 one (red) in simulated VBF \hbb (\cmsLeft) and \hhbb (\cmsRight) events, as a function of the generator-level Higgs boson \pt. Only events with ${\Delta R(\PQb,\PAQb)<0.8}$ are considered, selecting \hbb decays contained within a single AK8 jet. The Run~3 triggers reach near-full efficiency for ${\pt>350\GeV}$, approximately doubling the efficiency in the acceptance region relative to Run~2 for both benchmarks. This represents a major improvement for \hbb measurements in fully hadronic final states.

\section{Summary}\label{sec:summary}

The CMS trigger system plays a crucial role during data-taking, reducing the large collision rate delivered by the LHC to a few\unit{kHz} for data storage and subsequent offline analysis. The system aims at maintaining high selection efficiency, notably for processes involving jets from heavy-flavour (\PQb and \PQc) quarks which constitute a distinctive signature in many physics analyses. To achieve this in the challenging luminosity conditions of the LHC Run~3 while maintaining a sustainable trigger output rate, dedicated  jet flavour identification methods are developed and optimized for use in the CMS high-level trigger (HLT). This paper presents the design, performance, and commissioning of trigger algorithms based on novel deep-learning-based jet identification techniques deployed in the HLT during the LHC Run~3. The new algorithms enable high signal selection efficiency at fixed trigger rate for a broad range of physics targets, including non-resonant production of Higgs boson pairs decaying to four \PQb quarks, as well as Higgs boson production via vector boson fusion or in association with a \ttbar pair in the \hbb and \hcc decay channels.

\begin{acknowledgments}
We congratulate our colleagues in the CERN accelerator departments for the excellent performance of the LHC and thank the technical and administrative staffs at CERN and at other CMS institutes for their contributions to the success of the CMS effort. In addition, we gratefully acknowledge the computing centres and personnel of the Worldwide LHC Computing Grid and other centres for delivering so effectively the computing infrastructure essential to our analyses. Finally, we acknowledge the enduring support for the construction and operation of the LHC, the CMS detector, and the supporting computing infrastructure provided by the following funding agencies: SC (Armenia), BMFWF and FWF (Austria); FNRS and FWO (Belgium); CNPq, CAPES, FAPERJ, FAPERGS, and FAPESP (Brazil); MES and BNSF (Bulgaria); CERN; CAS, MoST, and NSFC (China); MINCIENCIAS (Colombia); MSES and CSF (Croatia); RIF (Cyprus); SENESCYT (Ecuador); ERC PRG and PSG, TARISTU24-TK10 and MoER TK202 (Estonia); Academy of Finland, MEC, and HIP (Finland); CEA and CNRS/IN2P3 (France); SRNSF (Georgia); BMFTR, DFG, and HGF (Germany); GSRI (Greece); MATE and NKFIH (Hungary); DAE and DST (India); IPM (Iran); SFI (Ireland); INFN (Italy); MSIT and NRF (Republic of Korea); MES (Latvia); LMTLT (Lithuania); MOE and UM (Malaysia); BUAP, CINVESTAV, CONACYT, LNS, SEP, and UASLP-FAI (Mexico); MOS (Montenegro); MBIE (New Zealand); PAEC (Pakistan); MSHE, NSC, and NAWA (Poland); FCT (Portugal);  MESTD (Serbia); MICIU/AEI and PCTI (Spain); MOSTR (Sri Lanka); Swiss Funding Agencies (Switzerland); MST (Taipei); MHESI (Thailand); TUBITAK and TENMAK (T\"{u}rkiye); NASU (Ukraine); STFC (United Kingdom); DOE and NSF (USA).

\begin{sloppypar}
\setlength\emergencystretch{\hsize}
\hyphenation{Rachada-pisek} Individuals have received support from the Marie-Curie programme and the European Research Council and Horizon 2020 Grant, contract Nos.\ 675440, 724704, 752730, 758316, 765710, 824093, 101115353, 101002207, 101001205, and COST Action CA16108 (European Union); the Leventis Foundation; the Alfred P.\ Sloan Foundation; the Alexander von Humboldt Foundation; the Science Committee, project no. 22rl-037 (Armenia); the Fonds pour la Formation \`a la Recherche dans l'Industrie et dans l'Agriculture (FRIA) and Fonds voor Wetenschappelijk Onderzoek contract No. 1228724N (Belgium); the Beijing Municipal Science \& Technology Commission, No. Z191100007219010, the Fundamental Research Funds for the Central Universities, the Ministry of Science and Technology of China under Grant No. 2023YFA1605804, the Natural Science Foundation of China under Grant No. 12535004, and USTC Research Funds of the Double First-Class Initiative No.\ YD2030002017 (China); the Ministry of Education, Youth and Sports (MEYS) of the Czech Republic; the Shota Rustaveli National Science Foundation (Georgia); the Deutsche Forschungsgemeinschaft (DFG), among others, under Germany's Excellence Strategy -- EXC 2121 ``Quantum Universe" -- 390833306, and under project number 400140256 - GRK2497; the Hellenic Foundation for Research and Innovation (HFRI), Project Number 2288 (Greece); the Hungarian Academy of Sciences, the New National Excellence Program - \'UNKP, the NKFIH research grants K 131991, K 138136, K 143460, K 143477, K 147557, K 146913, K 146914, K 147048, TKP2021-NKTA-64, and 2025-1.1.5-NEMZ\_KI-2025-00004, and MATE KKP and KKPCs Research Excellence and Flagship Research Groups grants (Hungary); the Council of Science and Industrial Research, India; ICSC -- National Research Centre for High Performance Computing, Big Data and Quantum Computing, FAIR -- Future Artificial Intelligence Research, and CUP I53D23001070006 (Mission 4 Component 1), funded by the NextGenerationEU program, the Italian Ministry of University and Research (MUR) under Bando PRIN 2022 -- CUP I53C24002390006, PRIN PRIMULA 2022RBYK7T (Italy); the Latvian Council of Science; the Ministry of Science and Higher Education, project no. 2022/WK/14, and the National Science Centre, contracts Opus 2021/41/B/ST2/01369, 2021/43/B/ST2/01552, 2023/49/B/ST2/03273, and the NAWA contract BPN/PPO/2021/1/00011 (Poland); the Funda\c{c}\~ao para a Ci\^encia e a Tecnologia (Portugal); the National Priorities Research Program by Qatar National Research Fund; MICIU/AEI/10.13039/501100011033, ERDF/EU, ``European Union NextGenerationEU/PRTR", projects PID2022-142604OB-C21, PID2022-139519OB-C21, PID2023-147706NB-I00, PID2023-148896NB-I00, PID2023-146983NB-I00, PID2023-147115NB-I00, PID2023-148418NB-C41, PID2023-148418NB-C42, PID2023-148418NB-C43, PID2023-148418NB-C44, PID2024-158190NB-C22, RYC2021-033305-I, RYC2024-048719-I, CNS2023-144781, CNS2024-154769 and Plan de Ciencia, Tecnolog{\'i}a e Innovaci{\'o}n de Asturias, Spain; the Chulalongkorn Academic into Its 2nd Century Project Advancement Project, the National Science, Research and Innovation Fund program IND\_FF\_68\_369\_2300\_097, and the Program Management Unit for Human Resources \& Institutional Development, Research and Innovation, grant B39G680009 (Thailand); the Eric \& Wendy Schmidt Fund for Strategic Innovation through the CERN Next Generation Triggers project under grant agreement number SIF-2023-004; the Kavli Foundation; the Nvidia Corporation; the SuperMicro Corporation; the Welch Foundation, contract C-1845; and the Weston Havens Foundation (USA).
\end{sloppypar}
\end{acknowledgments}\section*{Data availability} Release and preservation of data used by the CMS Collaboration as the basis for publications is guided by the  \href{https://doi.org/10.7483/OPENDATA.CMS.1BNU.8V1W}{CMS data preservation, re-use and open access policy}.

\bibliography{auto_generated}
\cleardoublepage \appendix\section{The CMS Collaboration \label{app:collab}}\begin{sloppypar}\hyphenpenalty=5000\widowpenalty=500\clubpenalty=5000\cmsinstitute{Yerevan Physics Institute, Yerevan, Armenia}
{\tolerance=6000
A.~Gevorgyan\cmsorcid{0000-0003-2751-9489}, A.~Hayrapetyan, V.~Makarenko\cmsorcid{0000-0002-8406-8605}, A.~Tumasyan\cmsAuthorMark{1}\cmsorcid{0009-0000-0684-6742}
\par}
\cmsinstitute{Institut f\"{u}r Hochenergiephysik, Vienna, Austria}
{\tolerance=6000
P.S.~Hussain\cmsorcid{0000-0002-4825-5278}, M.~Sonawane\cmsorcid{0000-0003-0510-7010}
\par}
\cmsinstitute{Marietta Blau Institute for Particle Physics, Vienna, Austria}
{\tolerance=6000
W.~Adam\cmsorcid{0000-0001-9099-4341}, L.~Benato\cmsorcid{0000-0001-5135-7489}, T.~Bergauer\cmsorcid{0000-0002-5786-0293}, M.~Dragicevic\cmsorcid{0000-0003-1967-6783}, S.~Gundacker\cmsorcid{0000-0003-2087-3266}, A.K.~Guven\cmsorcid{0009-0004-5670-5138}, M.~Jeitler\cmsAuthorMark{2}\cmsorcid{0000-0002-5141-9560}, N.~Krammer\cmsorcid{0000-0002-0548-0985}, A.~Li\cmsorcid{0000-0002-4547-116X}, D.~Liko\cmsorcid{0000-0002-3380-473X}, M.~Matthewman, A.~Pfeiffer\cmsorcid{0000-0001-5328-448X}, J.~Schieck\cmsAuthorMark{2}\cmsorcid{0000-0002-1058-8093}, R.~Sch\"{o}fbeck\cmsAuthorMark{2}\cmsorcid{0000-0002-2332-8784}, M.~Shooshtari\cmsorcid{0009-0004-8882-4887}, N.~Van~Den~Bossche\cmsorcid{0000-0003-2973-4991}, W.~Waltenberger\cmsorcid{0000-0002-6215-7228}, C.E.~Wulz\cmsAuthorMark{2}\cmsorcid{0000-0001-9226-5812}
% ...
\par}
$^{1}$Also at Yerevan State University, Yerevan, Armenia\\
$^{2}$Also at Technische Universit\"{a}t Wien, Vienna, Austria\\
% ...
\end{sloppypar}
\end{document}